\documentclass[journal=jctc,manuscript=article]{achemso}
\setkeys{acs}{maxauthors=0,articletitle=true}

\usepackage{achemso}
\usepackage{graphics}
\usepackage{amssymb,amsfonts}
\usepackage{graphicx}
\usepackage[table]{xcolor}
\usepackage{multirow}
\usepackage{caption}
\usepackage{subcaption}
\usepackage{booktabs}
\usepackage{colortbl}
\usepackage{amsmath}
\usepackage{amsopn}
\usepackage{bm}
\usepackage{siunitx}
\usepackage{bm}
\usepackage{color}
\usepackage{array}
\usepackage{lscape}
\usepackage{mciteplus}
\usepackage[version=3]{mhchem}
\usepackage{ulem}
\usepackage{listings}
\usepackage{enumerate}
\usepackage{lmodern}

\SectionNumbersOn
\DeclareMathOperator{\tr}{Tr}

\author{Janus J. Eriksen}
\email{janus.eriksen@bristol.ac.uk}
\affiliation[University of Bristol]
{School of Chemistry, University of Bristol, Cantock's Close, Bristol BS8 1TS, United Kingdom}

\title[TITLE]{Mean-Field Density Matrix Decompositions}

\begin{document}

%
%
\begin{abstract}

We introduce new and robust decompositions of mean-field Hartree-Fock (HF) and Kohn-Sham density functional theory (KS-DFT) relying on the use of localized molecular orbitals and physically sound charge population protocols. The new lossless property decompositions, which allow for partitioning 1-electron reduced density matrices into either bond-wise or atomic contributions, are compared to alternatives from the literature with regards to both molecular energies and dipole moments. Besides commenting on possible applications as an interpretative tool in the rationalization of certain electronic phenomena, we demonstrate how decomposed mean-field theory makes it possible to expose and amplify compositional features in the context of machine-learned quantum chemistry. This is made possible by improving upon the granularity of the underlying data. On the basis of our preliminary proof-of-concept results, we conjecture that many of the structure-property inferences in existence today may be further refined by efficiently leveraging an increase in dataset complexity and richness.

\end{abstract}

\newpage

%
%
\section{Introduction}\label{intro_sect}

The immense predictive powers of wave function-based quantum mechanics notwithstanding---as offered by, e.g., standard coupled cluster~\cite{cizek_1,cizek_2,paldus_cizek_shavitt,shavitt_bartlett_cc_book,mest} (CC) theory and the recent myriad of near-exact approximations to full configuration interaction~\cite{booth_alavi_fciqmc_jcp_2009,cleland_booth_alavi_fciqmc_jcp_2010,piecuch_monte_carlo_cc_prl_2017,sharma_umrigar_heat_bath_ci_jctc_2017,white_dmrg_prl_1992,white_dmrg_prb_1993,white_martin_dmrg_jcp_1999,chan_sharma_dmrg_review_arpc_2011,eriksen_mbe_fci_jpcl_2017,eriksen_mbe_fci_general_jpcl_2019,eriksen_benzene_jpcl_2020} (FCI) theory---the Kohn-Sham formulation of density functional theory~\cite{hohenberg_kohn_hk_theorem_phys_rev_1964,kohn_sham_ks_dft_phys_rev_1965,parr_yang_dft_book} (KS-DFT) has, by now, arguably manifested itself as the most pragmatic, versatile, and altogether functional tool in existence today, applicable for molecules~\cite{yang_dft_review_chem_rev_2012,becke_dft_review_jcp_2014,burke_dft_review_jcp_2012,burke_dft_review_arpc_2015,mardirossian_head_gordon_dft_review_mol_phys_2017} as well as solids~\cite{heine_2d_atlas_chem_soc_rev_2014,thygesen_2d_database_jpcc_2015,coudert_nanoporous_mat_chem_mater_2017,maurer_tkatchenko_dft_materials_ann_rev_mat_res_2019,schleder_fazzio_ml_materials_jpm_2019}. However, despite its reduced computational scaling~\cite{hernandez_gillan_prb_1995,bowler_gillan_j_phys_cond_matter_2002,skylaris_payne_onetep_cpc_2009}, the routine and reliable application of KS-DFT to extended systems remain challenging. This is true also for its uncorrelated mean-field sibling, Hartree-Fock (HF) theory, and although recent innovations have contributed positively towards enhancing the general application range of mean-field methods~\cite{hoyvik_multilevel_hf_jctc_2017,brandenburg_grimme_b97_3c_jcp_2018,brandenburg_simpl_dft_jpcm_2018,koch_multilevel_dft_arxiv_2020}, the past few years have seen an increasing interest in applying modern machine learning (ML) techniques as an alternative, in an attempt at mitigating the problems associated with the prohibitive scaling wall~\cite{bartol_csanyi_ml_qc_review_ijqc_2015,rupp_ml_qc_review_ijqc_2015,rupp_lilienfeld_burke_jcp_2018,lilienfeld_burke_nat_commun_2020,ceriotti_ml_qc_review_jcp_2019,dral_ml_qc_review_jpcl_2020}. In the course of the present work, a new and robust decomposition of mean-field theory will be introduced, which, we will argue, exhibits the potential to make KS-DFT (for a given choice of exchange-correlation ({\it{xc}}) functional) increasingly befitting to ML, with an aim at reducing its computational complexity even further. By explicitly incorporating electronic structure effects, we will further investigate to what extent the present decompositions may aid in the design and deployment of modern machine mappings between the geometrical arrangement of atoms alone and total inferred molecular quantities. As such, despite the overall concept of a marriage of ML and KS-DFT in common, the present work will not be concerned with the intelligent design of universal $xc$ functionals, which is another area of research that has experienced a true surge in interest in recent years~\cite{mueller_burke_ml_dft_jcp_2012,tuckerman_burke_mueller_ml_dft_nature_comm_2017,tuckerman_burke_mueller_ml_dft_nature_comm_2020,li_burke_ml_dft_arxiv_2020,hauser_ml_dft_jctc_2020,parkhill_kinetic_energy_jctc_2016,hegde_bowen_ml_dft_sci_reports_2017,tamblyn_ml_dft_pra_2019,fernandez_serra_ml_dft_jcp_2019,fernandez_serra_ml_dft_nat_commun_2020,lilienfeld_ml_dft_dispersion_jctc_2020,pande_ml_dft_arxiv_2018,pande_ml_dft_arxiv_2019,ryabov_zhilyaev_ml_dft_sci_rep_2020,lei_medford_ml_dft_phys_rev_mat_2019,ramprasad_ml_dft_npj_2019,chen_ml_dft_jpcl_2019,nagai_akashi_sugino_ml_dft_npj_2020,weinan_e_deepks_arxiv_2020}.\\

Starting from the mean-field functional shared between HF and KS-DFT in a basis of spatially localized molecular orbitals (MOs), we will describe and numerically illustrate how to partition the central 1-electron reduced density matrix at several different levels, by which total molecular properties may be decomposed into contributions associated with individual bonds and atomic centres. Upon introducing the new decompositions---in comparison with analogous models from the literature---our work will be concerned with how these succeed in the interpretation of physical properties (e.g., energies and dipole moments).\\

Besides stoichiometric composition (i.e., identity), the exercise of succinctly representing the immediate local structural neighbourhoods of all the atoms constituting a molecular system has been one of the most traversed areas of research in the application of ML to quantum chemistry (ML-QC) over the past decade~\cite{artrith_urban_ceder_ml_representation_prb_2017,lilienfeld_yaron_ml_representations_jcp_2018,schrier_ml_representation_jcim_2020,henkelman_ml_ann_jcp_2020,onat_ortner_kermode_ml_representation_jcp_2020,ceriotti_ml_representation_jcp_2020,hammer_ml_representation_jcp_2020}, for instance, in paving the way for its use as a tool to explore chemical compound space~\cite{ceriotti_chem_space_pccp_2016,isayev_roitberg_chem_space_jcp_2018,lilienfeld_qml_angew_chem_2018,lilienfeld_mueller_tkatchenko_qml_nat_rev_chem_2020,tkatchenko_ml_qc_nat_commun_2020}. Ultimately, the efficacy of any ML-QC model will be bound by how input data are passed to and manipulated by the underlying machine algorithm, and thus less so by exactly which regressor the data are processed. As modern structure-based descriptors operate by encoding the essential features of a molecular compound in a tensorial, machine-readable form, molecular similarity between different systems can be measured on the basis of so-called kernels of these, in the absence of any explicit simulations of the electronic structure. To that end, the ability of molecular representations to relay chemical information at a sufficient level of sophistication---considering the scale of the dimensionality reduction involved---follows foremost from a central uniqueness criterion, that is, an inductive bias, amounting to a number of required invariances with respect to index permutations, spatial rotations, as well as translations of same-element atoms, in addition to an overall requirement of smoothness~\cite{yang_gao_ml_qc_review_jpca_2020}. In the present work, we will further seek to test a hypothesis related to data granularity, namely, to what extent the compositional complexity and dimensionality of the data underpinning the mapping from structure to property may be refined to improve upon current- and next-generation ML-QC models.\\

Arguably the first successful example of ML-QC was and continues to be in its use as non-linear interpolations of interatomic potentials and for fitting potential energy surfaces~\cite{sumpter_noid_ml_pes_cpl_1992,ho_rabitz_ml_pes_jcp_1996,lorenz_gross_scheffler_ml_pes_cpl_2004,parrinello_ml_pes_prl_2007,braams_bowman_ml_pes_irpc_2009,agrawal_komanduri_ml_pes_jcp_2009,handley_popelier_ml_pes_jpca_2010,csanyi_ml_pes_prl_2010,carrington_ml_pes_jcp_2006,carrington_ml_pes_jcp_2007,carrington_ml_pes_ijqc_2015,carrington_ml_pes_chem_rev_2020,behler_ml_pes_prb_2011,behler_ml_pes_jpca_2013,behler_ml_qc_jcp_2016,behler_marx_ml_qc_jctc_2020,behler_ml_qc_arxiv_2020,marquetand_ml_wacsf_jcp_2018,christiansen_ml_pes_jcp_2019_1,christiansen_ml_pes_jcp_2019_2,christiansen_ml_pes_jcp_2020,sumpter_irle_ml_pes_mrs_comm_2019,glowacki_nn_pes_jpca_2019,dral_thiel_ml_pes_jcp_2017,dral_thiel_ml_pes_jpcl_2018,dral_csanyi_ml_pes_jcp_2020,bernstein_csanyi_deringer_npj_comp_mater_2019,csanyi_ortner_ml_pes_mlst_2020,csanyi_ortner_ml_pes_arxiv_2020,koner_meuwly_ml_pes_jctc_2020}. Limiting our discussion herein to descriptors that have been designed with chemical Hamiltonians in mind, rather than applications in the solid state or condensed-matter physics more generally~\cite{coudert_ml_zeolites_chem_mater_2017,isayev_curtarolo_crystals_chem_mater_2015,walsh_ml_mat_nature_2018,botu_ml_ff_jpcc_2017,lilienfeld_armiento_ml_atom_energy_prl_2016,tkatchenko_mueller_ml_ff_sci_adv_2017,mueller_tkatchenko_ml_ff_nature_comm_2018,tkatchenko_mueller_ml_ff_arxiv_2020,li_kermode_de_vita_ml_ff_prl_2015,day_ceriotti_ml_crystals_chem_sci_2018,csanyi_ml_ff_prx_2018,deringer_csanyi_silicon_jpcl_2018,csanyi_deringer_silicon_angew_chem_2019,csanyi_ml_pes_jctc_2019,podryabinkin_ml_crystals_prb_2019,csanyi_ml_ff_faraday_2020,kresse_ml_ff_prb_2019,heine_uff4mof_jctc_2016,heine_woell_mof_chem_eur_journ_2019,kresse_asahi_ml_qc_review_jpcl_2020,rupp_ml_qc_review_arxiv_2020}, recent examples of atomic representations include the atom-centered symmetry functions by Behler~\cite{behler_ml_acsf_jcp_2011}, the related many-body representations proposed by the von Lilienfeld, Tkatchenko, and M{\"u}ller groups~\cite{lilienfeld_ml_atom_energy_prl_2012,tkatchenko_mueller_lilienfeld_ml_atom_energy_njp_2013,lilienfeld_mueller_tkatchenko_ml_atom_energy_jpcl_2015,lilienfeld_fchl_jcp_2018,lilienfeld_fchl_jcp_2020,lilienfeld_ml_qc_mlst_2020}, the smooth overlap of atomic positions (SOAP) representation and its derivatives by Cs{\'a}nyi and Ceriotti~\cite{csanyi_ml_soap_prb_2013,ceriotti_ml_materials_sci_adv_2017,ceriotti_ml_long_range_jcp_2019,ceriotti_ml_qc_jcp_2019}, the overlap matrix (OM) representation by Goedecker {\textit{et al.}}~\cite{goedecker_ml_crystal_jcp_2016}, and the use of persistence images as an alternative for representing the homological features of a molecular system, as recently introduced by Vogiatzis and co-workers~\cite{vogiatzis_ml_qc_nat_commun_2020}. With the exception of the latter two examples, these modern molecular descriptors all seek to explicitly account for interatomic interactions by including physically motivated pairwise and many-body terms, preferably in as accurate, efficient, and transferrable a manner as possible~\cite{lilienfeld_qml_jcp_2016,tkatchenko_mueller_ml_many_body_jctc_2018}. As part of the present work, we will explore to which extent atom-based ML-QC representations succeed in capturing effects that are strictly quantum in nature, that is, not merely dependent on atomic positions and nuclear charges alone. Meanwhile, we will attempt to probe if the results of our proposed decompositions may serve to refine existing representations or even drive new developments.\\

Our proposed target of atom-centric ML-QC thus bears some resemblance to a number of contemporary endeavours in the scientific literature. In a recent approach by Huang and von Lilienfeld~\cite{huang_lilienfeld_ml_amons_nat_chem_2020,huang_lilienfeld_ml_amons_arxiv_2020}, the authors set out to harvest transferrable features between functional groups in different molecules, akin to traditional and machine-learned fragmentation schemes~\cite{gordon_slipchenko_chem_rev_2012,isayev_tropsha_crystals_nat_commun_2017,eckhoff_behler_mof_jctc_2019,skylaris_day_ml_crystals_jctc_2019}, but formulated on a foundation of Bayesian inference. The successful learning of such repetitive constituents that sum up to total properties admits not only a circumvention of the compositional scaling wall following from the combinatorial growth of chemical compound space, but also to achieve ML-QC within a more generalizable learning framework. Along a somewhat similar tangent, artificial neural networks (NNs) have been proposed as a means to provide a statistically rigorous partitioning of extensive molecular properties into atomic contributions~\cite{isayev_roitberg_ml_qc_chem_sci_2017,isayev_roitberg_ml_qc_nat_commun_2019,hammer_ml_qc_jctc_2018,hammer_ml_qc_jcp_2018,hammer_ml_qc_jcp_2019,unke_meuwly_ml_nn_jcp_2018}. In here, we will instead propose an alternative route towards this goal, namely, one that proceeds through an intermediate basis of localized MOs. The present work thus positions itself somewhere in-between the two approaches to transferrable ML-QC discussed here. On the basis of an atomic descriptor, we will seek to learn the magnitude of the corresponding atomic contribution, before adding these up to arrive at a final property. On par with standard kernel-based ML-QC, which constructs molecular kernels as a sum of pair-wise atomic kernels, we may further employ the finer granularity of our training data---as offered by the present decompositions---to learn componential rather than atomic contributions, whenever appropriate. The learning of such intensive, rather than extensive, contributions itself warrants a generalization of the learning process, reminiscent of a recent approach to the design of NN-based force fields where these are constructed from energy decompositions for homogeneous, solid-state systems~\cite{kang_wang_dft_decomp_prb_2017,huang_wang_nnff_prb_2019}. However, due to our formulation in a basis of spatially localized MOs, we will numerically illustrate how the present decompositions of the 1-electron reduced density matrix will allow for physically interpretable and transferrable atomic contributions for heterogeneous, multicomponent systems as well.\\

The present work will be organized as follows. In Section \ref{theory_sect}, we start by outlining the theory behind our mean-field decompositions, which are next numerically compared to alternatives from the literature in Section \ref{calibration_section}. In the course of these comparisons, we will further comment on applications of the theory outside its use in ML-QC, which is the topic of Section \ref{qml_section}. Finally, Section \ref{outlook_summary_section} presents a number of conclusive remarks as well as future prospects.

%
%
\section{Theory}\label{theory_sect}

In the following, we will discuss how to decompose total energies, while noting that other first-order properties may be treated likewise, as we will touch upon in Section \ref{calibration_water_subsection}. The starting point is the mean-field (MF) functional shared between HF and KS-DFT,
\begin{align}
E_{\text{MF}}(\bm{D}) = \sum_{\sigma=\alpha,\beta}(\tr[\bm{h}_{\text{core}}\bm{D}_{\sigma}] + \tfrac{1}{2}\tr[\bm{G}_{\sigma}(\bm{D})\bm{D}_{\sigma}]) + h_{\text{nuc}} \ (+ E_{xc}(\bm{D})) \ , \label{mf_energy_eq}
\end{align}
defined in terms of converged 1-electron reduced density matrices (RDM1s), $\bm{D}_{\sigma} = \bm{C}_{\sigma}\bm{C}_{\sigma}^T$, in turn obtained from the complete sets of $\mathcal{N}_{\alpha}$ and $\mathcal{N}_{\beta}$ occupied molecular spin-orbitals (MOs), $\bm{C}_{\sigma}$. When written without a spin subscript ($\sigma$), the RDM1 is assumed spin-summed, i.e., $\bm{D} = \bm{D}_{\alpha} + \bm{D}_{\beta}$. In Eq. \ref{mf_energy_eq}, $\bm{h}_{\text{core}} = \bm{T}_{\text{kin}} + \bm{V}_{\text{nuc}}$ is the core Hamiltonian (with the parametric dependence of the kinetic energy, $\bm{T}_{\text{kin}}$, and nuclear attraction, $\bm{V}_{\text{nuc}}$, operators on electronic and nuclear coordinates suppressed), $h_{\text{nuc}}$ is the scalar internuclear repulsion energy between a system's $\mathcal{M}_{\text{atom}}$ atoms, $E_{xc}(\bm{D})$ is the $xc$ energy exclusive to KS-DFT, while $\bm{G}_{\sigma}(\bm{D})$ denotes an effective Fock potential, $\bm{G}_{\sigma}(\bm{D}) = \bm{J}(\bm{D}_{\alpha}) + \bm{J}(\bm{D}_{\beta}) - \alpha\bm{K}(\bm{D}_{\sigma})$, in terms of Coulomb, $\bm{J}$, and exchange, $\bm{K}$, integrals. The exchange ratio is $\alpha \equiv 1.0$ in HF, while $0 < \alpha$ only for {\it{xc}} functionals at the hybrid level or higher~\cite{perdew_jacobs_ladder_aip_conf_proc_2001}. For brevity, attention will be focussed on HF theory in the subsections to follow. However, it is imperative to emphasize the comparability with KS-DFT, as the $xc$ energy in Eq. \ref{mf_energy_eq} may itself be decomposed in a similar manner by partitioning the total electronic density (see below).

\subsection{Bond Decompositions}\label{bonds_subsection}

It is now noted how the total energy may be decomposed into a sum of contributions specific to the individual occupied MOs ($\bm{C}_{i}$) of the system via orbital-specific RDM1s (orb-RDM1s), which are defined as $\bm{d}_{i,\sigma} = \bm{C}_{i,\sigma}\bm{C}^T_{i,\sigma}$. The HF energy from Eq. \ref{mf_energy_eq} then reads
\begin{align}
E_{\text{HF}}(\bm{D},\bm{d}) = \sum_{\sigma}\sum^{\mathcal{N}_{\sigma}}_{i}(\tr[\bm{h}_{\text{core}}\bm{d}_{i,\sigma}] + \tfrac{1}{2}\tr[\bm{G}_{\text{HF},\sigma}(\bm{D})\bm{d}_{i,\sigma}]) + h_{\text{nuc}} \ . \label{orb_decomp_mo_eq}
\end{align}
In KS-DFT, the $xc$ energy is expressed in terms of the associated energy density, $\epsilon_{xc}$, which is derived from the total electronic density, $\bm{\rho}$, and possibly its derivatives, depending on the chosen $xc$ functional. As $\bm{\rho}$ may, in turn, be computed from $\bm{D}$, we are free to express corresponding orbital-specific densities, $\{\bm{\varrho}\}$, in terms of $\{\bm{d}\}$, allowing for the $xc$ energy to be decomposed on par with the HF energy in Eq. \ref{orb_decomp_mo_eq},
\begin{align}
E_{xc}(\bm{\rho},\bm{\varrho}) = \sum_{\sigma}\sum^{\mathcal{N}_{\sigma}}_{i}\tr[\epsilon_{xc}(\bm{\rho})\bm{\varrho}_{i,\sigma}] \ . \label{xc_energy_eq}
\end{align}
In Eq. \ref{xc_energy_eq}, the dependencies of $\bm{\rho}$ and $\{\bm{\varrho}\}$ on $\bm{D}$ and $\{\bm{d}\}$, respectively, have been suppressed for notational conciseness. As an aside, it is noted that even with the exact $xc$ functional, the resulting RDM1 from a KS-DFT calculation will not equal that of FCI~\cite{mayer_nagy_fci_dft_rdm_jctc_2017}. However, approximate $xc$ functionals will usually yield accurate RDM1s as well as resonable densities~\cite{medvedev_perdew_lyssenko_density_errors_science_2017,sim_song_burke_density_errors_jpcl_2018}.\\

While the decompositions in Eqs. \ref{orb_decomp_mo_eq} and \ref{xc_energy_eq} are lossless (i.e., exact), orb-RDM1s in a conventional basis of canonical MOs will typically lie spanned completely delocalized over the entire system. As such, one cannot, in general, expect any degree of systematic grouping of the contributions to the above decomposition, amongst other things rendering a stringent mapping to molecular structure impossible. However, while total RDM1s and resulting properties in MF theory are invariant under rotations of the MOs, the orb-RDM1s are not. That is, one is free to perform a unitary transformation of the original set of canonical occupied MOs into some updated basis and repeat the decomposition in Eq. \ref{orb_decomp_mo_eq}.\\

In a basis of localized MOs~\cite{edmiston_ruedenberg_rev_mod_phys_1963,lehtola_jonsson_loc_orbs_jctc_2013}, decomposed MF results will indeed succeed in reflecting possible symmetries and corresponding degeneracies present in a standard Lewis depiction of a given molecule, which a basis of canonicalized MOs would otherwise fail to do so, despite these being symmetry-adapted. As an illustrative example (cf. Figure \ref{benzene_fig} of Section \ref{atoms_subsection}), localized results for the benzene molecule will show a grouping in accordance with its standard Kekul{\'e} representation---on par with standard MO diagram theory, the results successfully group into 6 contributions from the C(1s) orbitals, 3 from the carbon $\pi$-bonds, 3 from the carbon $\sigma$-bonds, and 6 from the C-H bonds, each of them arising from orb-RDM1s that are spatially local, correctly symmetric, and trivially degenerate.\\

While outside the objective of the present work, we note, in passing, how one might construct a deep NN (DNN) on the basis of the bond-wise contributions from Eqs. \ref{orb_decomp_mo_eq} and \ref{xc_energy_eq}, similar to the Bonds-in-Molecules Neural Network (BIM-NN) model by Parkhill {\textit{et al.}}~\cite{parkhill_intrinsic_bond_energies_jpcl_2017} where total molecular energies are summed up as an ensemble of bond energies using DNNs. However, in contrast to our hypothesized, decomposed model, the BIM-NN model relies critically on heuristics in learning different bond types and is thus somewhat devoid of the physical basis offered by the decompositions proposed in the present work.

\subsection{Atomic Partitionings}\label{atoms_subsection}

\begin{figure}[ht]
\vspace{-0.6cm}
\begin{center}
\includegraphics[width=1.0\textwidth]{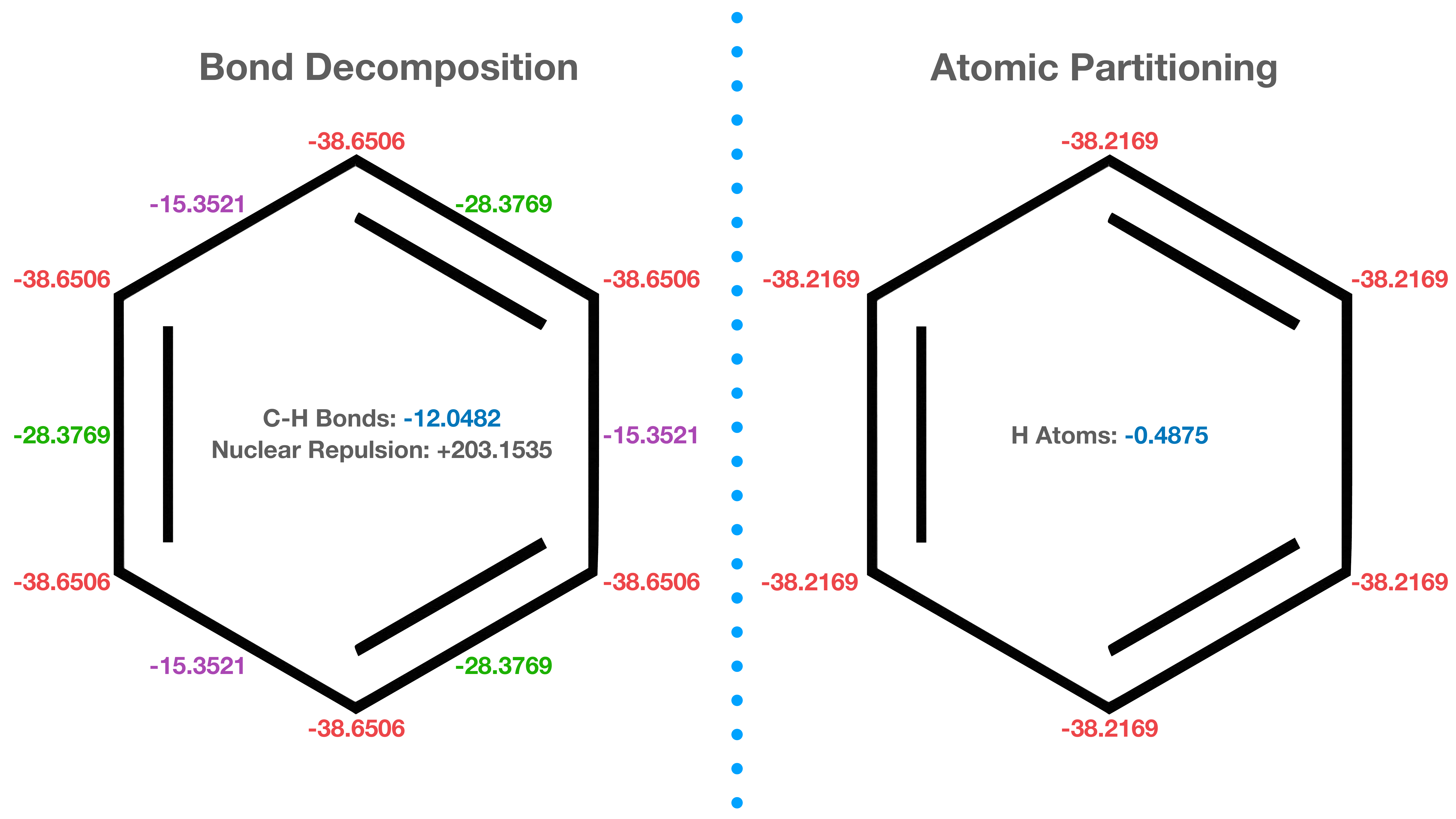}
\caption{Comparison of bond-decomposed (Eqs. \ref{orb_decomp_mo_eq} and \ref{xc_energy_eq}) and atom-partitioned (Eqs. \ref{atom_contr_eqs}) contributions (in units of $E_{\text{H}}$) to the total KS-DFT (TPSSh) energy of the benzene molecule. A combination of intrinsic atomic and bond orbitals (IAOs/IBOs) was used to obtain the contributions (see text for further details).}
\label{benzene_fig}
\end{center}
\vspace{-0.6cm}
\end{figure}

Up until this point, it has been illustrated how to decompose the electronic part of the MF functional, as exemplified for HF theory. While a grouping of certain orb-RDM1s on the basis of what bonds they represent is a perfectly valid option, any scheme that is bond- rather than atom-centric will---with an eye to the ML-QC applications to follow---necessitate the design of new, appropriate descriptors to facilitate the mapping between 2-dimensional Lewis bond structures and total MF energies. This reservation holds true regardless of how beneficial such models might prove to be going forward. In addition, only the electronic contributions to total molecular energies are decomposed by means of orb-RDM1s, thereby ignoring the intricate interplay that exists between true quantum and structural (steric) effects entirely. For this reason, and also given the amount of efforts that have been invested in designing atom-based representations over the years, cf. Section \ref{intro_sect}, it is advantageous---at least in the context of our present proof-of-concept study---to leverage these past endeavours. However, in order to do so, we will be required to further manipulate our decomposition into one that partitions the contributions over the individual atoms of a given molecule rather than into its constituent MOs (or bonds). For this purpose, the population of an underlying $i$th MO on all atoms of a given system, $\{\bm{p}_{i}\}$, is computed, before being employed---or rather the populations of all MOs on a given atom $K$, $\{\bm{p}^{K}\}$---as a relative weighting that allows us to rewrite the HF energy into the following, partitioned form:
\begin{align}
E_{\text{HF}} &= \sum^{\mathcal{M}_{\text{atom}}}_{K}E_{K}(\bm{D},\bm{\delta}_K) \nonumber \\
&= \sum^{\mathcal{M}_{\text{atom}}}_{K}E_{\text{elec},K}(\bm{D},\bm{\delta}_K) + E_{\text{nuc},K} \label{atom_decomp_eq}
\end{align}
in terms of atom-specific RDM1s (atom-RDM1s) defined as
\begin{align}
\bm{\delta}_K &= \sum_{\sigma}\bm{\delta}_{K,\sigma} \nonumber \\
&= \sum_{\sigma}\sum^{\mathcal{N}_{\sigma}}_{i}\bm{d}_{i,\sigma}\bm{p}^{K}_{i,\sigma} \ . \label{atom_rdm1_eq}
\end{align}
In Eq. \ref{atom_decomp_eq}, the nuclear and electronic contributions associated with atom $K$ are given as
\begin{subequations}
\label{atom_contr_eqs}
\begin{align}
E_{\text{nuc},K} &= Z_K\sum^{\mathcal{M}_{\text{atom}}}_{K<L}\frac{Z_L}{|\bm{r}_K - \bm{r}_L|} \label{nuc_contr_atom_eq} \\
E_{\text{elec},K} &= \tr[\bm{T}_{\text{kin}}\bm{\delta}_K] + \tfrac{1}{2}(\tr[\bm{V}_{K}\bm{D}] + \tr[\bm{V}_{\text{nuc}}\bm{\delta}_{K}]) + \tfrac{1}{2}\sum_{\sigma}\tr[\bm{G}_{\text{HF},\sigma}(\bm{D})\bm{\delta}_{K,\sigma}] \ . \label{elec_contr_atom_eq}
\end{align}
\end{subequations}
In Eq. \ref{nuc_contr_atom_eq}, $Z_L$ and $\bm{r}_L$ denote the nuclear charge and position of atom $L$, while $\bm{V}_{L}$ in Eq. \ref{elec_contr_atom_eq} denotes the attractive potential associated with this atom. Importantly, the above distribution of the total nuclear attraction energy among all of a system's $\mathcal{M}_{\text{atom}}$ atoms---as arising from {\bf{(i)}} the scaled Gaussian charge distribution representing a given atom and {\bf{(ii)}} the atom-RDM1 surrounding it---guarantees a systematic treatment of these effects in accordance with the manner in which the nuclear repulsion energy is partitioned, cf. Eq. \ref{nuc_contr_atom_eq}. In the case of KS-DFT, the $xc$ energy in Eq. \ref{xc_energy_eq} may be repartitioned in a manner similar to Eq. \ref{elec_contr_atom_eq}, again using the weights, $\{\bm{p}\}$, as the link between $\{\bm{\varrho}\}$ and the atomic centres of a given system. In turn, and as was the case with the orb-RDM1s of Section \ref{bonds_subsection}, the use of atom-RDM1s once again warrants a lossless decomposition of the underlying MF energy. In Figure \ref{benzene_fig}, the bond-decomposed results for the benzene molecule discussed in Section \ref{bonds_subsection} (using the TPSSh $xc$ functional~\cite{perdew_scuseria_tpssh_functional_prl_2003,scuseria_perdew_tpssh_functional_jcp_2003,scuseria_perdew_tpssh_functional_jcp_2004_erratum}) are compared to a corresponding atomic partitioning. These results clearly reflect the degenerate nature of the quintessential benzene molecule.\\

The partitioning in Eqs. \ref{atom_contr_eqs} may now be further contrasted with alternatives from the literature~\cite{li_parr_partitioning_jcp_1986,ichikawa_yoshida_partitioning_ijqc_1999,mayer_partitioning_cpl_2000,mayer_partitioning_cpl_2003,mandado_van_alsenoy_partitioning_cpc_2006,blanco_pendas_francisco_partitioning_jctc_2005,francisco_pendas_blanco_partitioning_jctc_2006,nalewajski_parr_partitioning_pnas_2000,parr_ayers_nalewajski_partitioning_jpca_2005,lilienfeld_alchemical_pt_jpcb_2019}, schemes which may be of one- or two-body nature and yielding either approximative or exact results. In the context of the present work, comparisons will be made to the equally lossless energy density analysis (EDA) partitioning by Nakai~\cite{nakai_eda_partitioning_cpl_2002}, in which the corresponding electronic contribution to $E_{\text{HF}}$ associated with atom $K$ is defined as
\begin{align}
E^{\text{EDA}}_{\text{elec},K} &= \tr_{\mu\in K}[\bm{T}_{\text{kin}}\bm{D}] + \tfrac{1}{2}(\tr[\bm{V}_{K}\bm{D}] + \tr_{\mu\in K}[\bm{V}_{\text{nuc}}\bm{D}]) + \tfrac{1}{2}\sum_{\sigma}\tr_{\mu\in K}[\bm{G}_{\text{HF},\sigma}(\bm{D})\bm{D}_{\sigma}] \ . \label{eda_elec_contr_atom_eq}
\end{align}
In Eq. \ref{eda_elec_contr_atom_eq}, the trace operations run over all atomic orbitals (AOs), $\{\bm{\mu}\}$, assigned to atom $K$. In the case of KS-DFT, the $xc$ energy may be treated in a similar manner, again partitioning the total density matrix solely on the basis of which atoms the individual AOs are localized on (that is, irrespective of any further population measure)~\cite{nakai_eda_partitioning_ijqc_2009}. In contrast to EDA, the present partitioning in Eqs. \ref{atom_contr_eqs} fundamentally operates on an MO rather than an AO level, which among other features offers tunability as we may employ different localization procedures and ways of defining the atom-RDM1s (cf. Section \ref{calibration_polyacetylenes_subsection}). In terms of the overall form of the two types of partitioning, however, obvious similarities are apparent, not least in the distribution of the nuclear attraction energy among the individual atomic centres. Subtle conceptual differences exist, though. For instance, EDA contributions will exhibit no complete basis set limit, and since the scheme fundamentally decomposes the KS-DFT electronic density into atom-associated contributions, these partial densities will not necessarily remain positive at each point in space, especially in extended basis sets. These issues, which might give rise to further problems in statistical analyses, are avoided in the present scheme.\\ 

By using population weights in the definition of $\{\bm{\delta}\}$ (Eq. \ref{atom_rdm1_eq}), a pronouncedly local orb-RDM1 $i$, i.e., one that arises from a core or lone-pair MO on atom $K$, will be assigned a weight of $p^{K}_{i} \approx 1.0$ (implying that $p^{L}_{i} \approx 0.0 \ \forall \ L \neq K$), while an orb-RDM1 that maps to a chemical bond between atoms $K$ and $L$ will be assigned a weight in the interval $0.0 < p^{K}_{i} < 1.0$ (and similarly so for $p^{L}_{i}$). In turn, these population weights may be computed on the basis of a number of metrics, as they, at best, function as proxies of the actual charge distribution within a given system. As a conventional choice, we may calculate weights as regular Mulliken populations, using $\{\bm{d}\}$ and the overlap matrix, $\bm{S}$, in the standard AO basis~\cite{mulliken_population_jcp_1955},
\begin{align}
\bm{p}^{K}_{i} = \underset{\mu \in K}{\tr}[\bm{d}_i\bm{S}] \ . \label{mulliken_pop_eq}
\end{align}
Alternatively, the weights may be based on Mulliken populations with $\bm{d}_i$ (or rather $\bm{C}_{i}$) recasted into an alternative AO basis where MOs are denoted by $\bm{B}$; in the present study, besides the occasional use of standard Mulliken populations, we will primarily compute weights from Knizia's intrinsic AO~\cite{knizia_iao_ibo_jctc_2013,knizia_visscher_iao_arxiv_2020} (IAO) population analyses (these were also used in Figure \ref{benzene_fig}). Specifically, the IAO transformation, $\bm{C}\mapsto\bm{B}$, proceeds through the initial construction of a reduced-dimension basis, the functions of which are constructed by a projection operation from a set of free-atom orbitals,
\begin{align}
\bm{B}_{i} = \bm{A}^{T}\bm{S}\bm{C}_i \ . \label{iao_trans_eq}
\end{align}
In Eq. \ref{iao_trans_eq}, $\bm{A}$ denotes the coefficient matrix of the (symmetrically orthogonalized) IAOs. Since the IAOs span the entire occupied space, this transformation is lossless as well. Importantly, while traditional Mulliken populations and the partial charges they give rise are extensively used in modern tight-binding, semi-empirical methods~\cite{christensen_elstner_semiempirical_qc_chem_rev_2016}, they have also been coined as mathematically ill-defined due to the fact that they exhibit a strong, explicit basis set dependence with no formal saturated limit~\cite{lehtola_jonsson_pm_jctc_2014}. These artefacts are avoided in the IAO scheme by the above projection onto the minimal basis, as this remains the same regardless of the choice of basis set to be used in the central MF calculation. However, for completeness, it is noted how a multitude of alternatives exist for determining such weights. Noteworthy examples include the Hirshfeld~\cite{hirshfeld_partitioning_tca_1977}, Becke~\cite{becke_partitioning_jcp_1988}, and Bader~\cite{bader_partitioning_book_1990} partitionings, methods which all apportion the electron density, rather than the RDM1 as in the present case, by weighting it among the atoms of a given system.

\subsection{Kernel Ridge Regression}\label{krr_subsection}

Despite the popularity surrounding NNs and their functions as regressors in ML-QC, cf. the earlier discussion in Section \ref{intro_sect}, we will here make use of kernel ridge regression~\cite{mueller_schoelkopf_ieee_trans_nn_2001} (KRR) for all of our present ML purposes due, first and foremost, to its ease of use and its technical transparency, that is, its relative simplicity in terms of interpretation and efficient implementation. In KRR, a property of interest, $\tilde{y}_K$, of an atom $K$ is estimated as a linear, weighted sum of kernels. These produce the similarity with an atom $P$ of a training dataset (of size $\mathcal{P}_{\text{train}}$), for which the corresponding property, $y_P$, is already known,
\begin{align}
\tilde{y}_K = \sum^{\mathcal{P}_{\text{train}}}_{P}\alpha_{P}\mathcal{K}(K,P) \ . \label{krr_eq}
\end{align}
Two identical atomic environments will give rise to a unit kernel similarity, while this measure will approach zero asymptotically for two atoms embedded in entirely different environmental settings. Typically, Gaussian or Laplacian kernel functions are used for this purpose, and we will here make use of the former,
\begin{align}
\mathcal{K}(K,P) = \exp\left(-\frac{||\mathcal{A}_K - \mathcal{A}_P||^{2}_{2}}{2\sigma^{2}}\right) \ , \label{gaussian_kernel_eq}
\end{align}
where $\sigma$ is the length scale, $\mathcal{A}$ an atomic representation, and $||\cdot||_{2}$ denotes the Euclidean ($L^2$) norm. The fitting coefficients in Eq. \ref{krr_eq}, $\{\bm{\alpha}\}$, are obtained through a regularized, linear least-squares optimization procedure,
\begin{align}
\bm{\alpha} = (\bm{\mathcal{K}} + \lambda\bm{I})^{-1}\bm{y} \ , \label{least_square_opt_eq}
\end{align}
where the regularizer, $\lambda$, is introduced to ensure numerical stability as well as to balance under- and overfitting~\cite{tikhonov_ill_posed_prob_1977}. In particular, by enforcing regularization of the training data, a choice of $\lambda > 0$ may be introduced to prevent the latter problem for moderately noisy data.

\section{Method Validation}\label{calibration_section}

Prior to gauging what merits, if any, the decomposed MF theories of Section \ref{theory_sect} may have in the context of ML-QC, we begin by comparing them to one another on the basis of whether or not they yield results that are physically intuitive, rigorous across various MF methods, and systematical with respect to an increase in problem size. In Section \ref{calibration_water_subsection}, we will use decomposed HF and KS-DFT to probe the electronic structure of water, while we will turn to polyacetylene chains in Section \ref{calibration_polyacetylenes_subsection} to investigate the size-intensive behaviour of the theories for this class of largely homogenous systems that are trivially increased in size. All decomposed results have been obtained using a new, open-source code named {\texttt{DECODENSE}}~\cite{decodense}, which is formulated around the {\texttt{PySCF}} electronic structure code~\cite{pyscf_wires_2018,pyscf_jcp_2020}.

\subsection{Water}\label{calibration_water_subsection}

\begin{figure}[ht]
\begin{center}
\includegraphics[width=0.85\textwidth]{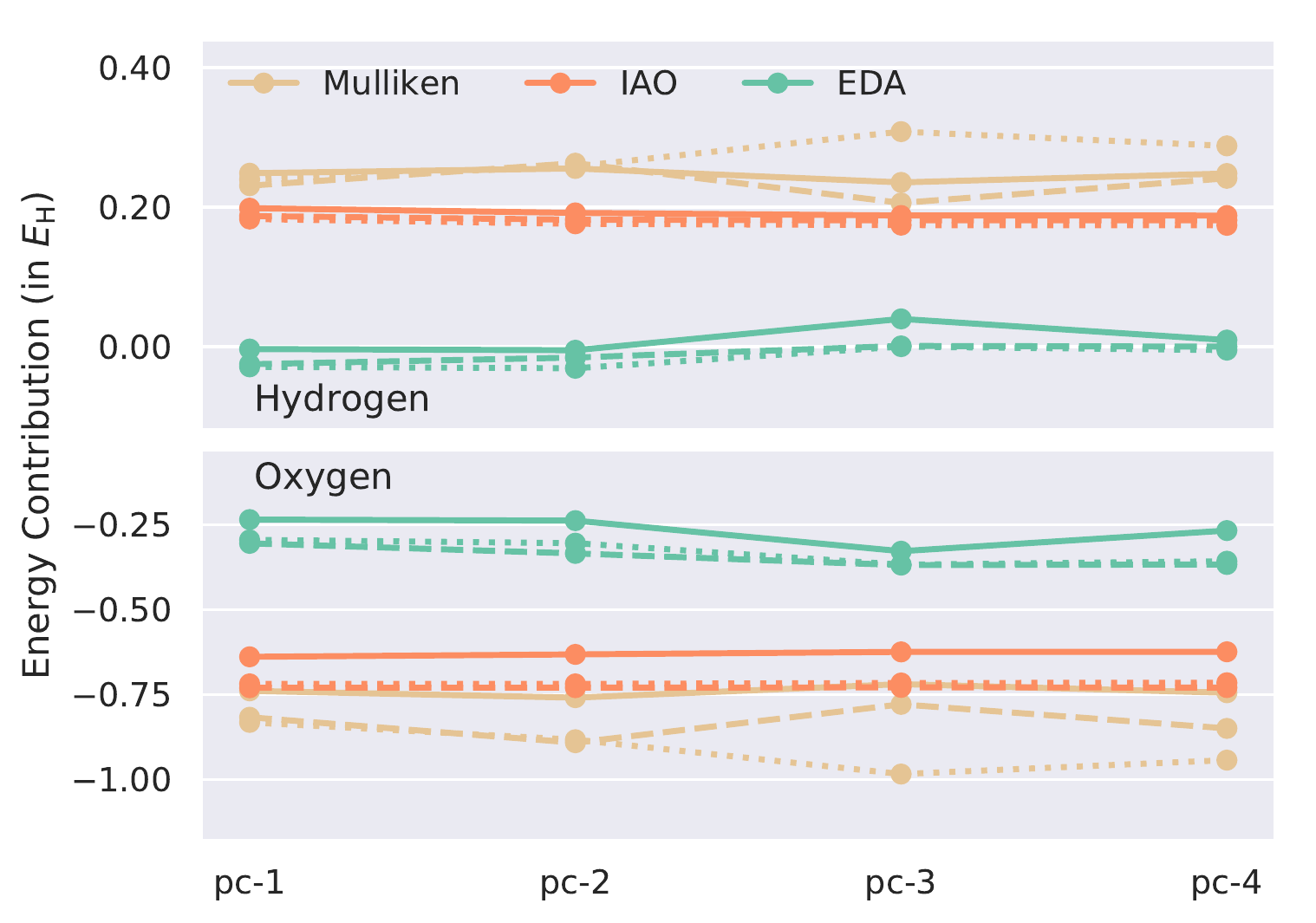}
\caption{HF (solid lines), $\omega$B97X-D (dashed lines), and M06-2X (dotted lines) atomization energies of H$_2$O in the pc-$n$ basis sets, as partitioned into contributions from the oxygen and each of the hydrogen atoms by means of IBOs and either IAO or Mulliken weights using Eqs. \ref{atom_contr_eqs} or the EDA scheme in Eq. \ref{eda_elec_contr_atom_eq}.}
\label{h2o_energy_basis_ibo2_fig}
\end{center}
\vspace{-0.6cm}
\end{figure}
As a simple, yet illustrative example of the basis set (in)dependence of IAO and Mulliken populations and the effects these artefacts, in turn, have on the resulting decompositions, Figure \ref{h2o_energy_basis_ibo2_fig} shows the magnitude of the contributions to the HF, $\omega$B97X-D~\cite{chai_head_gordon_wb97x_d_functional_pccp_2008}, and M06-2X~\cite{zhao_truhlar_m06_functional_tca_2008} atomization energies of the water molecule from the oxygen and each of the hydrogen atoms in Jensen's pc-$n$ basis sets~\cite{jensen_pc_basis_sets_jcp_2001} (double- through pentuple-$\zeta$ quality). Results are presented using weights of either type, and these are, in turn, augmented by corresponding EDA results obtained using Eq. \ref{eda_elec_contr_atom_eq}. $\omega$B97X-D and M06-2X were chosen upon as two modern, yet non-related $xc$ functionals. The results---obtained in a localized MO basis of intrinsic bond orbitals~\cite{knizia_iao_ibo_jctc_2013} (IBOs) in all cases, except for the EDA results, which are orbital-invariant---show how a partitioning into contributions from the O and H atoms appears to be most consistently achieved by means of the current scheme in combination with IAO weights. The IAO-based results are observed to vary the least upon enlarging the employed basis set, and the profiles of the HF, $\omega$B97X-D, and M06-2X curves are all identical (bar an expected vertical shift in the KS-DFT curves), unlike the results obtained using Mulliken populations or the EDA scheme, which both show some variance with an enlargement of the one-electron basis set. In addition, we note how the EDA partitioning yields contributions associated with the hydrogen atoms that are vanishing or even slightly negative in most cases. The physical interpretation of these results thus contradicts expectation on the basis of the known polarity of the water molecule, in the sense that the difference in energy between an isolated hydrogen atom and one embedded in the water molecule is negligible or even negative, unlike what is observed for the Mulliken- and IAO-based results of the present work.\\

These observations may be further strengthened by comparing decomposed molecular dipole moments, which we compute---irrespective of the employed level of MF theory---as
\begin{align}
\bm{\mu}_{\text{MF}} = \sum^{\mathcal{M}_{\text{atom}}}_{K}\bm{\mu}_{\text{elec},K}(\bm{\delta}_K) + \bm{\mu}_{\text{nuc},K} \ . \label{atom_decomp_dipmom_eq}
\end{align}
In Eq. \ref{atom_decomp_dipmom_eq}, the nuclear and electronic contributions read
\begin{subequations}
\label{atom_contr_dipmom_eqs}
\begin{align}
\bm{\mu}_{\text{nuc},K} &= Z_{K}\bm{r}_{K} \label{nuc_contr_atom_dipmom_eq} \\
\bm{\mu}_{\text{elec},K} &= -\sum_{r}\tr[\bm{\mu}_{r}\bm{\delta}_{K}] \label{elec_contr_atom_dipmom_eq}
\end{align}
\end{subequations}
in terms of dipole integrals, $\bm{\mu}_{r}$, in the AO basis for each of the three Cartesian components, $r=x,y,z$. It is worth noting that this decomposition is once again lossless and hence different from the corresponding dipole moment computed from partial charges alone~\bibnote{The reason for this is that the centre of the electronic density associated with a given atom (of which the partial charge is the essential measured quantity) will not necessarily coincide with the corresponding nuclear coordinates, e.g., when constituent atoms carry lone pairs.}. In the EDA partitioning, the electronic contributions are herein defined---theoretically on par with Nakai's original definition of the energetic analogues in Eq. \ref{eda_elec_contr_atom_eq}---as follows
\begin{align}
\bm{\mu}^{\text{EDA}}_{\text{elec},K} &= -\sum_{r}\tr_{\mu\in K}[\bm{\mu}_{r}\bm{D}] \ . \label{eda_elec_contr_atom_dipmom_eq}
\end{align}
\begin{figure}[ht]
\vspace{-0.8cm}
\begin{center}
\includegraphics[width=0.85\textwidth]{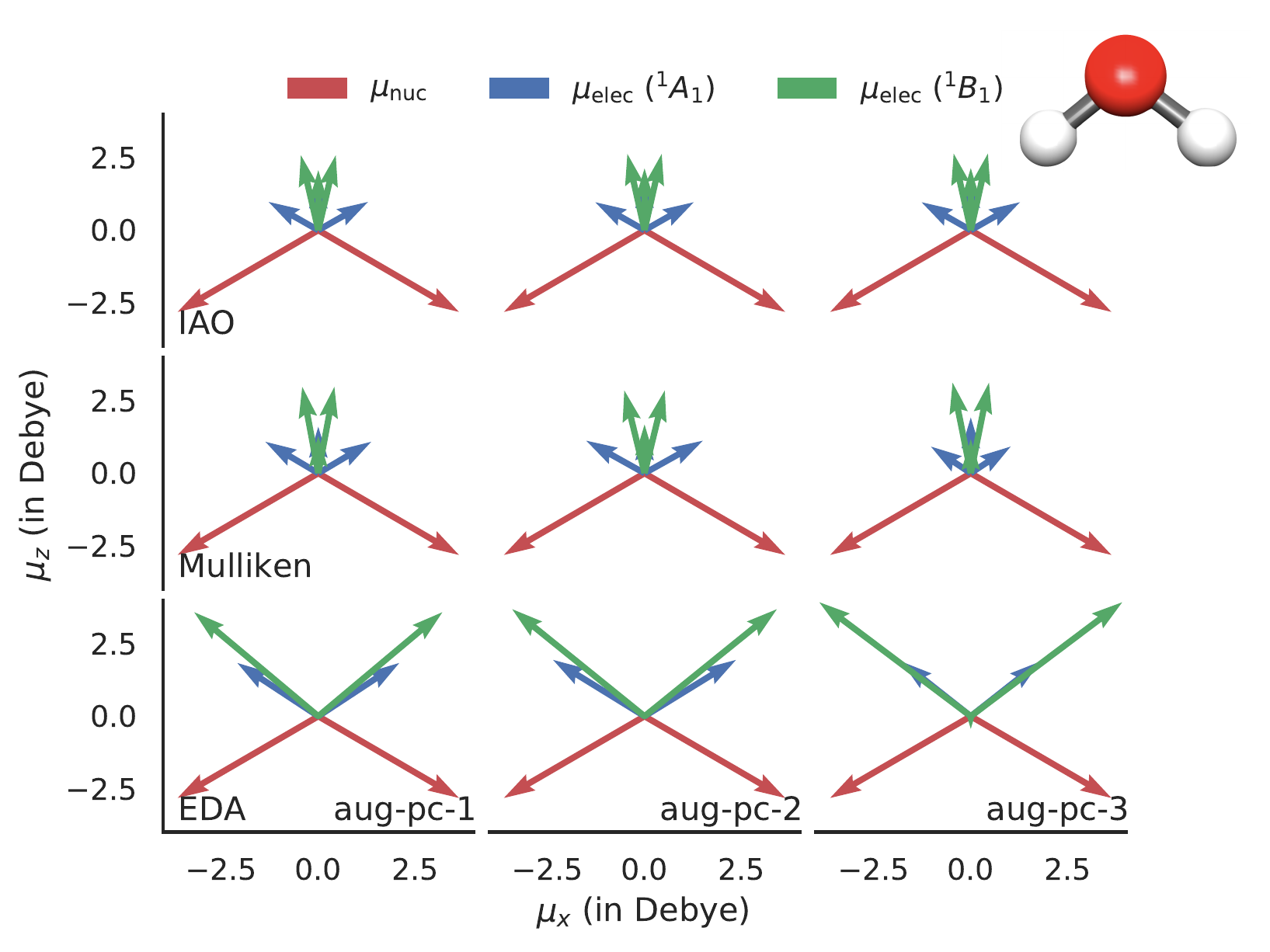}
\caption{$\omega$B97X-D ground (gs) and excited (ex) state molecular dipole moments of H$_2$O in the aug-pc-$n$ basis sets (in units of Debye), as partitioned into nuclear and electronic contributions, with the gauge origin at the position of the oxygen atom, $\bm{R}_{\text{O}} = (0, 0, 0)$. The latter of the contributions ($\mu_{\text{elec}}$) have been decomposed by means of IBOs and either IAO or Mulliken weights using Eqs. \ref{atom_contr_dipmom_eqs} or the EDA scheme in Eq. \ref{eda_elec_contr_atom_dipmom_eq}.}
\label{h2o_dipmom_basis_fig}
\end{center}
\vspace{-0.4cm}
\end{figure}
We will here use decomposed dipole moments to further scrutinize how the different schemes allow for a detailed description of the electronic structure of the ubiquitous water molecule. Specifically, we present results in Figure \ref{h2o_dipmom_basis_fig} for decomposed ground- and excited-state dipole moments obtained using the $\omega$B97X-D $xc$ functional in the aug-pc-$n$ basis sets, with corresponding HF and M06-2X results presented in Figure S1 of the Supporting Information (SI). The excited-state results are computed for the ${^{1}}B_1$ state ($1b_1\rightarrow4a_1$ transition), as obtained by (unrestricted) $\Delta$-SCF theory in combination with Gill's maximum overlap method~\cite{delta_scf_mom_gill_jpca_2008}. In all cases, the gauge origin coincides with the position of the oxygen atom.\\

While no unambiguous measure of the correctness of an atomic decomposition of a molecular property such as dipole moments exists, and keeping in mind that the (vertical Frank-Condon) excited state dipole moment is decomposed at the ground-state equilibrium geometry, our results in Figure \ref{h2o_dipmom_basis_fig} may still be evaluated on the basis of what might be expected {\textit{a priori}} of these dipole moments as a measure of the separation of positive and negative (partial) charges in said states. As discussed to great extent by Urban and co-workers over the years~\cite{urban_sadlej_h2o_dipole_tca_1990,urban_h2o_dipole_mol_phys_2008}, the most characteristic feature adherent to these two dipole moments is the change in orientation following upon the electronic transition. As the nuclear components remain the same between the ground and excited states, this change along the $z$-axis is mediated by an alteration of the molecular polarity. Interestingly, this redistribution of the electronic structure is not fully reflected in the partial atomic IAO charges alone; along the transition, these are observed to change from $-0.743$ ($+0.371$) to $-0.263$ ($+0.132$) for the oxygen (hydrogen) atoms in the ground and excited states, respectively, at the $\omega$B97X-D/aug-pc-3 level of theory. Instead, the change in the dipole moment is attributed predominantly to a stronger occupation of the oxygen lone pair MOs perpendicular to the molecular plane in the ${^{1}}B_1$ state, the $3s$-Rydberg character of which is also found to increase~\cite{urban_h2o_dipole_mol_phys_2008}.\\

From the results in Figure \ref{h2o_dipmom_basis_fig}, a number of observations may be made with regards as to how the different decompositions generally describe a polar molecule like water and how they reflect the reorganization of the electronic distribution involved in said transition. First, weak basis set dependencies are observed for all three decompositions---as for the case of the ground-state energy in Figure \ref{h2o_energy_basis_ibo2_fig}, the IAO-based results vary marginally, followed by the Mulliken- and EDA-based results (in that order). Second, despite the reduction in partial charge of the oxygen atom, the Mulliken- and IAO-based results both illustrate how the change in orientation of the molecular dipole moment is indeed attributed to an increase in its electronic contributions along the positive $z$-axis, not only of the contributions associated with the oxygen atom, but also of those associated with the two hydrogen atoms. In contrast, the EDA-based results have a negative oxygen component along the $z$-axis in both states, amplified from $-0.011$ to $-0.436$ Debye at the $\omega$B97X-D/aug-pc-3 level of theory. Not only do these results again contradict expectation, but the overall invariance of the ${^{1}}A_1$ and ${^{1}}B_1$ results, for instance, in the direction of the hydrogen components, seems to hint at a deeper issue with the EDA partitioning. Namely, its components appear to once again be insensitive to underlying changes in the electronic structure; by solely partitioning electronic properties on the basis of the localization of AOs in a system, their distribution onto its atoms become largely predetermined. This is unlike the present decompositions, particularly so when these are formulated in terms of the physically sound IAO populations, as our scheme succeeds in capturing any such changes related to electronic effects. We will return to this point, and its specific consequences for water, later on in Section \ref{qml_water_subsection}.

\subsection{Polyacetylenes}\label{calibration_polyacetylenes_subsection}

\begin{figure}[ht]
\vspace{-0.8cm}
\begin{center}
\includegraphics[width=0.9\textwidth]{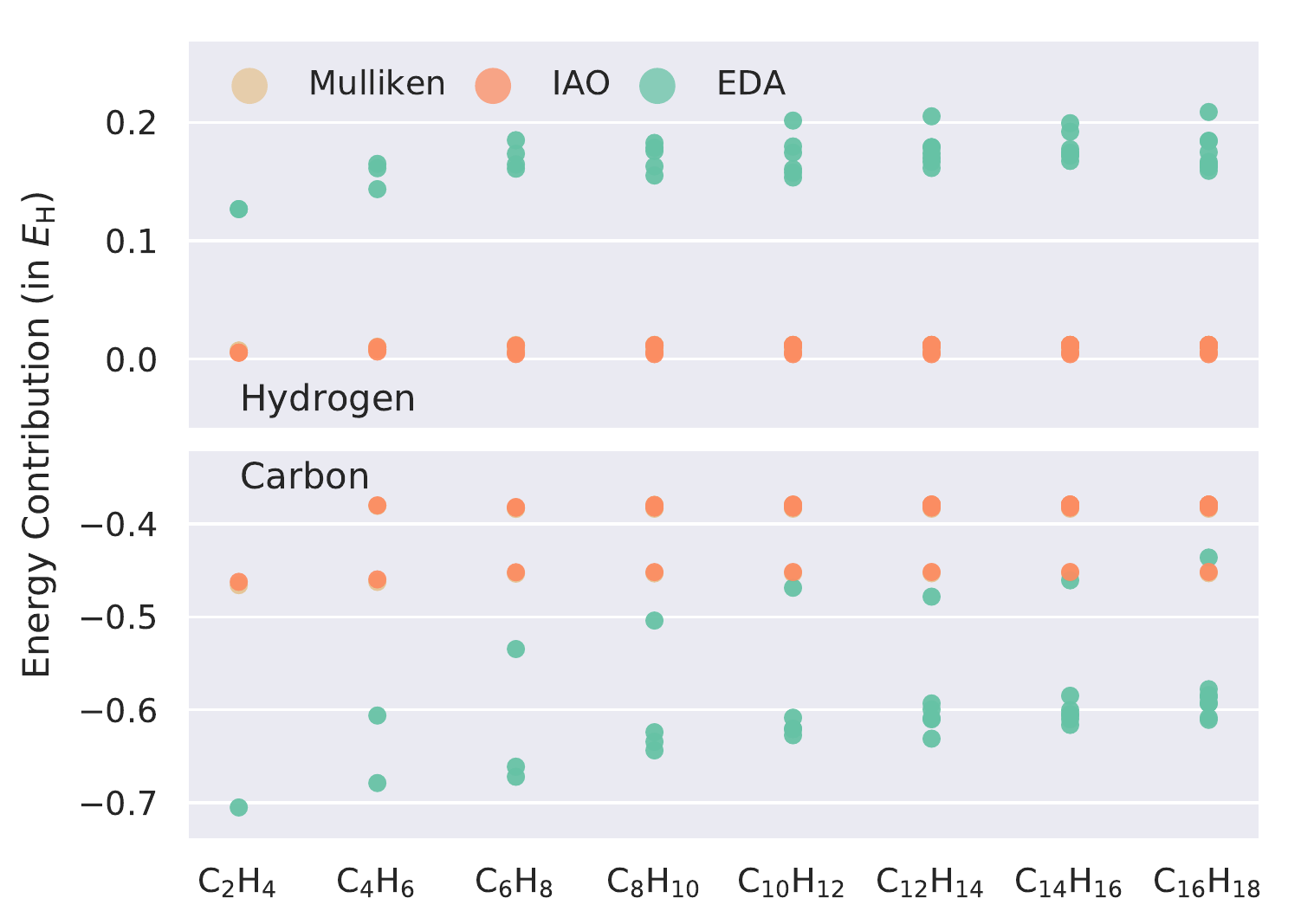}
\caption{PBE/6-31G atomization energies of a series of polyacetylenes, as partitioned into contributions from the each of the carbon and hydrogen atoms by means of IBOs and either IAO or Mulliken weights using Eqs. \ref{atom_contr_eqs} or the EDA scheme in Eq. \ref{eda_elec_contr_atom_eq}.}
\label{polyacetylenes_pbe_631g_ibo2_fig}
\end{center}
\vspace{-0.6cm}
\end{figure}

Having compared the different decompositions for a single system in a selection of basis sets, we will next look at how these yield results for a specific class of systems of increasing composition. Figure \ref{polyacetylenes_pbe_631g_ibo2_fig} presents results for atomization energies of a series of polyacetylenes~\cite{gagliardi_polyacetylenes_chem_sci_2019} (C$_n$H$_{n+2}$ for $2 \leq n \leq 16$) at the PBE/6-31G level of theory~\cite{perdew_burke_ernzerhof_pbe_functional_prl_1996,pople_1}, again obtained using localized IBOs. Polyacetylenes belong to a wider class of polyene compounds that have long been favoured examples of conjugated systems at an extended scale due to their alternating single and double bonds arranged in a chain along a single dimension. Furthermore, polyacetylenes have been used as valuable model systems to understand the electronic properties of more complicated biological systems and conjugated polymers in general~\cite{burke_polyacetylenes_nat_chem_2014,chan_polyacetylenes_jcp_2008}.\\

From the results in Figure \ref{polyacetylenes_pbe_631g_ibo2_fig}, striking differences in-between the different types of atomic partitioning are once again obvious. While the Mulliken- and IAO-based results coincide almost exactly, to the point where they are hardly distinguishable from one another, those based on the EDA partitioning are found to vary significantly upon an increase in system size. The results of all of the decompositions are observed to split into contributions from the terminal groups and all other carbon centres, but whereas the latter class of contributions are practically degenerate---both for a given system size and across the series---in the decompositions of the present work, this is not the case in the EDA-based counterpart. Given the homogeneity of the polyacetylenes, it is fair to expect the contributions associated with individual carbon and hydrogen atoms to converge onto a system-specific value early on in the series, and this is exactly the type of size intensivity observed in the Mulliken- and IAO-based results of the present work.\\

\begin{figure}[ht]
\vspace{-0.8cm}
\begin{center}
\includegraphics[width=0.9\textwidth]{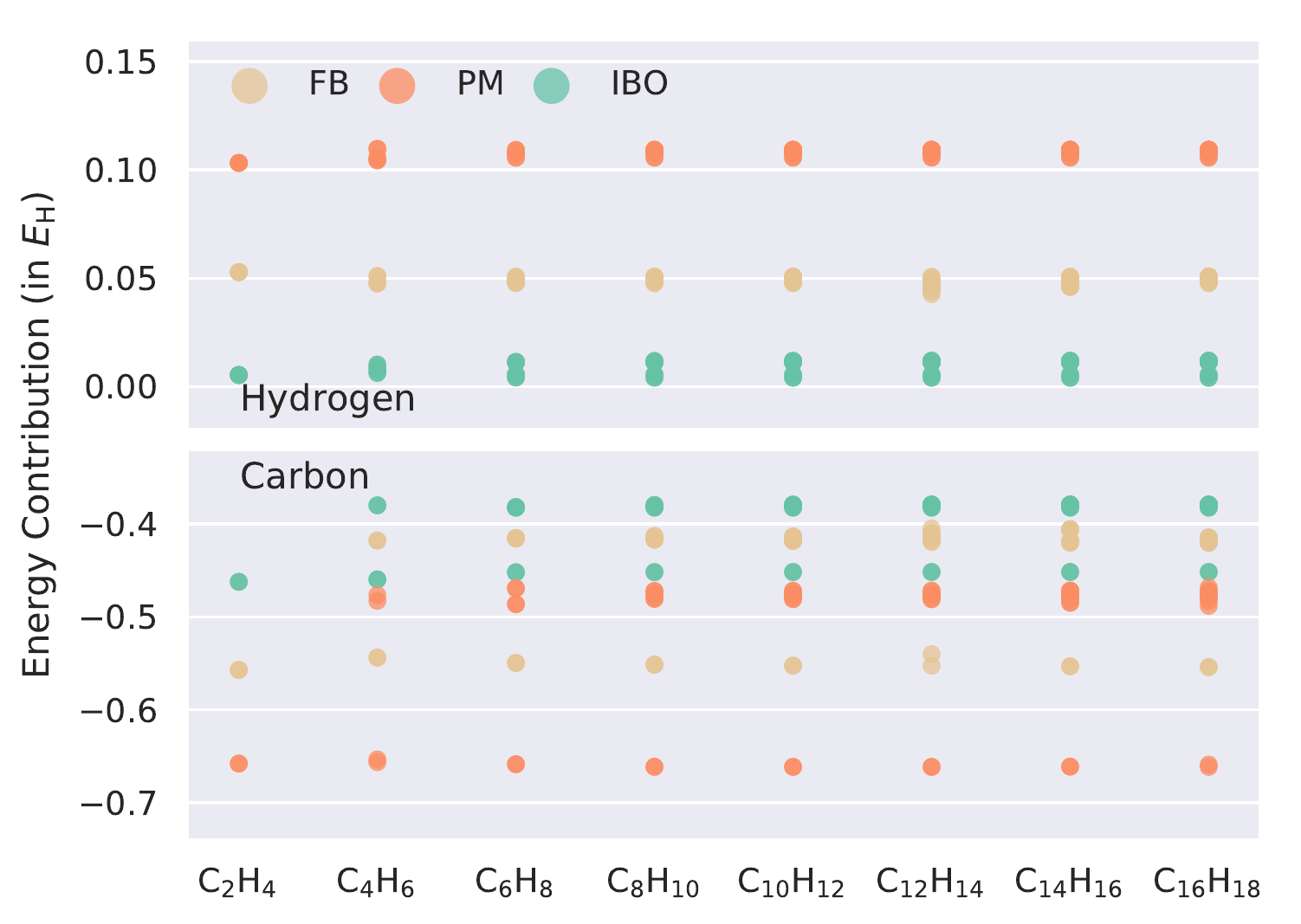}
\caption{PBE/6-31G atomization energies of the polyacetylenes, as partitioned into contributions from the each of the carbon and hydrogen atoms by means of IAO weights (Eqs. \ref{atom_contr_eqs}) and either intrinsic bond orbitals (IBOs), Foster-Boys (FB), or Pipek-Mezey (PM) MOs.}
\label{polyacetylenes_pbe_631g_local_fig}
\end{center}
\vspace{-0.6cm}
\end{figure}
Besides the choice of charge population protocol, the behaviour and performance of the atom-wise partitioning in Eqs. \ref{atom_contr_eqs} are generally governed by the choice of localization procedure. In Figure \ref{polyacetylenes_pbe_631g_local_fig}, we compare results obtained with IBOs (Figure \ref{polyacetylenes_pbe_631g_ibo2_fig}) to corresponding IAO-based results obtained using either Foster-Boys~\cite{foster_boys_rev_mod_phys_1960} (FB) or Pipek-Mezey~\cite{pipek_mezey_jcp_1989} (PM) localized MOs. The latter of these (PM) is inherently similar to the IBO procedure used here~\bibnote{In the iterative optimization of IBOs used in the present work, a PM localization power ($p=2$) has been used throughout, cf. Appendix D of Ref. \citenum{knizia_iao_ibo_jctc_2013}}, but differing in the way atomic charges are estimated in the optimization scheme. While all of the results in Figure \ref{polyacetylenes_pbe_631g_local_fig} are observed to be more consistent than the corresponding EDA results in Figure \ref{polyacetylenes_pbe_631g_ibo2_fig}, differences still exist due to the different MOs from which they are computed. For instance, the splitting between contributions from (non-)terminal carbon centres increases in moving from IBOs over FB to PM, whereas the variations of the hydrogen and carbon contributions across the series are largest for FB and PM and smallest in the case of IBOs. Most noteworthy, however, is the fact that the IBO-based decomposition yields the least polarized set of results, aligning well with the small IAO partial charges and overall expectation. As an example, for the largest system (C$_{16}$H$_{18}$), the terminal carbons (hydrogens) have a partial charge of $-0.312$ ($+0.155$), whereas charges for all of the other carbon (hydrogen) centres fall within a narrow interval from $-0.141$ to $-0.135$ ($+0.140$ to $+0.147$).\\

For this reason---and further supported by similar results for a dataset of thermalized water molecules in Figures S4 and S5 of the SI (cf. Section \ref{qml_water_subsection})---we are led to conclude that IBOs generally yield consistent results, in accordance with chemical and physical intuition, and we will thus use these throughout the remainder of the present study. In the following Section \ref{qml_section}, we wish to investigate if the results of decomposed MF theory may be used in the calibration of molecular force fields, as parametrized by ML techniques. In particular, we will assess to what degree the consistency of atomic decompositions will influence such attempts. A somewhat unrelated, yet pertinent question in this context, which we will postpone for future studies, is whether or not decomposed MF theory and its orb-/atom-RDM1s may be used to scrutinize the complex electronic structure of carbon-rich systems without resorting to inspections of the involved MO basis alone~\cite{hoffmann_solomon_helical_orbs_acs_cent_sci_2018,solomon_helical_orbs_chem_sci_2019,solomon_elec_trans_jpcc_2020}, cf. the discussion in Section \ref{outlook_summary_section}.

\section{Preliminary Applications in Machine Learning}\label{qml_section}

In the following, all results are obtained with either of the B3LYP~\cite{becke_b3lyp_functional_jcp_1993,frisch_b3lyp_functional_jpc_1994} and PBE0~\cite{adamo_barone_pbe0_functional_jcp_1999} $xc$ functionals, again in combination with the standard pc-1 (double-$\zeta$) basis set. As we will be concerned with results computed from training sets that have been randomly drawn from a main set of geometries, we will only compare results in terms of general trends and distinct differences, not explicitly on the basis of mean absolute errors, as such measures are not particularly meaningful in the present context. Prediction errors will instead be visualized by means of kernel density estimations (KDEs), as implemented in the {\texttt{seaborn}} Python module~\cite{seaborn}. KDEs represent the errors using continuous probability density curves, analogous to a histogram. A total of 10 contour levels have been used in all KDE plots, and the warmest color indicates the greatest density for every color palette. All alkane and water geometries have been generated by molecular dynamics (MD) simulations $@$ 350 K, extracted from Refs. \citenum{miller_ml_qc_data_2019} and \citenum{lilienfeld_ml_qc_data_2019}, and the datasets have been pre-randomized to avoid any possible autocorrelation between the training and test sets. As such, all ensembles will cover molecular structures that lie in close proximity to their given equilibrium geometries.\\

We will exclusively present results in terms of out-of-sample errors, using identical length scales and regularizers throughout, that is, without recourse to a proper cross-validation of these parameters. However unconventional this choice may seem, the objective herein is not to compute benchmark numbers, but rather to compare the performance of the individual decompositions against one another in an unbiased manner free of any external parameters (i.e., freedom in choice of $\sigma$ and $\lambda$ in Eqs. \ref{gaussian_kernel_eq} and \ref{least_square_opt_eq}, respectively). As our structure encoder, we have chosen upon the FCHL representation~\cite{lilienfeld_fchl_jcp_2018,lilienfeld_fchl_jcp_2020} developed by von Lilienfeld's group in Switzerland due to its atom-centric nature, its reported performance (also for atomic properties~\cite{glowacki_butts_impression_chem_sci_2020,ramakrishnan_ml_nmr_arxiv_2020}), and its availability in the open-source {\texttt{QML}} software~\cite{qml_prog}, which is used for all ML-QC calculations to follow. FCHL is also used as our KRR reference, in which case molecular kernel similarities are computed simply as sums over those for the constituent atomic kernels. Traditional FCHL results will be denoted by a {\textbf{`Reference'}} label throughout, while we will denote decomposed FCHL results by the label {\textbf{`DECODENSE'}}.\\

Since all parameters, both for atomic and molecular learning, have been kept fixed ($\sigma = 5.0$ and $\lambda = \num{1.e-10}$, if not noted otherwise, and default FCHL hyperparameters except for an increased cut-off radius of 10 \AA), these two approaches (atomic and molecular) represent alternative paths toward a common target; by summing over individual kernel similarities, the composite structural fingerprint of a molecule is effectively folded into the learning of its scalar energy, while in the present approach, individual atomic environments are fitted and the final molecular energy assembled as a sum of learned contributions. Clearly, for the present approach to yield theoretically reasonable results, electronic effects ideally need to be accounted for in addition to whatever immediate chemical environment an atom is embedded in (hybridization, electronegativity, etc.). This necessarily places a lot of emphasis on the sensitivity of the employed representation (cf. Section \ref{outlook_summary_section}). However, as important will be the correctness of the underlying atomic decomposition, gauged, for instance, through stress tests involving structural and compositional changes. In the following, we will complement the results of Section \ref{calibration_section} by directly contrasting the various decompositions of Section \ref{theory_sect} with one another and compare them with traditional, molecular FCHL. Not only will this allow us to evaluate potential pros and cons of an atomic learning procedure, but also the overall suitability of the FCHL representation with respect to the present purpose.

\subsection{Hydrocarbons}\label{qml_hydrocarbons_subsection}

\begin{figure}[ht!]
\begin{center}
\includegraphics[width=0.5\textwidth]{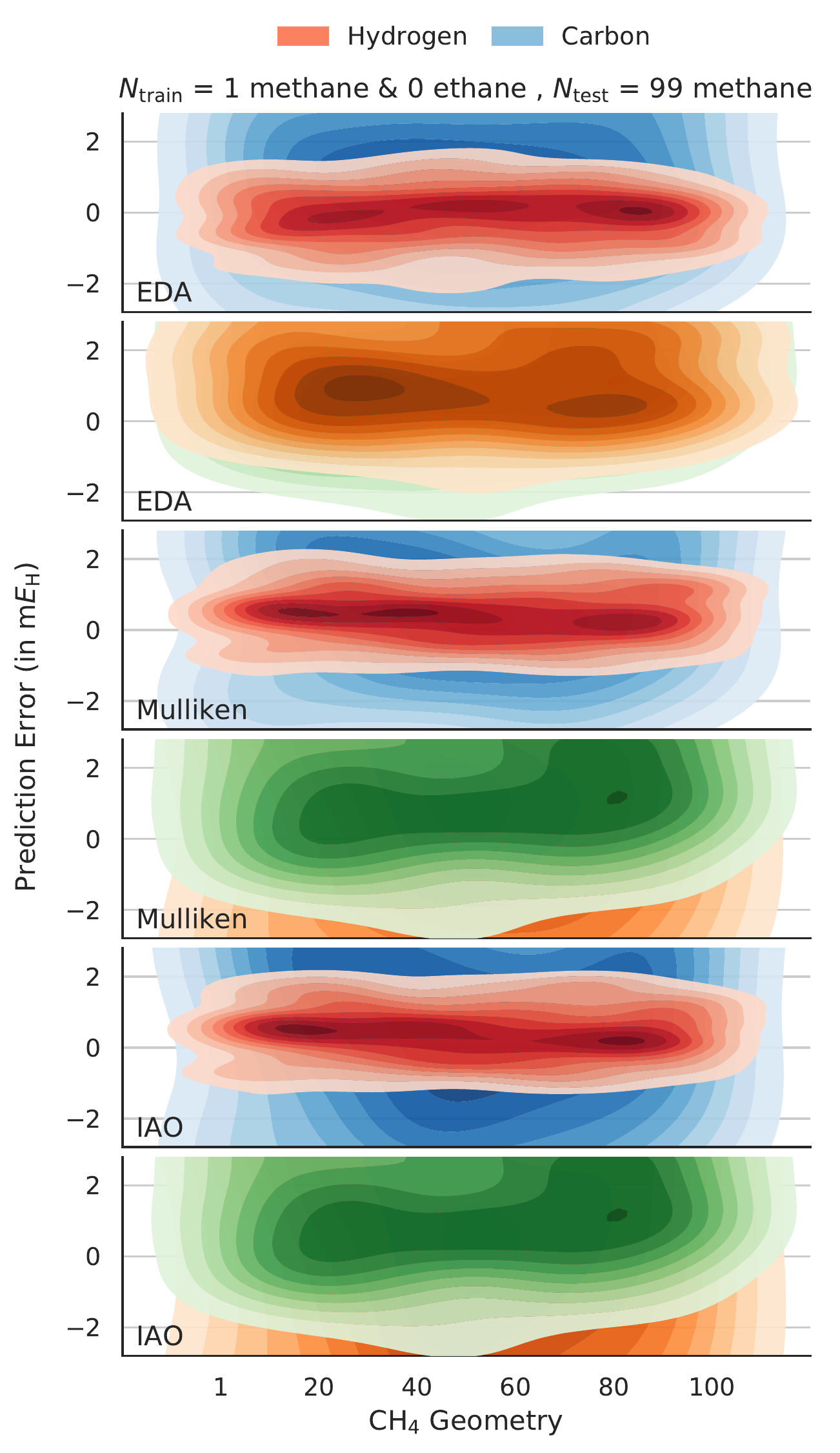}%
\includegraphics[width=0.5\textwidth]{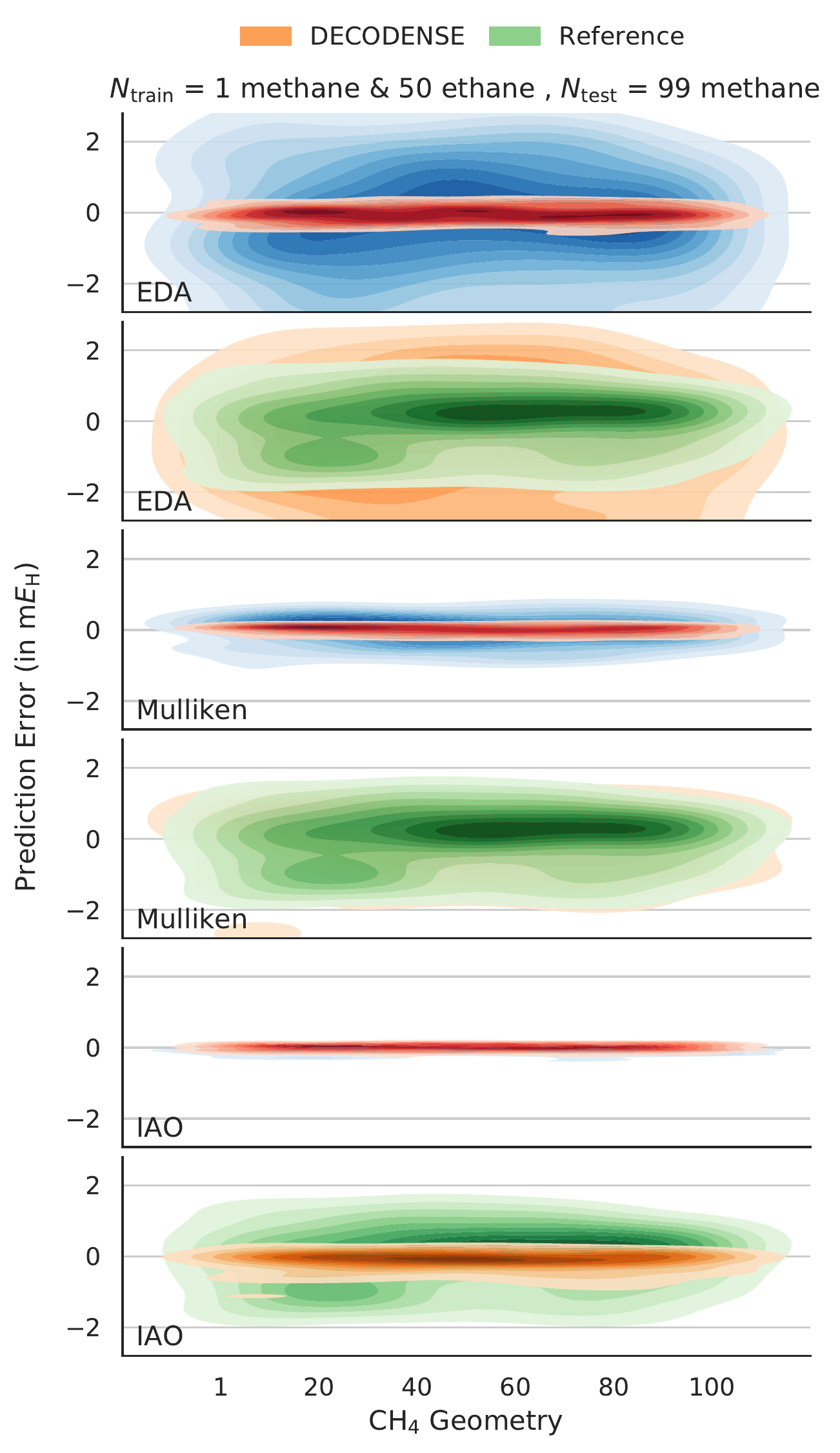}
\caption{Atomic (first-, third-, and fifth-row panels) and corresponding molecular (second-, fourth-, and sixth-row panels) predictions errors for the thermalized ground state of methane, as visualized by means of kernel density estimations (KDEs). As noted in the text, KDEs represent continuous probability density curves (analogous to a histogram), and the warmest colors thus indicate the areas of greatest density in the underlying scatter plots of errors.}
\label{methane_therm_100_fchl_b3lyp_ibo2_pc1_fig}
\end{center}
\vspace{-0.6cm}
\end{figure}
In Figure \ref{methane_therm_100_fchl_b3lyp_ibo2_pc1_fig}, out-of-sample prediction errors for the thermally accessible potential energy surface (PES) of methane at the B3LYP/pc-1 level of theory are compared from training on a single, randomly chosen methane geometry (left panel) or the same methane geometry alongside 50 randomly chosen ethane geometries (right panel). Comparing first the individual plots on the left panel, the Mulliken- and IAO-based results are observed to differ from the EDA-based results, primarily in the predictions of total energies and less so in those of individual hydrogen and carbon contributions. However, among themselves, the Mulliken- and IAO-based decompositions yield very similar results as expected on the basis of the polyacetylene results in Section \ref{calibration_polyacetylenes_subsection}. Common to all three sets of results is the observation that the prediction error for each of the four hydrogen contributions associated with a given geometry is lower than that for the corresponding carbon contribution.\\

By comparing the left and right panels of Figure \ref{methane_therm_100_fchl_b3lyp_ibo2_pc1_fig}, it is clear how prediction errors of the hydrogen and carbon contributions are greatly lowered with the inclusion of the ethane geometries in the training set, as ultimately evidenced from the fact that the total errors in the prediction of the methane molecular energies are correspondingly lowered as well. This improvement is observed, despite the fact that information on secondary rather than primary carbon atoms is added to the training set in moving from the left to the right panels of Figure \ref{methane_therm_100_fchl_b3lyp_ibo2_pc1_fig}. In comparison, the effect on traditional, molecular FCHL from this augmentation of the training set is much smaller, indicating a possible faster rate of learning new compositional diversity when training occurs on dedicated atomic contributions.\\

\begin{figure}[ht!]
\begin{center}
\includegraphics[width=0.5\textwidth]{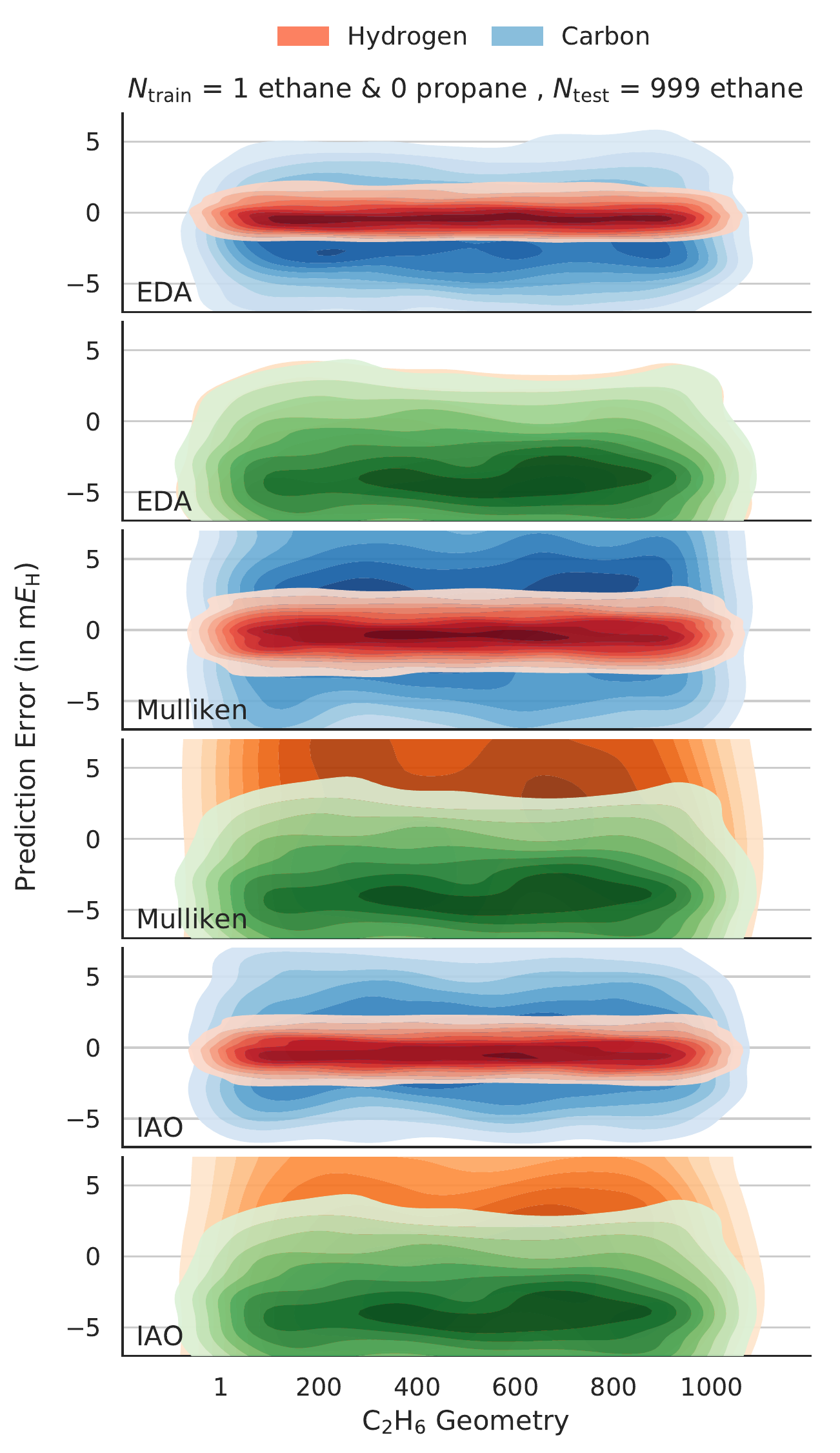}%
\includegraphics[width=0.5\textwidth]{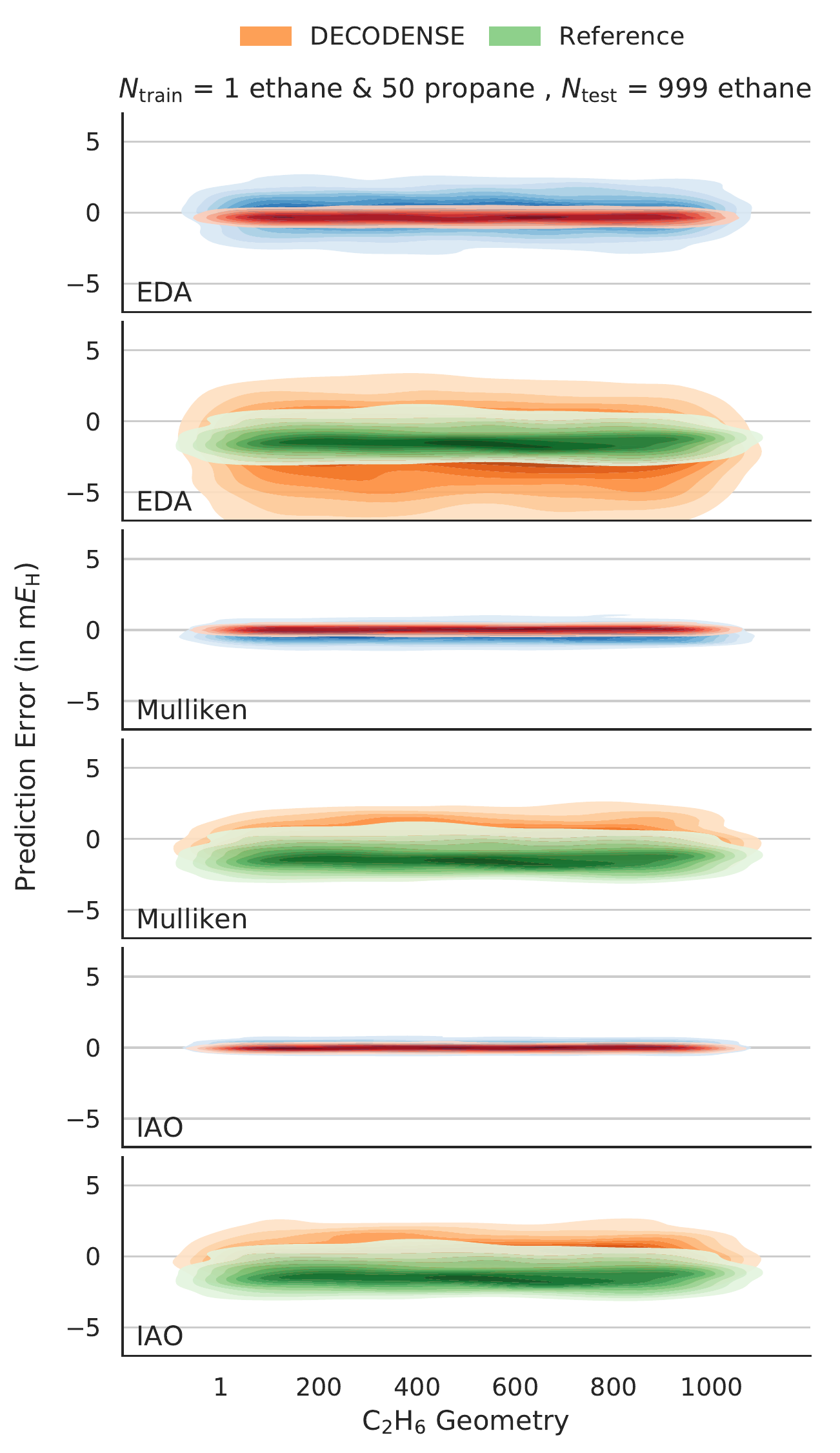}
\caption{KDEs of the atomic (first-, third-, and fifth-row panels) and molecular (second-, fourth-, and sixth-row panels) predictions errors for the thermalized ground state of ethane.}
\label{ethane_therm_1000_fchl_b3lyp_ibo2_pc1_fig}
\end{center}
\vspace{-0.6cm}
\end{figure}
\begin{figure}[ht!]
\begin{center}
\includegraphics[width=0.5\textwidth]{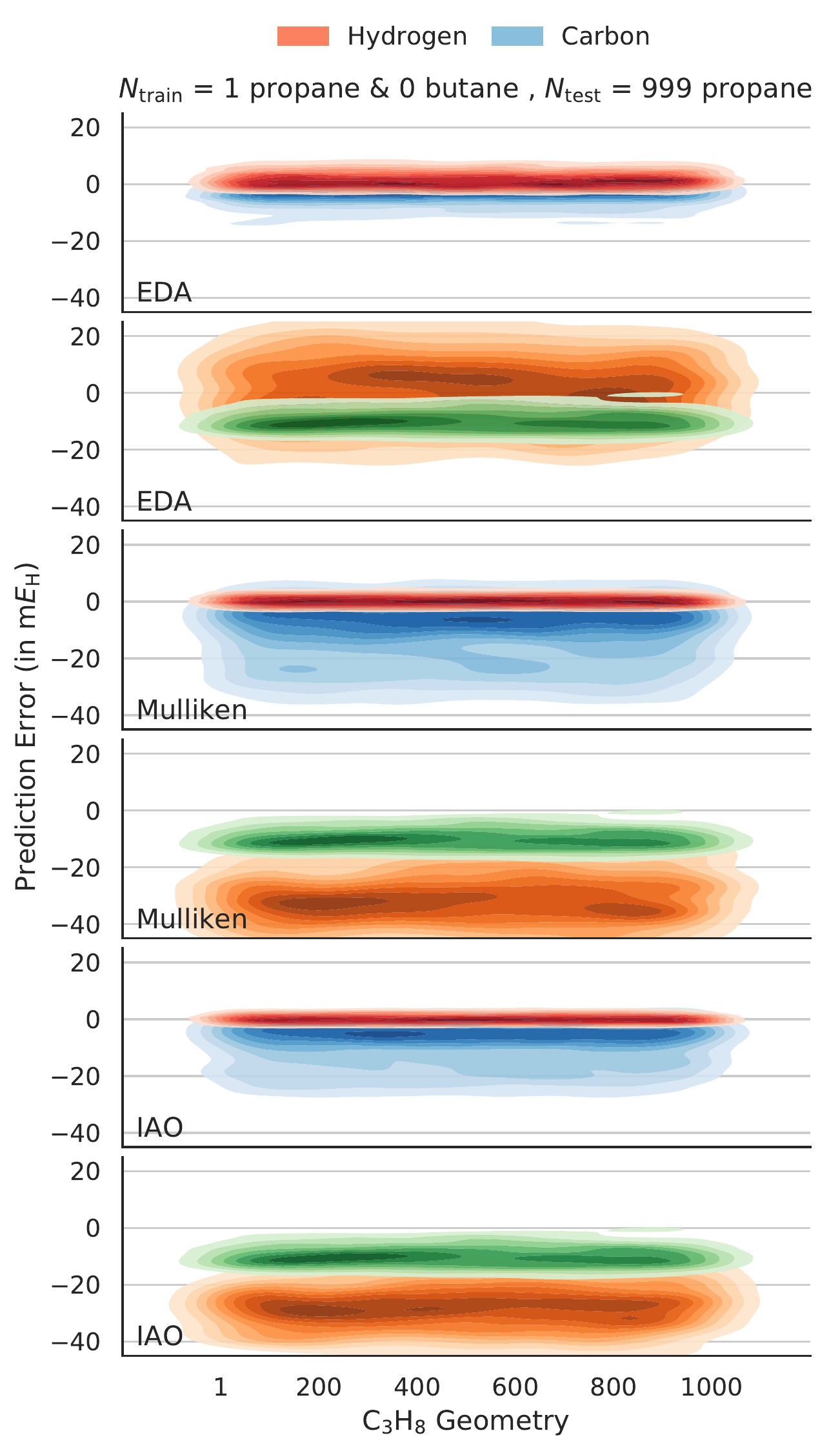}%
\includegraphics[width=0.5\textwidth]{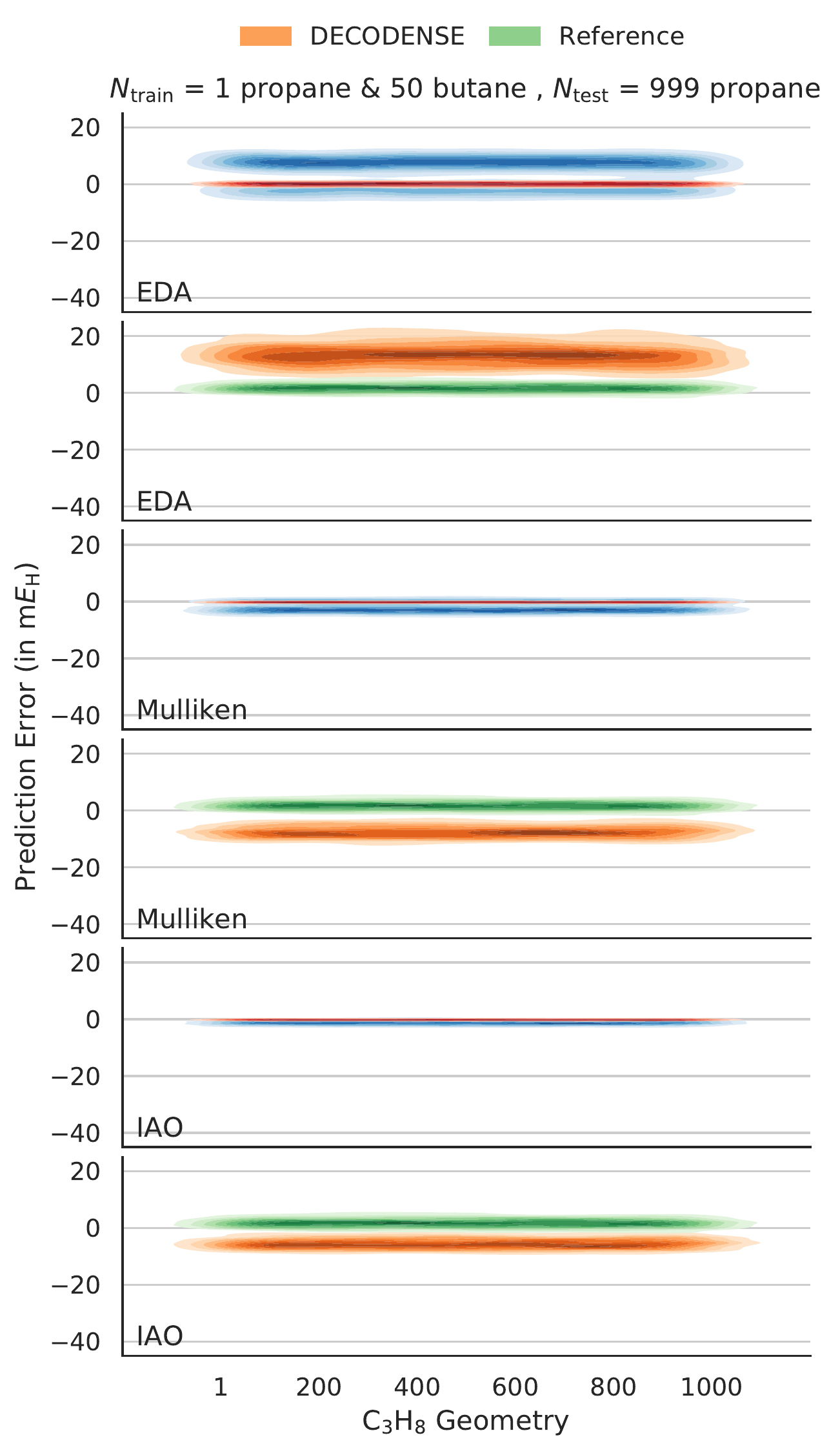}
\caption{KDEs of the atomic (first-, third-, and fifth-row panels) and molecular (second-, fourth-, and sixth-row panels) predictions errors for the thermalized ground state of propane.}
\label{propane_therm_1000_fchl_b3lyp_ibo2_pc1_fig}
\end{center}
\vspace{-0.6cm}
\end{figure}
To verify that the observations made for methane are not unique, we present similar results for ethane and propane in Figures \ref{ethane_therm_1000_fchl_b3lyp_ibo2_pc1_fig} and \ref{propane_therm_1000_fchl_b3lyp_ibo2_pc1_fig}, augmenting a single, random ethane (propane) geometry by 50 random propane (butane) geometries, respectively. In both cases, the general trends observed from Figure \ref{methane_therm_100_fchl_b3lyp_ibo2_pc1_fig} are replicated, namely, that the proposed atomic learning performs best on the basis of an IAO-based decomposition, less so on the basis of a Mulliken-based decomposition, and worst when based on an EDA partitioning. In the case of IAO-based molecular errors (generally observed to be on par with the reference results), these may be almost solely attributed to the errors associated with the prediction of individual carbon contributions, which are, in turn, observed to increase upon moving to larger species. This is only true to a lesser degree for the hydrogen contributions. We speculate this to be, in part, due to the smaller magnitude of these as well as the decreased heterogeneity of the hydrogen atoms with respect to the corresponding carbon centres. This hypothesis is further supported by Figures S2 and S3 of the SI, which show the distribution of the individual contributions associated with the hydrogen and carbon atoms in the butane results behind Figure \ref{propane_therm_1000_fchl_b3lyp_ibo2_pc1_fig} as well as how these correlate with the underlying partial IAO charges. From these results, it is further confirmed that the EDA partitioning fails to recognize the diversity between different primary and secondary carbon centres by erroneously assigning them similar contributions to the total molecular energy. The same holds true for the corresponding hydrogen atoms. This apparent lack of responsiveness to different chemical settings renders the EDA partitioning inept for the present purpose, in contrast to the Mulliken- and IAO-based decompositions which successfully make these distinctions.

\subsection{Water}\label{qml_water_subsection}

\begin{figure}[ht!]
\begin{center}
\includegraphics[width=0.5\textwidth]{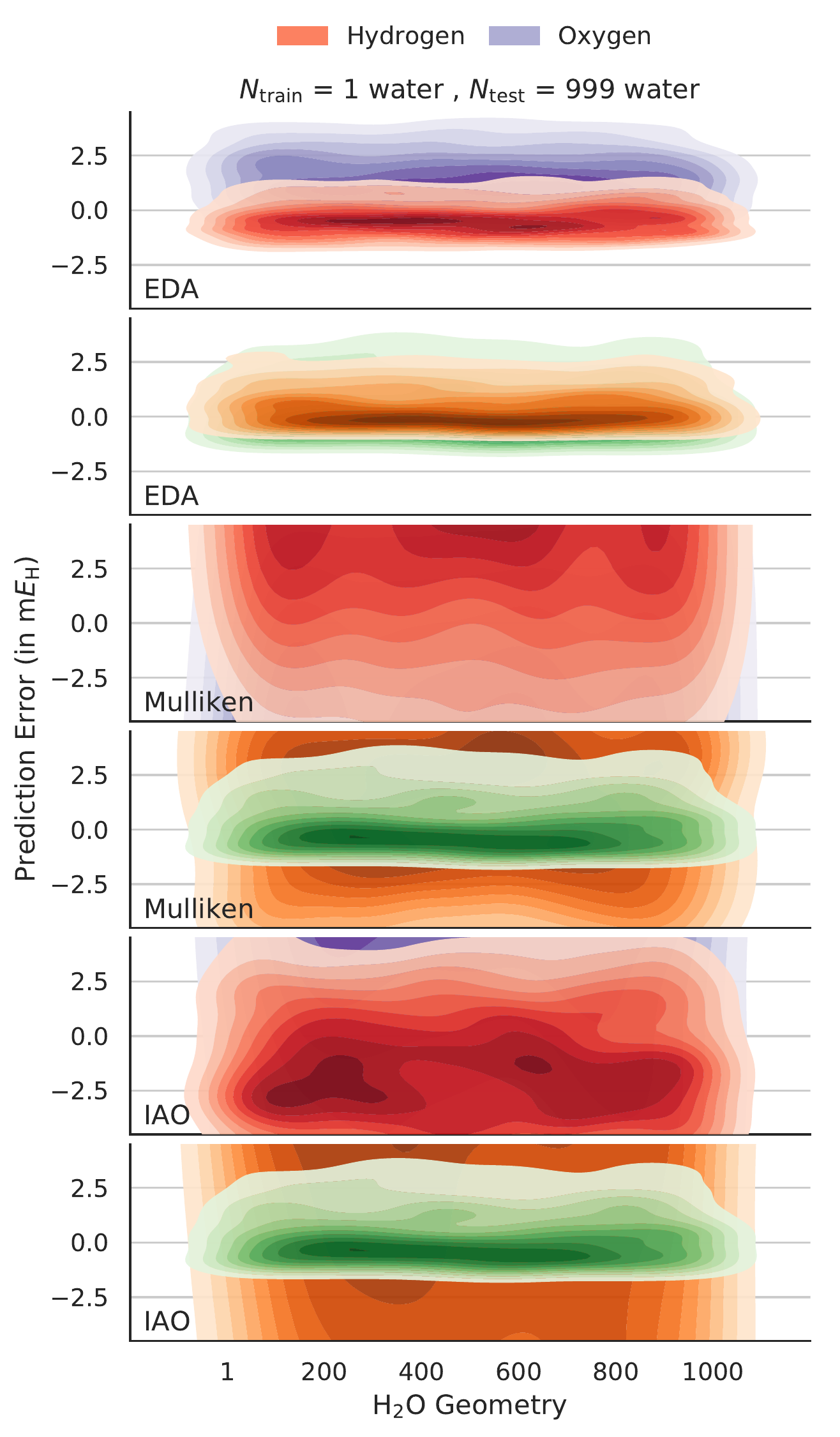}%
\includegraphics[width=0.5\textwidth]{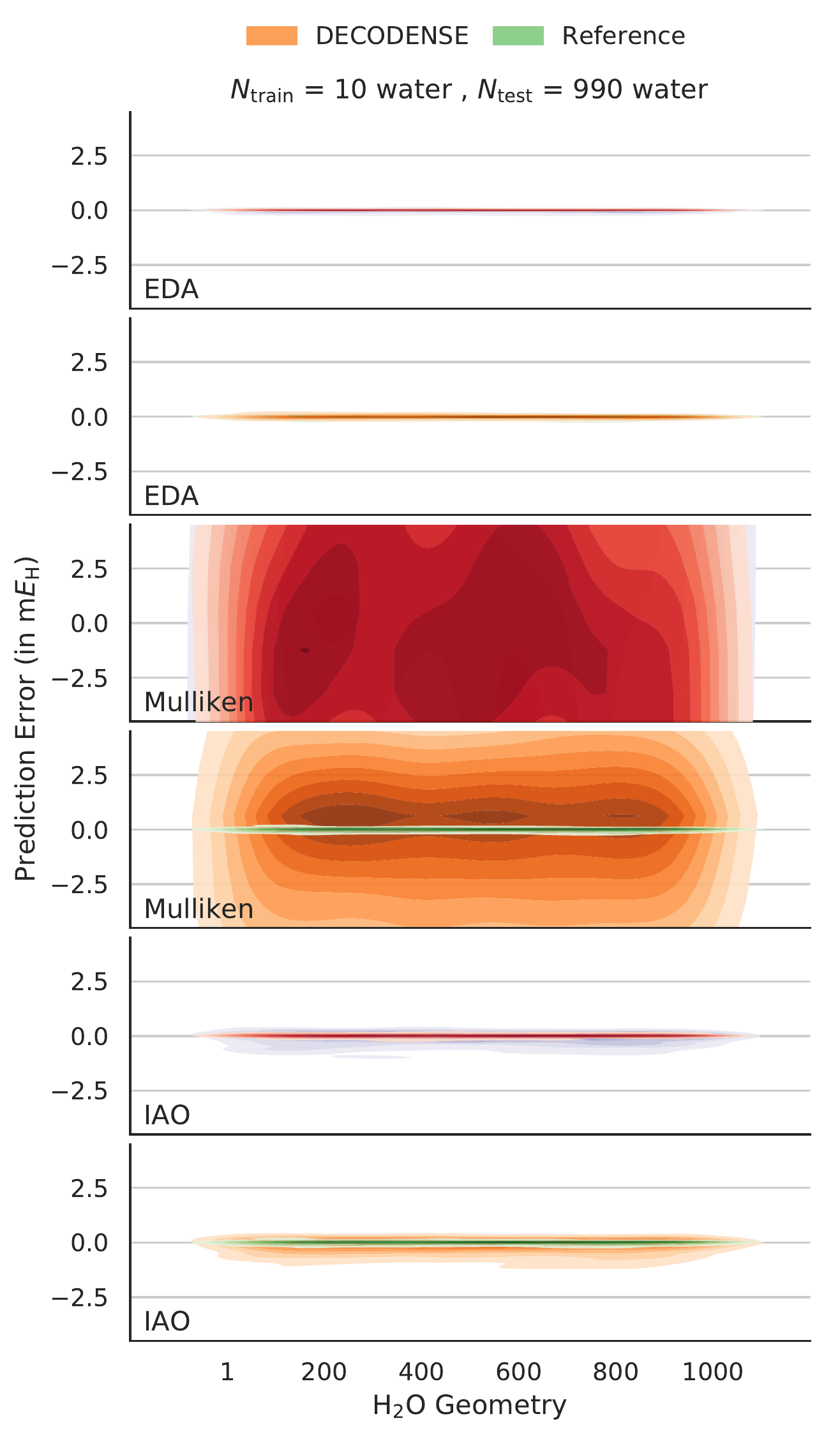}
\caption{KDEs of the atomic (first-, third-, and fifth-row panels) and molecular (second-, fourth-, and sixth-row panels) predictions errors for the thermalized ground state of water.}
\label{water_therm_1000_fchl_pbe0_ibo2_pc1_fig}
\end{center}
\vspace{-0.6cm}
\end{figure}
\begin{figure}[ht!]
\begin{center}
\includegraphics[scale=0.8]{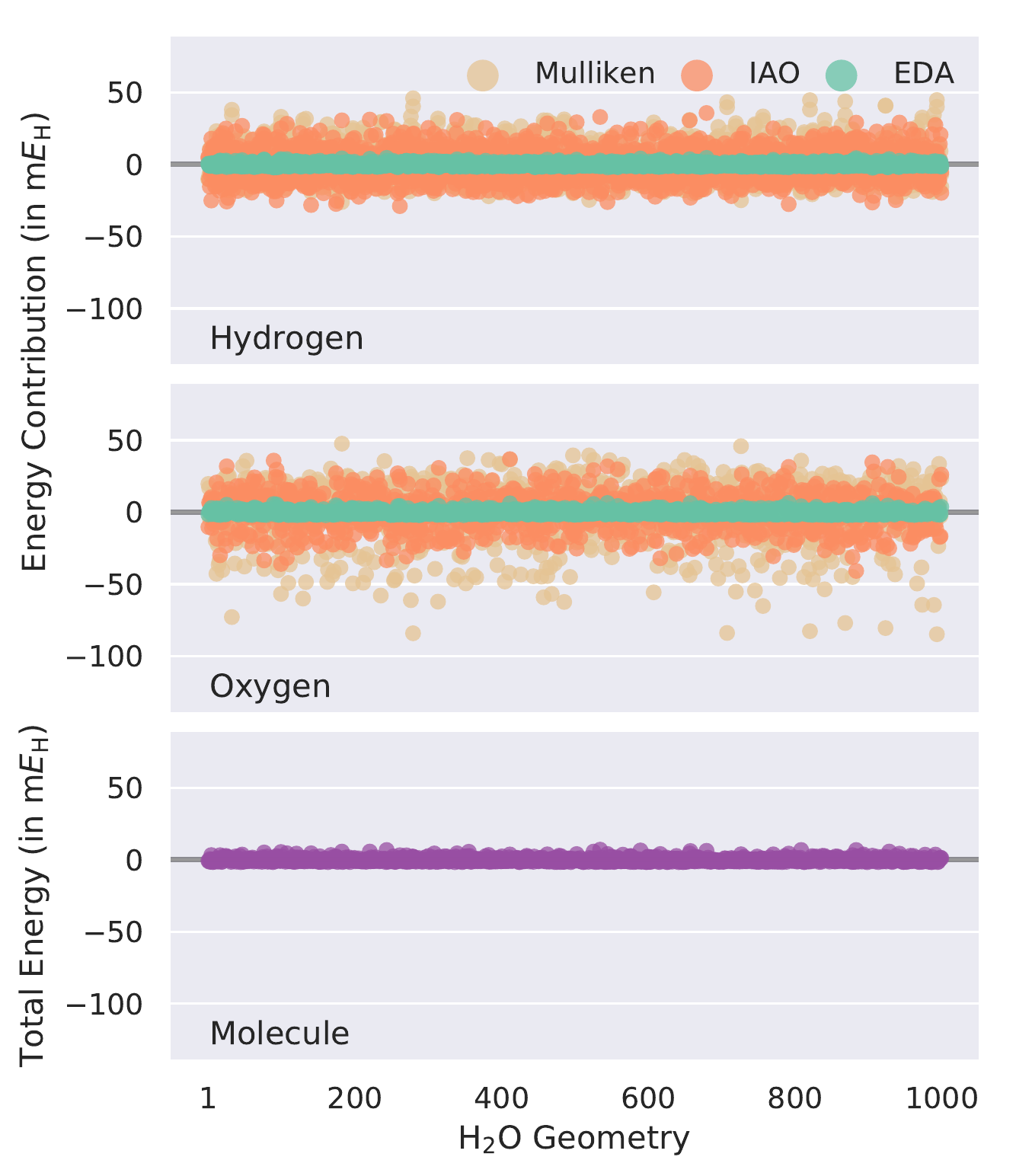}
\caption{Inspection of the atomic energy contributions and molecular energies behind Figure \ref{water_therm_1000_fchl_pbe0_ibo2_pc1_fig}, as measured against the respective mean values.}
\label{water_therm_1000_inspect_pbe0_ibo2_pc1_fig}
\end{center}
\vspace{-0.6cm}
\end{figure}
We next return to the case of water in Figure \ref{water_therm_1000_fchl_pbe0_ibo2_pc1_fig}, which presents prediction errors for a set of 1,000 thermalized water geometries at the PBE0/pc-1 level of theory. In contrast to the hydrocarbons in Section \ref{qml_hydrocarbons_subsection}, the prediction errors for water are remarkably different. When training on only a single water geometry, the Mulliken- and IAO-based results are noticeably worse off than the corresponding EDA-based results, which are slightly under-/overestimated for the hydrogen and oxygen contributions, respectively. Upon enlarging the training set to 10 random geometries, the differences between the IAO- and EDA-based results largely diminish, while the Mulliken-based results remain entirely unsystematic. In probing the cause of these differences in behaviour, Figure \ref{water_therm_1000_inspect_pbe0_ibo2_pc1_fig} presents the distribution of the atomic energy contributions and molecular energies behind Figure \ref{water_therm_1000_fchl_pbe0_ibo2_pc1_fig}. By plotting these as deviations around the respective mean values, the EDA partitioning is observed to reflect the overall invariance (on the scale of $<10$ m$E_{\text{H}}$) of the total energy on the thermally accessible PES of water. The results of the Mulliken- and IAO-based decompositions, on the other hand, are observed to fluctuate significantly around the mean, most severely so in the case of the former of the two.\\

\begin{figure}[ht!]
\begin{center}
\includegraphics[scale=0.8]{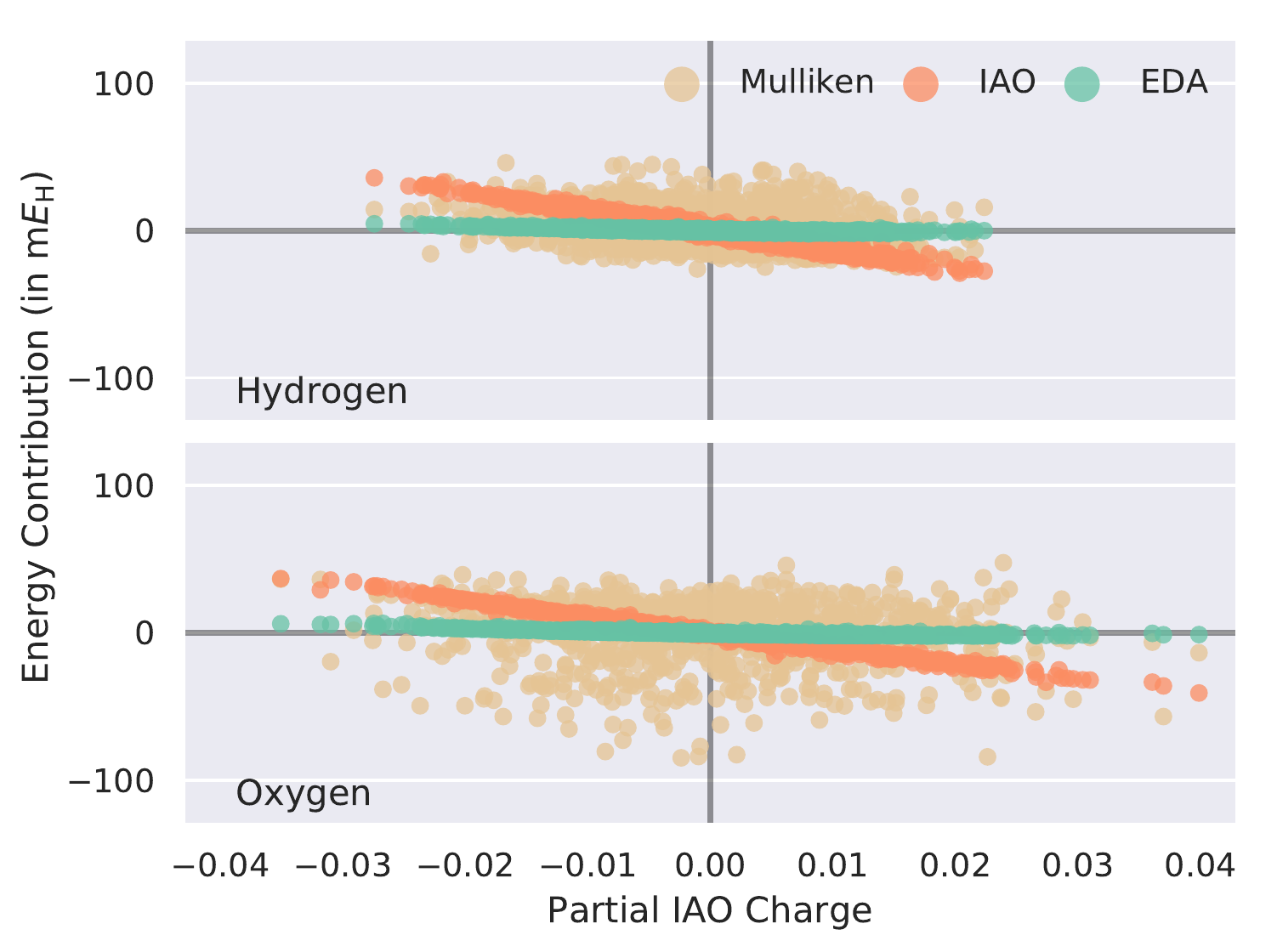}
\caption{Correlation of the atomic energy contributions with the corresponding IAO partial charges, as plotted around the respective mean values.}
\label{water_therm_1000_charge_pbe0_ibo2_pc1_fig}
\end{center}
\vspace{-0.6cm}
\end{figure}
In light of our previous decomposed results for water in Section \ref{calibration_water_subsection}, the above results now beg the question of whether or not atomic contributions should be expected to vary with structural distortion, on the basis of not only the underlying physics but also the extent to which the electronic structure is perturbed on the relatively narrow interval of the PES explored in Figures \ref{water_therm_1000_fchl_pbe0_ibo2_pc1_fig} and \ref{water_therm_1000_inspect_pbe0_ibo2_pc1_fig}. In Figure \ref{water_therm_1000_charge_pbe0_ibo2_pc1_fig}, the sensitivity of the atomic contributions with respect to changes in molecular structure is depicted, the latter of which is compositely represented by changes in IAO partial charges~\bibnote{The choice of IAO over Mulliken partial charges are justified in Figure S6 of the SI, in which the former are shown to vary continuously along the symmetric stretch in H$_2$O while also giving rise to smoothly varying atom-RDM1s}. The results in Figure \ref{water_therm_1000_charge_pbe0_ibo2_pc1_fig} are once again plotted as deviations from the respective mean values, implying that for the hydrogen contributions, positive (negative) relative charges correspond to compressed (expanded) geometries with respect to an average form, and {\textit{vice versa}} for the oxygen contributions. From the correlations in Figure \ref{water_therm_1000_charge_pbe0_ibo2_pc1_fig} or the lack hereof, we are in a position to rationalize the three different profiles observed in Figure \ref{water_therm_1000_fchl_pbe0_ibo2_pc1_fig}. First, the Mulliken-based results again appear erratic, failing to manifest any kind of sensible regularity. On the other hand, a clear correlation is observed for the IAO-based contributions, ultimately governed by the subtle interplay between quantum and steric effects. Upon compressing a water molecule, the electronic contributions associated with the involved hydrogen and oxygen atoms generally increase (in absolute terms), but this effect is naturally counterbalanced by corresponding increases in their repulsive nuclear contributions to the extent where the total oxygen energy contributions become influenced in a manner opposite to the electronic-only effects. Finally, the EDA-based results are observed to be systematically uncorrelated, in the sense that this partitioning yields atomic contributions of the same magnitude regardless of any perturbation to the molecular and electronic structures. In addition, the EDA-based results are again significantly less polarized on average than the corresponding IAO-based results, as may also be noted by comparing the atomization energy results in Figures S6 and S7 of the SI.\\

The results in Figure \ref{water_therm_1000_charge_pbe0_ibo2_pc1_fig} thus agree well with the previous results for the ground- and excited-state dipole moments of water in Section \ref{calibration_water_subsection}. Due to its formulation in the AO basis, the EDA partitioning fails to reflect the known change to the electronic structure of water that follows upon distortion away from its equilibrium geometry~\cite{olsen_bond_break_h2o_jcp_1996,chan_bond_break_h2o_jcp_2003}. As the archetype of a polar molecule, the imbalance in the response to structural perturbations that permeates the EDA partitioning now sheds some light on the reasons behind the machine-learned results obtained from it. In the left panel of Figure \ref{water_therm_1000_fchl_pbe0_ibo2_pc1_fig}, significant discrepancies in the results are to be expected since only a single (random) geometry is available for training an ML-QC model, while these prediction errors will necessarily get reduced upon transferring increasingly more knowledge of the configurational space into the model. Such a pattern is indeed observed for the IAO-based results, but not for those based on Mulliken charges nor the EDA partitioning. The former of these types of partitioning fails due to unpredictable shortcomings of the Mulliken charges themselves. The stellar performance of the EDA partitioning, on the other hand, for this system (unlike for the unpolar alkanes in Section \ref{qml_hydrocarbons_subsection}) is deemed to be due to data fitting alone and, as such, positively benefitting from the relative invariance of the total energy results in the lower panel of Figure \ref{water_therm_1000_inspect_pbe0_ibo2_pc1_fig}, rather than because the partitioning at its core succeeds in capturing the complexity of the underlying physics at play.\\

Turned on its head, and disregarding the Mulliken-based results from hereon, the disagreement between the IAO- and EDA-based results---coupled with the performance of the latter on par with standard FCHL for this system as well as the remaining discrepancies of the IAO-based model---furthermore calls into question the level at which state-of-the-art atom-based representations (e.g., FCHL) are sensitive and flexible enough to capture true quantum effects such as the redistribution of electronic density upon a change in geometry.\\

\begin{figure}[ht!]
\vspace{-0.6cm}
\begin{center}
\includegraphics[width=0.5\textwidth]{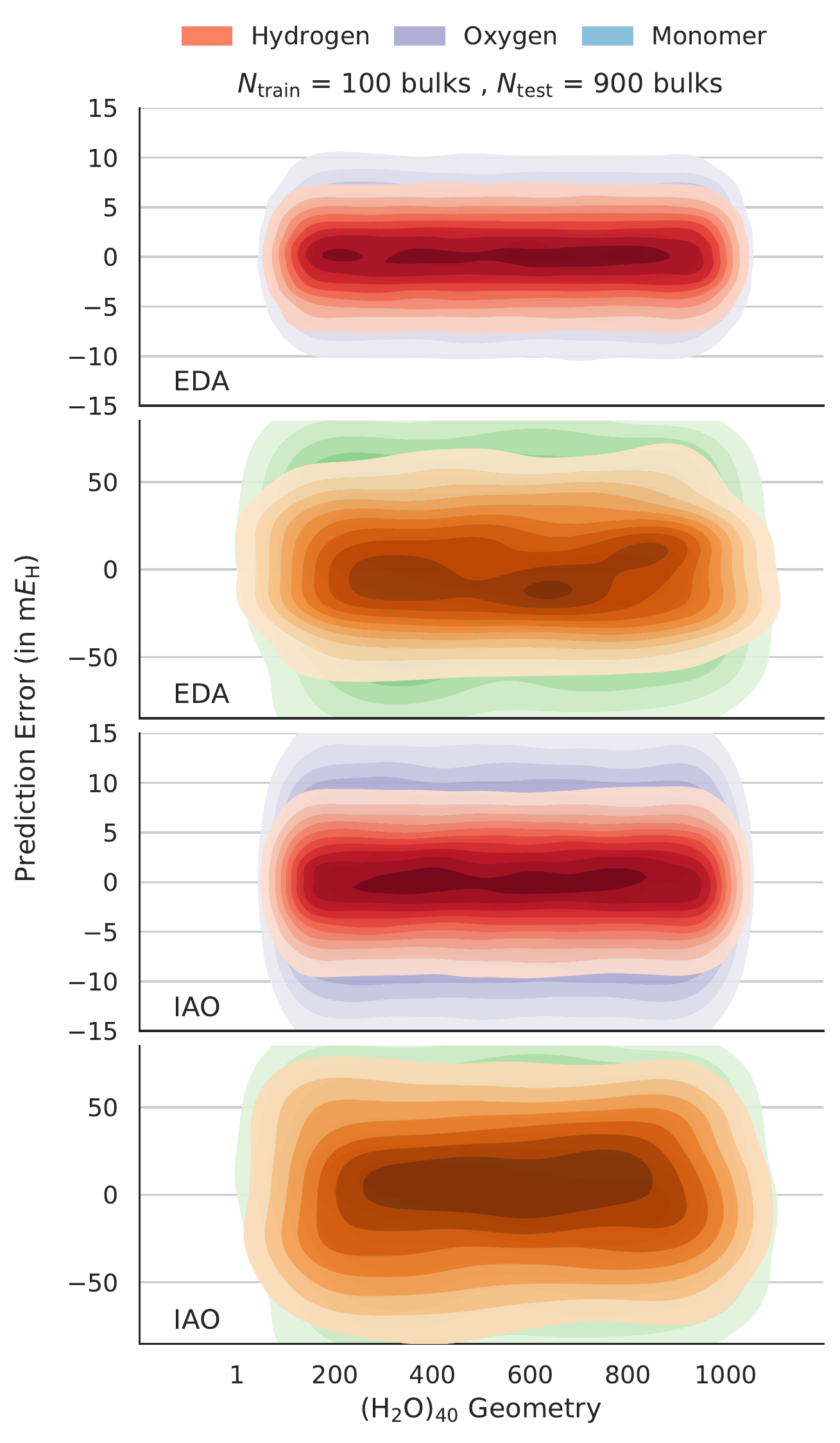}%
\includegraphics[width=0.5\textwidth]{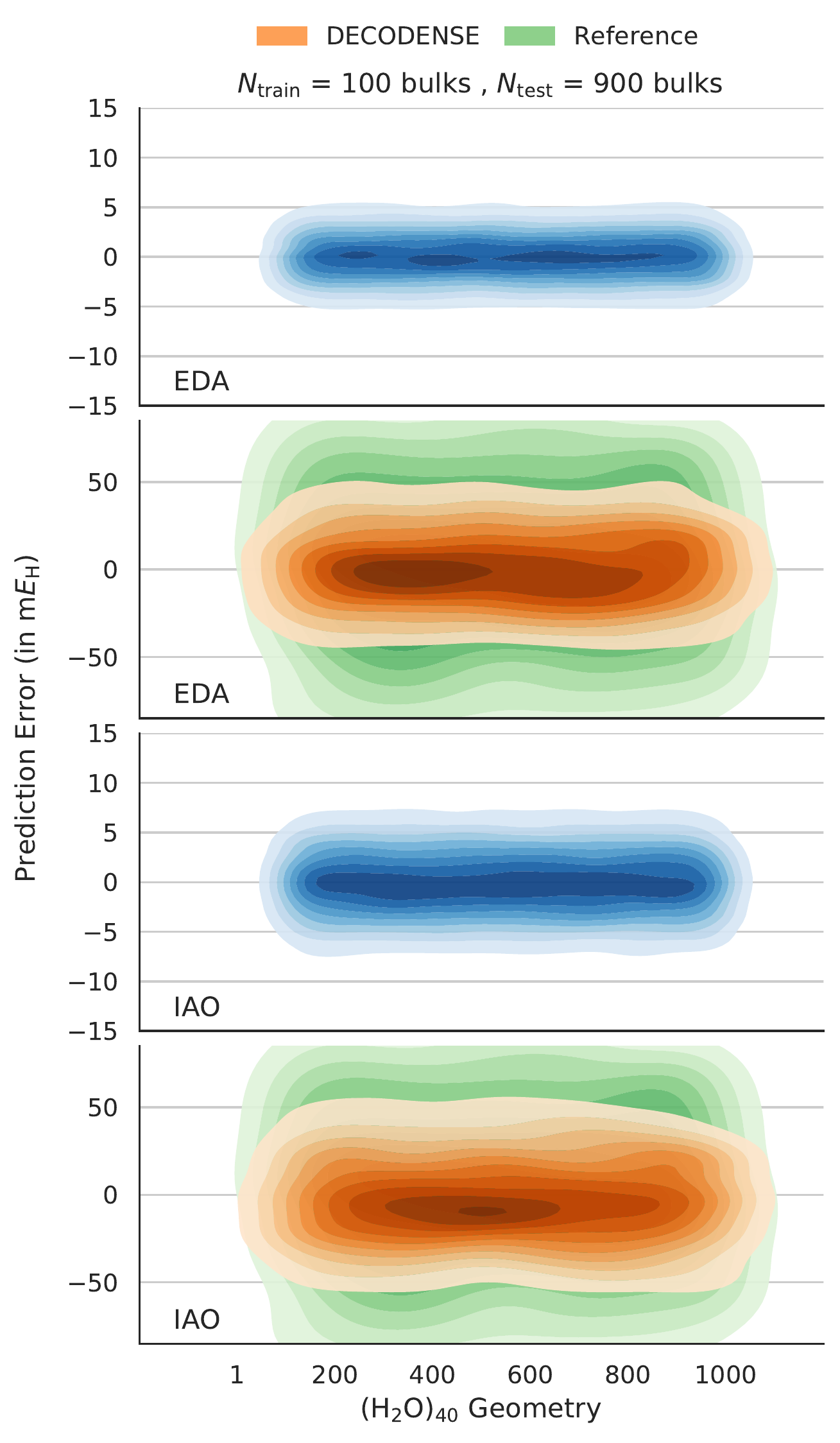}
\caption{KDEs of the atomic or monomeric (first- and third-row left or right panels, respectively) and molecular (second- and fourth-row panels) predictions errors for (H$_2$O)$_{40}$.}
\label{water_40_1000_fchl_pbe0_ibo2_pc1_fig}
\end{center}
\vspace{-0.6cm}
\end{figure}
Finally, we move from a single water molecule to the bulk phase by machine learning PBE0/pc-1 energies of 1,000 (H$_2$O)$_{40}$ clusters, randomly drawn from a larger set of 10,000 geometries (using $\lambda = \num{1.e-9}$). In this case, we may probe how the theory performs for a condensed-phase rather than an isolated system by training a model on the same data, but resolved on different scales; in particular, besides traditional FCHL resolved at a bulk level, one may take advantage of the flexibility of the present approach and group individual atomic contributions at an H$_2$O monomer level, constituting the smallest repetitive units in the bulk. The left and right parts of Figure \ref{water_40_1000_fchl_pbe0_ibo2_pc1_fig} show out-of-sample results for the atomic and monomeric learning models, respectively, when trained and tested on 100 and 900 bulk geometries, respectively. Once again, the merits of the present approach are obvious and this example thus illustrates how decomposed MF theory, in general, allows for ML-QC to be performed on data resolved at various resolutions. For instance, one might consider grouping the atoms of a large training set by their hybridization, oxidation level, or the chemical functional group (alcohols, thiols, esters, etc.) to which they belong, to name just a few examples. This type of guided ML-QC will hence contrast itself with the ongoing alchemical trend where one attempts to make wide-ranging extrapolations across the entirety of chemical space. Moving forward, we speculate that the closer ties to chemical intuition offered by the present approach might prove beneficial in applications to more general and heterogenous datasets where individual physical features can be exposed and amplified.

\section{Summary and Outlook}\label{outlook_summary_section}

We have introduced new decompositions of mean-field electronic structure theory, encompassing both HF and KS-DFT, which allow for total molecular first-order properties to be decomposed into contributions associated with individual bonds and atomic centres. Through numerical comparisons with alternative partitioning schemes from the literature, and by means of tests using a suite of different orbital localization and charge population protocols, we have shown how optimal choices of these admit the determination of consistent, robust, and physically intuitive atomic contributions to molecular energies and dipole moments. Besides alluding to the possibility of elucidating various electronic and bonding phenomena on the basis of the present decompositions, we have presented proof-of-concept results in support of their future applications in the context of machine-learned quantum chemistry. More in-depth applications along this research direction (e.g., to more diverse and heterogenous molecular datasets with corresponding learning curves) will be the topic of future work.\\

In addition to the present use of KRR, modern compute graph-based NN implementations of ML-QC represent an alternative application avenue, in particular, due to the fact that these models are more scalable and less limited by expansive training sets. In addition, NN implementations typically offer automatic differentiation engines and the extension to non-energetic properties may be significantly streamlined as a consequence. As examples, the open-source {\texttt{TensorMol}}~\cite{parkhill_tensormol_chem_sci_2018} and {\texttt{TorchANI}}~\cite{isayev_roitberg_torchani_jcim_2020} codes offer regression architectures and dataset utilities for preparing new, general-purpose atomistic potentials of this type for use in molecular modelling. We further envision that results of the present decompositions may be employed as a guiding tool for calibrating new representations and general ML models, particularly also when concerned with the learning of dipole moments for the efficient simulation of infrared spectroscopy~\cite{marquetand_ml_infrared_chem_sci_2017,pereira_aires_de_sousa_ml_dipole_jc_2018,lilienfeld_ml_qc_jcp_2019,ceriotti_ml_resp_prop_prl_2018,ceriotti_ml_polarizability_pnas_2019,ceriotti_ml_dipole_jcp_2020}, bypassing the need for predicting environment-dependent charges that seek to reproduce dipole moments in the best possible way.\\

However, descriptor-based ML-QC models will ultimately remain bound by the quality of the local representations on which they rely, and for any of these to be successful in the present context, they will need to be sensitive enough for detectable variances to be observed with respect to infinitesimal geometry displacements. In the recent Ref. \citenum{behler_lilienfeld_goedecker_ml_representation_mlst_2020}, based on a physically motivated measure of this (in)variance, relatively few modes were found to have a strong influence on the present FCHL representation for a selection of C$_{60}$ structures and it was generally observed to be inferior in this respect to other alternatives from the literature. For instance, the aforementioned SOAP~\cite{csanyi_ml_soap_prb_2013} and OM~\cite{goedecker_ml_crystal_jcp_2016} counterparts were found to exhibit a more satisfactory structural resolution. Geometric movements along displacement modes associated with large and small values of this proxy were furthermore found to influence the force on a reference atom in essentially the same way in the case of FCHL. Be that as it may, these deficiencies were not found to give rise to apparent
errors in the prediction of extensive properties, due, in part, to error cancellations. The results of the present work go some way to confirm these observations, however, viewed from a completely different angle. To that end, and in contrast to popular belief~\cite{de_vita_ml_represent_prb_2018,kresse_ml_represent_jcp_2020}, it has been demonstrated in Ref. \citenum{ceriotti_ml_representation_prl_2020} (and earlier in Ref. \citenum{lilienfeld_ml_representation_ijqc_2015}) that the existence of degenerate geometrical arrangements (in terms of the representations they give rise to) may introduce a pronounced distortion of the molecular feature space, in turn, resulting in a deterioration of the overall regression of, e.g., molecular energies.\\

Thus, the results of Ref. \citenum{behler_lilienfeld_goedecker_ml_representation_mlst_2020} may appear to suggest the OM and SOAP representations as potential better matches for the present decompositions. However, like FCHL, the latter of these is still constructed from two- and three-body features, which have been shown to be incapable of differentiating sizeable manifolds of atomic environments even for the simple, prototypical case of methane (cf. Section \ref{qml_hydrocarbons_subsection})~\cite{ceriotti_ml_representation_prl_2020,ceriotti_ml_representation_arxiv_2020}. Perhaps even more appealing are thus future combinations of our theory with DNNs capable of sophisticating the actual, dressed molecular representation as an integrated part of the learning process~\cite{carleo_troyer_ml_qc_science_2017,carleo_nomura_imada_ml_qc_nat_commun_2018,mueller_tkatchenko_ml_qc_nature_comm_2017,tkatchenko_mueller_maurer_ml_qc_nat_commun_2019,gastegger_maurer_ml_qc_jcp_2020,lubbers_smith_barros_ml_hip_nn_jcp_2018}, e.g., the {\texttt{SchNet}}~\cite{tkatchenko_mueller_ml_qc_adv_neural_2017,tkatchenko_mueller_ml_qc_jcp_2018}, {\texttt{PhysNet}}~\cite{unke_meuwly_ml_physnet_jctc_2019}, and {\texttt{AIMNet}}~\cite{isayev_ml_qc_sci_adv_2019} models. In particular, since the present decompositions are intimately tied to partial nuclear charges, as computed by means of physically sound population protocols, the manner in which meaningful representations of local chemical environments are learned may be improved through relaxation with respect to reference charges. Despite the notion put forward in Ref. \citenum{unke_meuwly_ml_physnet_jctc_2019} that decomposition schemes are essentially arbitrary ({\textit{sic}}), it is our expectation that the firm localized basis proposed in the present work, as opposed to a reliance on statistics only, might ameliorate these concerns. To that end, we consider the coupling of decomposed MF theory to DNNs an interesting future area of application, e.g., via an interface to the {\texttt{SchNetPack}} code~\cite{tkatchenko_mueller_ml_qc_jctc_2019}, operating under the assumption that the finer granularity following from a decomposition of MF data can help make high-dimensional DNNs more quantum and, thereby, more realistic and increasingly accurate, also in conjunction with recent rotationally equivariant enhancements~\cite{anderson_hy_kondor_cormorant_nips_2019,klicpera_gross_gruennemann_dimenet_2020,noe_e3nn_arxiv_2020,smidt_equivariant_chemrxiv_2020}.\\

As an aside, we note that rather than designing ML representations in terms of atom- or structure-specific attributes, the featurization of molecular systems may also be constructed directly in terms of quantum chemical quantities~\cite{miller_ml_qc_jctc_2018,miller_ml_qc_jcp_2019,miller_ml_qc_jctc_2019,miller_ml_qc_arxiv_2020,weinan_e_deephf_jpca_2020}. Most recently, feature vectors have been proposed that build upon the properties of low-cost, semi-empirical electronic structure calculations~\cite{grimme_gfn1_xtb_jctc_2017,grimme_gfn2_xtb_jctc_2019,grimme_gfn_xtb_review_wires_2020}, thereby partly mitigating the penalty associated with the necessity of a preceding and potentially inhibiting MF calculation to assemble the required MO basis. For instance, the {\texttt{OrbNet}} model has been introduced as a graph-convolution DNN-based approach to predict the difference between KS-DFT and semi-empirical total energies~\cite{miller_orbnet_jcp_2020}. The idea of somehow adapting the decompositions of the present work to semi-empirical theory, so as to facilitate a basis for featurized ML-QC models, also makes up a potential application.\\

Finally, given the classifications of individual bond-wise and atomic contributions offered by our take on decomposed MF theory, we further conjecture that IAO-based decompositions may serve future purposes as an interpretative tool when probing various physicochemical effects in molecular and extended systems. For instance, the underlying sources governing the relative stability of polyynic and cumulenic structural forms of pure carbon networks may be probed~\cite{almlof_feyereisen_c18_jacs_1991,plattner_houk_c18_jacs_1995,neiss_gorling_c18_chem_phys_chem_2014,kaiser_gross_anderson_c18_science_2019,bremond_sancho_garcia_c18_jcp_2019,baryshnikov_aagren_c18_jpcl_2019,stasyuk_voityuk_c18_chem_commun_2020,mazziotti_c18_pccp_2020}, and new light may potentially be shed on the many ways in which electronic structure influences the transmission of charges through molecules~\cite{solomon_ratner_interference_jcp_2008,solomon_ratner_current_nat_chem_2010,solomon_interference_nat_chem_2015,grozema_interference_chem_sci_2015}. Hitherto, quantum transport and interference phenomena have typically been rationalized from symmetry arguments in the underlying MO basis, but it is our belief that the shapes and relative contributions of orb- and atom-RDM1s have the potential to further aid in the understanding of a variety of complex relationships between structure and property~\cite{troisi_orb_sym_acie_2013,walsh_helical_orbs_chem_sci_2013,januszewskia_tykwinski_cumulenes_chem_soc_rev_2014,hoffmann_bond_trans_acs_nano_2015,hoffmann_solomon_interference_pnas_2016,solomon_elec_trans_jpcc_2018,solomon_current_dens_jpcc_2019,solomon_elec_trans_jpcl_2020}.

%
%
\section*{Acknowledgments}

The Independent Research Fund Denmark is gratefully acknowledged for financial support. The author further thanks Kieron Burke (UC Irvine) and Fred Manby (University of Bristol) for providing fruitful comments to an earlier draft of the present work.

%
%
\section*{Supporting Information}

The SI presents relevant molecular geometries (Tables S1 and S2), HF and M06-2X results similar to Figure \ref{h2o_dipmom_basis_fig} (Figure S1), plots similar to Figures \ref{water_therm_1000_inspect_pbe0_ibo2_pc1_fig} and \ref{water_therm_1000_charge_pbe0_ibo2_pc1_fig} for C$_4$H$_{10}$ (Figures S2 and S3) and H$_2$O (Figures S4 and S5), using different localized MOs in the latter case. Finally, plots showing results for the H$_2$O symmetric stretch PES are presented in Figures S6 and S7, alongside corresponding partial charges.

%
%
\section*{Data Availability}

Data in support of the findings of this study are available within the article and its SI.

\newpage

\providecommand{\latin}[1]{#1}
\makeatletter
\providecommand{\doi}
  {\begingroup\let\do\@makeother\dospecials
  \catcode`\{=1 \catcode`\}=2 \doi@aux}
\providecommand{\doi@aux}[1]{\endgroup\texttt{#1}}
\makeatother
\providecommand*\mcitethebibliography{\thebibliography}
\csname @ifundefined\endcsname{endmcitethebibliography}
  {\let\endmcitethebibliography\endthebibliography}{}


\begin{mcitethebibliography}{278}
\providecommand*\natexlab[1]{#1}
\providecommand*\mciteSetBstSublistMode[1]{}
\providecommand*\mciteSetBstMaxWidthForm[2]{}
\providecommand*\mciteBstWouldAddEndPuncttrue
  {\def\EndOfBibitem{\unskip.}}
\providecommand*\mciteBstWouldAddEndPunctfalse
  {\let\EndOfBibitem\relax}
\providecommand*\mciteSetBstMidEndSepPunct[3]{}
\providecommand*\mciteSetBstSublistLabelBeginEnd[3]{}
\providecommand*\EndOfBibitem{}
\mciteSetBstSublistMode{f}
\mciteSetBstMaxWidthForm{subitem}{(\alph{mcitesubitemcount})}
\mciteSetBstSublistLabelBeginEnd
  {\mcitemaxwidthsubitemform\space}
  {\relax}
  {\relax}

\bibitem[{\v{C}}{\'{i}}{\v{z}}ek(1966)]{cizek_1}
{\v{C}}{\'{i}}{\v{z}}ek,~J. {On the Correlation Problem in Atomic and Molecular
  Systems. Calculation of Wavefunction Components in Ursell-Type Expansion
  Using Quantum-Field Theoretical Methods}. \emph{{J}. {C}hem. {P}hys.}
  \textbf{1966}, \emph{45}, 4256\relax
\mciteBstWouldAddEndPuncttrue
\mciteSetBstMidEndSepPunct{\mcitedefaultmidpunct}
{\mcitedefaultendpunct}{\mcitedefaultseppunct}\relax
\EndOfBibitem
\bibitem[{\v{C}}{\'{i}}{\v{z}}ek(1969)]{cizek_2}
{\v{C}}{\'{i}}{\v{z}}ek,~J. {On the Use of the Cluster Expansion and the
  Technique of Diagrams in Calculations of Correlation Effects in Atoms and
  Molecules}. \emph{{A}dv. {C}hem. {P}hys.} \textbf{1969}, \emph{14}, 35\relax
\mciteBstWouldAddEndPuncttrue
\mciteSetBstMidEndSepPunct{\mcitedefaultmidpunct}
{\mcitedefaultendpunct}{\mcitedefaultseppunct}\relax
\EndOfBibitem
\bibitem[Paldus \latin{et~al.}(1972)Paldus, {\v{C}}{\'{i}}{\v{z}}ek, and
  Shavitt]{paldus_cizek_shavitt}
Paldus,~J.; {\v{C}}{\'{i}}{\v{z}}ek,~J.; Shavitt,~I. {Correlation Problems in
  Atomic and Molecular Systems. IV. Extended Coupled-Pair Many-Electron Theory
  and Its Application to the BH$_3$ Molecule}. \emph{Phys. Rev. A}
  \textbf{1972}, \emph{5}, 50\relax
\mciteBstWouldAddEndPuncttrue
\mciteSetBstMidEndSepPunct{\mcitedefaultmidpunct}
{\mcitedefaultendpunct}{\mcitedefaultseppunct}\relax
\EndOfBibitem
\bibitem[Shavitt and Bartlett(2009)Shavitt, and
  Bartlett]{shavitt_bartlett_cc_book}
Shavitt,~I.; Bartlett,~R.~J. \emph{{M}any-{B}ody {M}ethods in {C}hemistry and
  {P}hysics: {M}any-{B}ody {P}erturbation {T}heory and {C}oupled-{C}luster
  {T}heory}; Cambridge University Press: Cambridge, UK, 2009\relax
\mciteBstWouldAddEndPuncttrue
\mciteSetBstMidEndSepPunct{\mcitedefaultmidpunct}
{\mcitedefaultendpunct}{\mcitedefaultseppunct}\relax
\EndOfBibitem
\bibitem[Helgaker \latin{et~al.}(2000)Helgaker, J{\o}rgensen, and Olsen]{mest}
Helgaker,~T.; J{\o}rgensen,~P.; Olsen,~J. \emph{{M}olecular
  {E}lectronic-{S}tructure {T}heory}, 1st ed.; Wiley \& Sons, Ltd.: West
  Sussex, UK, 2000\relax
\mciteBstWouldAddEndPuncttrue
\mciteSetBstMidEndSepPunct{\mcitedefaultmidpunct}
{\mcitedefaultendpunct}{\mcitedefaultseppunct}\relax
\EndOfBibitem
\bibitem[Booth \latin{et~al.}(2009)Booth, Thom, and
  Alavi]{booth_alavi_fciqmc_jcp_2009}
Booth,~G.~H.; Thom,~A. J.~W.; Alavi,~A. {Fermion Monte Carlo Without Fixed
  Nodes: A Game of Life, Death, and Annihilation in Slater Determinant Space}.
  \emph{{J}. {C}hem. {P}hys.} \textbf{2009}, \emph{131}, 054106\relax
\mciteBstWouldAddEndPuncttrue
\mciteSetBstMidEndSepPunct{\mcitedefaultmidpunct}
{\mcitedefaultendpunct}{\mcitedefaultseppunct}\relax
\EndOfBibitem
\bibitem[Cleland \latin{et~al.}(2010)Cleland, Booth, and
  Alavi]{cleland_booth_alavi_fciqmc_jcp_2010}
Cleland,~D.; Booth,~G.~H.; Alavi,~A. {Communications: Survival of the Fittest:
  Accelerating Convergence in Full Configuration-Interaction Quantum Monte
  Carlo}. \emph{{J}. {C}hem. {P}hys.} \textbf{2010}, \emph{132}, 041103\relax
\mciteBstWouldAddEndPuncttrue
\mciteSetBstMidEndSepPunct{\mcitedefaultmidpunct}
{\mcitedefaultendpunct}{\mcitedefaultseppunct}\relax
\EndOfBibitem
\bibitem[Deustua \latin{et~al.}(2017)Deustua, Shen, and
  Piecuch]{piecuch_monte_carlo_cc_prl_2017}
Deustua,~J.~E.; Shen,~J.; Piecuch,~P. {Converging High-Level Coupled-Cluster
  Energetics by Monte Carlo Sampling and Moment Expansions}. \emph{Phys. Rev.
  Lett.} \textbf{2017}, \emph{119}, 223003\relax
\mciteBstWouldAddEndPuncttrue
\mciteSetBstMidEndSepPunct{\mcitedefaultmidpunct}
{\mcitedefaultendpunct}{\mcitedefaultseppunct}\relax
\EndOfBibitem
\bibitem[Sharma \latin{et~al.}(2017)Sharma, Holmes, Jeanmairet, Alavi, and
  Umrigar]{sharma_umrigar_heat_bath_ci_jctc_2017}
Sharma,~S.; Holmes,~A.~A.; Jeanmairet,~G.; Alavi,~A.; Umrigar,~C.~J.
  {Semistochastic Heat-Bath Configuration Interaction Method: Selected
  Configuration Interaction with Semistochastic Perturbation Theory}.
  \emph{{J}. {C}hem. {T}heory {C}omput.} \textbf{2017}, \emph{13}, 1595\relax
\mciteBstWouldAddEndPuncttrue
\mciteSetBstMidEndSepPunct{\mcitedefaultmidpunct}
{\mcitedefaultendpunct}{\mcitedefaultseppunct}\relax
\EndOfBibitem
\bibitem[White(1992)]{white_dmrg_prl_1992}
White,~S.~R. {Density Matrix Formulation for Quantum Renormalization Groups}.
  \emph{{P}hys. {R}ev. {L}ett.} \textbf{1992}, \emph{69}, 2863\relax
\mciteBstWouldAddEndPuncttrue
\mciteSetBstMidEndSepPunct{\mcitedefaultmidpunct}
{\mcitedefaultendpunct}{\mcitedefaultseppunct}\relax
\EndOfBibitem
\bibitem[White(1993)]{white_dmrg_prb_1993}
White,~S.~R. {Density-Matrix Algorithms for Quantum Renormalization Groups}.
  \emph{{P}hys. {R}ev. {B}} \textbf{1993}, \emph{48}, 10345\relax
\mciteBstWouldAddEndPuncttrue
\mciteSetBstMidEndSepPunct{\mcitedefaultmidpunct}
{\mcitedefaultendpunct}{\mcitedefaultseppunct}\relax
\EndOfBibitem
\bibitem[White and Martin(1999)White, and Martin]{white_martin_dmrg_jcp_1999}
White,~S.~R.; Martin,~R.~L. {{\it{Ab Initio}} Quantum Chemistry using the
  Density Matrix Renormalization Group}. \emph{{J}. {C}hem. {P}hys.}
  \textbf{1999}, \emph{110}, 4127\relax
\mciteBstWouldAddEndPuncttrue
\mciteSetBstMidEndSepPunct{\mcitedefaultmidpunct}
{\mcitedefaultendpunct}{\mcitedefaultseppunct}\relax
\EndOfBibitem
\bibitem[Chan and Sharma(2011)Chan, and
  Sharma]{chan_sharma_dmrg_review_arpc_2011}
Chan,~G. K.-L.; Sharma,~S. {The Density Matrix Renormalization Group in Quantum
  Chemistry}. \emph{{A}nnu. {R}ev. {P}hys. {C}hem.} \textbf{2011}, \emph{62},
  465\relax
\mciteBstWouldAddEndPuncttrue
\mciteSetBstMidEndSepPunct{\mcitedefaultmidpunct}
{\mcitedefaultendpunct}{\mcitedefaultseppunct}\relax
\EndOfBibitem
\bibitem[Eriksen \latin{et~al.}(2017)Eriksen, Lipparini, and
  Gauss]{eriksen_mbe_fci_jpcl_2017}
Eriksen,~J.~J.; Lipparini,~F.; Gauss,~J. {Virtual Orbital Many-Body Expansions:
  A Possible Route towards the Full Configuration Interaction Limit}.
  \emph{{J}. {P}hys. {C}hem. {L}ett.} \textbf{2017}, \emph{8}, 4633\relax
\mciteBstWouldAddEndPuncttrue
\mciteSetBstMidEndSepPunct{\mcitedefaultmidpunct}
{\mcitedefaultendpunct}{\mcitedefaultseppunct}\relax
\EndOfBibitem
\bibitem[Eriksen and Gauss(2019)Eriksen, and
  Gauss]{eriksen_mbe_fci_general_jpcl_2019}
Eriksen,~J.~J.; Gauss,~J. {Generalized Many-Body Expanded Full Configuration
  Interaction Theory}. \emph{{J}. {P}hys. {C}hem. {L}ett.} \textbf{2019},
  \emph{10}, 7910\relax
\mciteBstWouldAddEndPuncttrue
\mciteSetBstMidEndSepPunct{\mcitedefaultmidpunct}
{\mcitedefaultendpunct}{\mcitedefaultseppunct}\relax
\EndOfBibitem
\bibitem[Eriksen \latin{et~al.}(2020)Eriksen, Anderson, Deustua, Ghanem, Hait,
  Hoffmann, Lee, Levine, Magoulas, Shen, Tubman, Whaley, Xu, Yao, Zhang, Alavi,
  Chan, Head-Gordon, Liu, Piecuch, Sharma, Ten-no, Umrigar, and
  Gauss]{eriksen_benzene_jpcl_2020}
Eriksen,~J.~J.; Anderson,~T.~A.; Deustua,~J.~E.; Ghanem,~K.; Hait,~D.;
  Hoffmann,~M.~R.; Lee,~S.; Levine,~D.~S.; Magoulas,~I.; Shen,~J.;
  Tubman,~N.~M.; Whaley,~K.~B.; Xu,~E.; Yao,~Y.; Zhang,~N.; Alavi,~A.; Chan,~G.
  K.-L.; Head-Gordon,~M.; Liu,~W.; Piecuch,~P.; Sharma,~S.; Ten-no,~S.~L.;
  Umrigar,~C.~J.; Gauss,~J. {The Ground State Electronic Energy of Benzene}.
  \emph{{J}. {P}hys. {C}hem. {L}ett.} \textbf{2020}, \emph{11}, 8922\relax
\mciteBstWouldAddEndPuncttrue
\mciteSetBstMidEndSepPunct{\mcitedefaultmidpunct}
{\mcitedefaultendpunct}{\mcitedefaultseppunct}\relax
\EndOfBibitem
\bibitem[Hohenberg and Kohn(1964)Hohenberg, and
  Kohn]{hohenberg_kohn_hk_theorem_phys_rev_1964}
Hohenberg,~P.; Kohn,~W. {Inhomogeneous Electron Gas}. \emph{Phys. Rev.}
  \textbf{1964}, \emph{136}, B864\relax
\mciteBstWouldAddEndPuncttrue
\mciteSetBstMidEndSepPunct{\mcitedefaultmidpunct}
{\mcitedefaultendpunct}{\mcitedefaultseppunct}\relax
\EndOfBibitem
\bibitem[Kohn and Sham(1965)Kohn, and Sham]{kohn_sham_ks_dft_phys_rev_1965}
Kohn,~W.; Sham,~L.~J. {Self-Consistent Equations Including Exchange and
  Correlation Effects}. \emph{Phys. Rev.} \textbf{1965}, \emph{140},
  A1133\relax
\mciteBstWouldAddEndPuncttrue
\mciteSetBstMidEndSepPunct{\mcitedefaultmidpunct}
{\mcitedefaultendpunct}{\mcitedefaultseppunct}\relax
\EndOfBibitem
\bibitem[Parr and Yang(1994)Parr, and Yang]{parr_yang_dft_book}
Parr,~R.~G.; Yang,~W. \emph{{Density-Functional Theory of Atoms and
  Molecules}}; Oxford University Press: Oxford, UK, 1994\relax
\mciteBstWouldAddEndPuncttrue
\mciteSetBstMidEndSepPunct{\mcitedefaultmidpunct}
{\mcitedefaultendpunct}{\mcitedefaultseppunct}\relax
\EndOfBibitem
\bibitem[Cohen \latin{et~al.}(2012)Cohen, Mori-S{\'a}nchez, and
  Yang]{yang_dft_review_chem_rev_2012}
Cohen,~A.~J.; Mori-S{\'a}nchez,~P.; Yang,~W. {Challenges for Density Functional
  Theory}. \emph{Chem. Rev.} \textbf{2012}, \emph{112}, 289\relax
\mciteBstWouldAddEndPuncttrue
\mciteSetBstMidEndSepPunct{\mcitedefaultmidpunct}
{\mcitedefaultendpunct}{\mcitedefaultseppunct}\relax
\EndOfBibitem
\bibitem[Becke(2014)]{becke_dft_review_jcp_2014}
Becke,~A.~D. {Perspective: Fifty Years of Density-Functional Theory in Chemical
  Physics}. \emph{J. Chem. Phys.} \textbf{2014}, \emph{140}, 18A301\relax
\mciteBstWouldAddEndPuncttrue
\mciteSetBstMidEndSepPunct{\mcitedefaultmidpunct}
{\mcitedefaultendpunct}{\mcitedefaultseppunct}\relax
\EndOfBibitem
\bibitem[Burke(2012)]{burke_dft_review_jcp_2012}
Burke,~K. {Perspective on Density Functional Theory}. \emph{J. Chem. Phys.}
  \textbf{2012}, \emph{136}, 150901\relax
\mciteBstWouldAddEndPuncttrue
\mciteSetBstMidEndSepPunct{\mcitedefaultmidpunct}
{\mcitedefaultendpunct}{\mcitedefaultseppunct}\relax
\EndOfBibitem
\bibitem[Pribram-Jones \latin{et~al.}(2015)Pribram-Jones, Gross, and
  Burke]{burke_dft_review_arpc_2015}
Pribram-Jones,~A.; Gross,~D.~A.; Burke,~K. {DFT: A Theory Full of Holes?}
  \emph{Ann. Rev. Phys. Chem.} \textbf{2015}, \emph{66}, 283\relax
\mciteBstWouldAddEndPuncttrue
\mciteSetBstMidEndSepPunct{\mcitedefaultmidpunct}
{\mcitedefaultendpunct}{\mcitedefaultseppunct}\relax
\EndOfBibitem
\bibitem[Mardirossian and Head-Gordon(2017)Mardirossian, and
  Head-Gordon]{mardirossian_head_gordon_dft_review_mol_phys_2017}
Mardirossian,~N.; Head-Gordon,~M. {Thirty Years of Density Functional Theory in
  Computational Chemistry: An Overview and Extensive Assessment of 200 Density
  Functionals}. \emph{Mol. Phys.} \textbf{2017}, \emph{115}, 2315\relax
\mciteBstWouldAddEndPuncttrue
\mciteSetBstMidEndSepPunct{\mcitedefaultmidpunct}
{\mcitedefaultendpunct}{\mcitedefaultseppunct}\relax
\EndOfBibitem
\bibitem[Mir{\'o} \latin{et~al.}(2014)Mir{\'o}, Audiffred, and
  Heine]{heine_2d_atlas_chem_soc_rev_2014}
Mir{\'o},~P.; Audiffred,~M.; Heine,~T. {An Atlas of Two-Dimensional Materials}.
  \emph{Chem. Soc. Rev.} \textbf{2014}, \emph{43}, 6537\relax
\mciteBstWouldAddEndPuncttrue
\mciteSetBstMidEndSepPunct{\mcitedefaultmidpunct}
{\mcitedefaultendpunct}{\mcitedefaultseppunct}\relax
\EndOfBibitem
\bibitem[Rasmussen and Thygesen(2015)Rasmussen, and
  Thygesen]{thygesen_2d_database_jpcc_2015}
Rasmussen,~F.~A.; Thygesen,~K.~S. {Computational 2D Materials Database:
  Electronic Structure of Transition-Metal Dichalcogenides and Oxides}.
  \emph{J. Phys. Chem. C} \textbf{2015}, \emph{119}, 13169\relax
\mciteBstWouldAddEndPuncttrue
\mciteSetBstMidEndSepPunct{\mcitedefaultmidpunct}
{\mcitedefaultendpunct}{\mcitedefaultseppunct}\relax
\EndOfBibitem
\bibitem[Evans \latin{et~al.}(2017)Evans, Fraux, Gaillac, Kohen, Trousselet,
  Vanson, and Coudert]{coudert_nanoporous_mat_chem_mater_2017}
Evans,~J.~D.; Fraux,~G.; Gaillac,~R.; Kohen,~D.; Trousselet,~F.; Vanson,~J.-M.;
  Coudert,~F.-X. {Computational Chemistry Methods for Nanoporous Materials}.
  \emph{Chem. Mater.} \textbf{2017}, \emph{29}, 199\relax
\mciteBstWouldAddEndPuncttrue
\mciteSetBstMidEndSepPunct{\mcitedefaultmidpunct}
{\mcitedefaultendpunct}{\mcitedefaultseppunct}\relax
\EndOfBibitem
\bibitem[Maurer \latin{et~al.}(2019)Maurer, Freysoldt, Reilly, Brandenburg,
  Hofmann, Bj{\"o}rkman, Leb{\`e}gue, and
  Tkatchenko]{maurer_tkatchenko_dft_materials_ann_rev_mat_res_2019}
Maurer,~R.~J.; Freysoldt,~C.; Reilly,~A.~M.; Brandenburg,~J.~G.;
  Hofmann,~O.~T.; Bj{\"o}rkman,~T.; Leb{\`e}gue,~S.; Tkatchenko,~A. {Advances
  in Density-Functional Calculations for Materials Modeling}. \emph{Annu. Rev.
  Mater. Res.} \textbf{2019}, \emph{49}, 1\relax
\mciteBstWouldAddEndPuncttrue
\mciteSetBstMidEndSepPunct{\mcitedefaultmidpunct}
{\mcitedefaultendpunct}{\mcitedefaultseppunct}\relax
\EndOfBibitem
\bibitem[Schleder \latin{et~al.}(2019)Schleder, Padilha, Acosta, Costa, and
  Fazzio]{schleder_fazzio_ml_materials_jpm_2019}
Schleder,~G.~R.; Padilha,~A. C.~M.; Acosta,~C.~M.; Costa,~M.; Fazzio,~A. {From
  DFT to Machine Learning: Recent Approaches to Materials Science--A Review}.
  \emph{J. Phys.: Mater.} \textbf{2019}, \emph{2}, 032001\relax
\mciteBstWouldAddEndPuncttrue
\mciteSetBstMidEndSepPunct{\mcitedefaultmidpunct}
{\mcitedefaultendpunct}{\mcitedefaultseppunct}\relax
\EndOfBibitem
\bibitem[Hern{\'a}ndez and Gillan(1995)Hern{\'a}ndez, and
  Gillan]{hernandez_gillan_prb_1995}
Hern{\'a}ndez,~E.; Gillan,~M.~J. {Self-Consistent First-Principles Technique
  With Linear Scaling}. \emph{Phys. Rev. B} \textbf{1995}, \emph{51},
  10157\relax
\mciteBstWouldAddEndPuncttrue
\mciteSetBstMidEndSepPunct{\mcitedefaultmidpunct}
{\mcitedefaultendpunct}{\mcitedefaultseppunct}\relax
\EndOfBibitem
\bibitem[Bowler \latin{et~al.}(2002)Bowler, Miyazaki, and
  Gillan]{bowler_gillan_j_phys_cond_matter_2002}
Bowler,~D.~R.; Miyazaki,~T.; Gillan,~M.~J. {Recent Progress in Linear Scaling
  {\textit{Ab Initio}} Electronic Structure Techniques}. \emph{J. Phys.:
  Condens. Matter.} \textbf{2002}, \emph{14}, 2781\relax
\mciteBstWouldAddEndPuncttrue
\mciteSetBstMidEndSepPunct{\mcitedefaultmidpunct}
{\mcitedefaultendpunct}{\mcitedefaultseppunct}\relax
\EndOfBibitem
\bibitem[Hine \latin{et~al.}(2009)Hine, Haynes, Mostofi, Skylaris, and
  Payne]{skylaris_payne_onetep_cpc_2009}
Hine,~N. D.~M.; Haynes,~P.~D.; Mostofi,~A.~A.; Skylaris,~C.-K.; Payne,~M.~C.
  {Linear-Scaling Density-Functional Theory with Tens of Thousands of Atoms:
  Expanding the Scope and Scale of Calculations with {\texttt{ONETEP}}}.
  \emph{Comput. Phys. Commun.} \textbf{2009}, \emph{180}, 1041\relax
\mciteBstWouldAddEndPuncttrue
\mciteSetBstMidEndSepPunct{\mcitedefaultmidpunct}
{\mcitedefaultendpunct}{\mcitedefaultseppunct}\relax
\EndOfBibitem
\bibitem[S{\ae}ther \latin{et~al.}(2017)S{\ae}ther, Kj{\ae}rgaard, Koch, and
  H{\o}yvik]{hoyvik_multilevel_hf_jctc_2017}
S{\ae}ther,~S.; Kj{\ae}rgaard,~T.; Koch,~H.; H{\o}yvik,~I.-M. {Density-Based
  Multilevel Hartree-Fock Model}. \emph{J. Chem. Theory Comput.} \textbf{2017},
  \emph{13}, 5282\relax
\mciteBstWouldAddEndPuncttrue
\mciteSetBstMidEndSepPunct{\mcitedefaultmidpunct}
{\mcitedefaultendpunct}{\mcitedefaultseppunct}\relax
\EndOfBibitem
\bibitem[Brandenburg \latin{et~al.}(2018)Brandenburg, Bannwarth, Hansen, and
  Grimme]{brandenburg_grimme_b97_3c_jcp_2018}
Brandenburg,~J.~G.; Bannwarth,~C.; Hansen,~A.; Grimme,~S. {B97-3c: A Revised
  Low-Cost Variant of the B97-D Density Functional Method}. \emph{{J}. {C}hem.
  {P}hys.} \textbf{2018}, \emph{148}, 064104\relax
\mciteBstWouldAddEndPuncttrue
\mciteSetBstMidEndSepPunct{\mcitedefaultmidpunct}
{\mcitedefaultendpunct}{\mcitedefaultseppunct}\relax
\EndOfBibitem
\bibitem[Caldeweyher and Brandenburg(2018)Caldeweyher, and
  Brandenburg]{brandenburg_simpl_dft_jpcm_2018}
Caldeweyher,~E.; Brandenburg,~J.~G. {Simplified DFT Methods for Consistent
  Structures and Energies of Large Systems}. \emph{{J}. {P}hys.: {C}ondens.
  {M}atter.} \textbf{2018}, \emph{30}, 213001\relax
\mciteBstWouldAddEndPuncttrue
\mciteSetBstMidEndSepPunct{\mcitedefaultmidpunct}
{\mcitedefaultendpunct}{\mcitedefaultseppunct}\relax
\EndOfBibitem
\bibitem[Marrazzini \latin{et~al.}(2020)Marrazzini, Giovannini, Scavino, Egidi,
  Cappelli, and Koch]{koch_multilevel_dft_arxiv_2020}
Marrazzini,~G.; Giovannini,~T.; Scavino,~M.; Egidi,~F.; Cappelli,~C.; Koch,~H.
  {Multilevel Density Functional Theory}. \emph{arXiv:2009.05333}
  \textbf{2020}\relax
\mciteBstWouldAddEndPunctfalse
\mciteSetBstMidEndSepPunct{\mcitedefaultmidpunct}
{\mcitedefaultendpunct}{\mcitedefaultseppunct}\relax
\EndOfBibitem
\bibitem[Bart{\'o}k and Cs{\'a}nyi(2015)Bart{\'o}k, and
  Cs{\'a}nyi]{bartol_csanyi_ml_qc_review_ijqc_2015}
Bart{\'o}k,~A.~P.; Cs{\'a}nyi,~G. {Gaussian Approximation Potentials: A Brief
  Tutorial Introduction}. \emph{Int. J. Quantum Chem.} \textbf{2015},
  \emph{115}, 1051\relax
\mciteBstWouldAddEndPuncttrue
\mciteSetBstMidEndSepPunct{\mcitedefaultmidpunct}
{\mcitedefaultendpunct}{\mcitedefaultseppunct}\relax
\EndOfBibitem
\bibitem[Rupp(2015)]{rupp_ml_qc_review_ijqc_2015}
Rupp,~M. {Machine Learning for Quantum Mechanics in a Nutshell}. \emph{Int. J.
  Quantum Chem.} \textbf{2015}, \emph{115}, 1058\relax
\mciteBstWouldAddEndPuncttrue
\mciteSetBstMidEndSepPunct{\mcitedefaultmidpunct}
{\mcitedefaultendpunct}{\mcitedefaultseppunct}\relax
\EndOfBibitem
\bibitem[Rupp \latin{et~al.}(2018)Rupp, von Lilienfeld, and
  Burke]{rupp_lilienfeld_burke_jcp_2018}
Rupp,~M.; von Lilienfeld,~O.~A.; Burke,~K. {Guest Editorial: Special Topic on
  Data-Enabled Theoretical Chemistry}. \emph{J. Chem. Phys.} \textbf{2018},
  \emph{148}, 241401\relax
\mciteBstWouldAddEndPuncttrue
\mciteSetBstMidEndSepPunct{\mcitedefaultmidpunct}
{\mcitedefaultendpunct}{\mcitedefaultseppunct}\relax
\EndOfBibitem
\bibitem[von Lilienfeld and Burke(2020)von Lilienfeld, and
  Burke]{lilienfeld_burke_nat_commun_2020}
von Lilienfeld,~O.~A.; Burke,~K. {Retrospective on a Decade of Machine Learning
  for Chemical Discovery}. \emph{Nat. Commun.} \textbf{2020}, \emph{11},
  4895\relax
\mciteBstWouldAddEndPuncttrue
\mciteSetBstMidEndSepPunct{\mcitedefaultmidpunct}
{\mcitedefaultendpunct}{\mcitedefaultseppunct}\relax
\EndOfBibitem
\bibitem[Ceriotti(2019)]{ceriotti_ml_qc_review_jcp_2019}
Ceriotti,~M. {Unsupervised Machine Learning in Atomistic Simulations, Between
  Predictions and Understanding}. \emph{J. Chem. Phys.} \textbf{2019},
  \emph{150}, 150901\relax
\mciteBstWouldAddEndPuncttrue
\mciteSetBstMidEndSepPunct{\mcitedefaultmidpunct}
{\mcitedefaultendpunct}{\mcitedefaultseppunct}\relax
\EndOfBibitem
\bibitem[Dral(2020)]{dral_ml_qc_review_jpcl_2020}
Dral,~P.~O. {Quantum Chemistry in the Age of Machine Learning}. \emph{J. Phys.
  Chem. Lett.} \textbf{2020}, \emph{11}, 2336\relax
\mciteBstWouldAddEndPuncttrue
\mciteSetBstMidEndSepPunct{\mcitedefaultmidpunct}
{\mcitedefaultendpunct}{\mcitedefaultseppunct}\relax
\EndOfBibitem
\bibitem[Snyder \latin{et~al.}(2012)Snyder, Rupp, Hansen, M{\"u}ller, and
  Burke]{mueller_burke_ml_dft_jcp_2012}
Snyder,~J.~C.; Rupp,~M.; Hansen,~K.; M{\"u}ller,~K.-R.; Burke,~K. {Finding
  Density Functionals with Machine Learning}. \emph{J. Chem. Phys.}
  \textbf{2012}, \emph{108}, 253002\relax
\mciteBstWouldAddEndPuncttrue
\mciteSetBstMidEndSepPunct{\mcitedefaultmidpunct}
{\mcitedefaultendpunct}{\mcitedefaultseppunct}\relax
\EndOfBibitem
\bibitem[Brockherde \latin{et~al.}(2017)Brockherde, Vogt, Li, Tuckerman, Burke,
  and M{\"u}ller]{tuckerman_burke_mueller_ml_dft_nature_comm_2017}
Brockherde,~F.; Vogt,~L.; Li,~L.; Tuckerman,~M.~E.; Burke,~K.;
  M{\"u}ller,~K.-R. {Bypassing the Kohn-Sham Equations with Machine Learning}.
  \emph{Nat. Commun.} \textbf{2017}, \emph{8}, 872\relax
\mciteBstWouldAddEndPuncttrue
\mciteSetBstMidEndSepPunct{\mcitedefaultmidpunct}
{\mcitedefaultendpunct}{\mcitedefaultseppunct}\relax
\EndOfBibitem
\bibitem[Bogojeski \latin{et~al.}(2020)Bogojeski, Vogt-Maranto, Tuckerman,
  M{\"u}ller, and Burke]{tuckerman_burke_mueller_ml_dft_nature_comm_2020}
Bogojeski,~M.; Vogt-Maranto,~L.; Tuckerman,~M.~E.; M{\"u}ller,~K.-R.; Burke,~K.
  {Quantum Chemical Accuracy from Density Functional Approximations via Machine
  Learning}. \emph{Nat. Commun.} \textbf{2020}, \emph{11}, 5223\relax
\mciteBstWouldAddEndPuncttrue
\mciteSetBstMidEndSepPunct{\mcitedefaultmidpunct}
{\mcitedefaultendpunct}{\mcitedefaultseppunct}\relax
\EndOfBibitem
\bibitem[Li \latin{et~al.}(2020)Li, Hoyer, Pederson, Sun, Cubuk, Riley, and
  Burke]{li_burke_ml_dft_arxiv_2020}
Li,~L.; Hoyer,~S.; Pederson,~R.; Sun,~R.; Cubuk,~E.~D.; Riley,~P.; Burke,~K.
  {Kohn-Sham Equations as Regularizer: Building Prior Knowledge into
  Machine-Learned Physics}. \emph{arXiv: 2009.08551} \textbf{2020}\relax
\mciteBstWouldAddEndPunctfalse
\mciteSetBstMidEndSepPunct{\mcitedefaultmidpunct}
{\mcitedefaultendpunct}{\mcitedefaultseppunct}\relax
\EndOfBibitem
\bibitem[Meyer \latin{et~al.}(2020)Meyer, Weichselbaum, and
  Hauser]{hauser_ml_dft_jctc_2020}
Meyer,~R.; Weichselbaum,~M.; Hauser,~A.~W. {Machine Learning Approaches toward
  Orbital-free Density Functional Theory: Simultaneous Training on the Kinetic
  Energy Density Functional and Its Functional Derivative}. \emph{J. Chem.
  Theory Comput.} \textbf{2020}, \emph{16}, 5685\relax
\mciteBstWouldAddEndPuncttrue
\mciteSetBstMidEndSepPunct{\mcitedefaultmidpunct}
{\mcitedefaultendpunct}{\mcitedefaultseppunct}\relax
\EndOfBibitem
\bibitem[Yao and Parkhill(2016)Yao, and
  Parkhill]{parkhill_kinetic_energy_jctc_2016}
Yao,~K.; Parkhill,~J. {Kinetic Energy of Hydrocarbons as a Function of Electron
  Density and Convolutional Neural Networks}. \emph{J. Chem. Theory Comput.}
  \textbf{2016}, \emph{12}, 1139\relax
\mciteBstWouldAddEndPuncttrue
\mciteSetBstMidEndSepPunct{\mcitedefaultmidpunct}
{\mcitedefaultendpunct}{\mcitedefaultseppunct}\relax
\EndOfBibitem
\bibitem[Hegde and Bowen(2017)Hegde, and
  Bowen]{hegde_bowen_ml_dft_sci_reports_2017}
Hegde,~G.; Bowen,~R.~C. {Machine-Learned Approximations to Density Functional
  Theory Hamiltonians}. \emph{Sci. Rep.} \textbf{2017}, \emph{7}, 42669\relax
\mciteBstWouldAddEndPuncttrue
\mciteSetBstMidEndSepPunct{\mcitedefaultmidpunct}
{\mcitedefaultendpunct}{\mcitedefaultseppunct}\relax
\EndOfBibitem
\bibitem[Ryczko \latin{et~al.}(2019)Ryczko, Strubbe, and
  Tamblyn]{tamblyn_ml_dft_pra_2019}
Ryczko,~K.; Strubbe,~D.; Tamblyn,~I. {Deep Learning and Density Functional
  Theory}. \emph{Phys. Rev. A} \textbf{2019}, \emph{100}, 022512\relax
\mciteBstWouldAddEndPuncttrue
\mciteSetBstMidEndSepPunct{\mcitedefaultmidpunct}
{\mcitedefaultendpunct}{\mcitedefaultseppunct}\relax
\EndOfBibitem
\bibitem[Dick and Fernandez-Serra(2019)Dick, and
  Fernandez-Serra]{fernandez_serra_ml_dft_jcp_2019}
Dick,~S.; Fernandez-Serra,~M. {Learning from the Density to Correct Total
  Energy and Forces in First Principle Simulations}. \emph{J. Chem. Phys.}
  \textbf{2019}, \emph{151}, 144102\relax
\mciteBstWouldAddEndPuncttrue
\mciteSetBstMidEndSepPunct{\mcitedefaultmidpunct}
{\mcitedefaultendpunct}{\mcitedefaultseppunct}\relax
\EndOfBibitem
\bibitem[Dick and Fernandez-Serra(2020)Dick, and
  Fernandez-Serra]{fernandez_serra_ml_dft_nat_commun_2020}
Dick,~S.; Fernandez-Serra,~M. {Machine Learning Accurate Exchange and
  Correlation Functionals of the Electronic Density}. \emph{Nat. Commun.}
  \textbf{2020}, \emph{11}, 3509\relax
\mciteBstWouldAddEndPuncttrue
\mciteSetBstMidEndSepPunct{\mcitedefaultmidpunct}
{\mcitedefaultendpunct}{\mcitedefaultseppunct}\relax
\EndOfBibitem
\bibitem[Mezei and von Lilienfeld(2020)Mezei, and von
  Lilienfeld]{lilienfeld_ml_dft_dispersion_jctc_2020}
Mezei,~P.~D.; von Lilienfeld,~O.~A. {Non-Covalent Quantum Machine Learning
  Corrections to Density Functionals}. \emph{J. Chem. Theory Comput.}
  \textbf{2020}, \emph{16}, 2647\relax
\mciteBstWouldAddEndPuncttrue
\mciteSetBstMidEndSepPunct{\mcitedefaultmidpunct}
{\mcitedefaultendpunct}{\mcitedefaultseppunct}\relax
\EndOfBibitem
\bibitem[Sinitskiy and Pande(2018)Sinitskiy, and
  Pande]{pande_ml_dft_arxiv_2018}
Sinitskiy,~A.~V.; Pande,~V.~S. {Deep Neural Network Computes Electron Densities
  and Energies of a Large Set of Organic Molecules Faster than Density
  Functional Theory (DFT)}. \emph{arXiv:1809.02723} \textbf{2018}\relax
\mciteBstWouldAddEndPunctfalse
\mciteSetBstMidEndSepPunct{\mcitedefaultmidpunct}
{\mcitedefaultendpunct}{\mcitedefaultseppunct}\relax
\EndOfBibitem
\bibitem[Sinitskiy and Pande(2019)Sinitskiy, and
  Pande]{pande_ml_dft_arxiv_2019}
Sinitskiy,~A.~V.; Pande,~V.~S. {Physical Machine Learning Outperforms ``Human
  Learning" in Quantum Chemistry}. \emph{arXiv:1908.00971} \textbf{2019}\relax
\mciteBstWouldAddEndPunctfalse
\mciteSetBstMidEndSepPunct{\mcitedefaultmidpunct}
{\mcitedefaultendpunct}{\mcitedefaultseppunct}\relax
\EndOfBibitem
\bibitem[Ryabov and Zhilyaev(2020)Ryabov, and
  Zhilyaev]{ryabov_zhilyaev_ml_dft_sci_rep_2020}
Ryabov,~A.; Zhilyaev,~P. {Neural Network Interpolation of Exchange-Correlation
  Functional}. \emph{Sci. Rep.} \textbf{2020}, \emph{10}, 8000\relax
\mciteBstWouldAddEndPuncttrue
\mciteSetBstMidEndSepPunct{\mcitedefaultmidpunct}
{\mcitedefaultendpunct}{\mcitedefaultseppunct}\relax
\EndOfBibitem
\bibitem[Lei and Medford(2019)Lei, and
  Medford]{lei_medford_ml_dft_phys_rev_mat_2019}
Lei,~X.; Medford,~A.~J. {Design and Analysis of Machine Learning
  Exchange-Correlation Functionals via Rotationally Invariant Convolutional
  Descriptors}. \emph{Phys. Rev. Materials} \textbf{2019}, \emph{3},
  063801\relax
\mciteBstWouldAddEndPuncttrue
\mciteSetBstMidEndSepPunct{\mcitedefaultmidpunct}
{\mcitedefaultendpunct}{\mcitedefaultseppunct}\relax
\EndOfBibitem
\bibitem[Chandrasekaran \latin{et~al.}(2019)Chandrasekaran, Kamal, Batra, Kim,
  Chen, and Ramprasad]{ramprasad_ml_dft_npj_2019}
Chandrasekaran,~A.; Kamal,~D.; Batra,~R.; Kim,~C.; Chen,~L.; Ramprasad,~R.
  {Solving the Electronic Structure Problem with Machine Learning}. \emph{npj
  Comput. Mater.} \textbf{2019}, \emph{5}, 22\relax
\mciteBstWouldAddEndPuncttrue
\mciteSetBstMidEndSepPunct{\mcitedefaultmidpunct}
{\mcitedefaultendpunct}{\mcitedefaultseppunct}\relax
\EndOfBibitem
\bibitem[Zhou \latin{et~al.}(2019)Zhou, Wu, Chen, and
  Chen]{chen_ml_dft_jpcl_2019}
Zhou,~Y.; Wu,~J.; Chen,~S.; Chen,~G. {Toward the Exact Exchange-Correlation
  Potential: A Three-Dimensional Convolutional Neural Network Construct}.
  \emph{J. Phys. Chem. Lett.} \textbf{2019}, \emph{10}, 7264\relax
\mciteBstWouldAddEndPuncttrue
\mciteSetBstMidEndSepPunct{\mcitedefaultmidpunct}
{\mcitedefaultendpunct}{\mcitedefaultseppunct}\relax
\EndOfBibitem
\bibitem[Nagai \latin{et~al.}(2020)Nagai, Akashi, and
  Sugino]{nagai_akashi_sugino_ml_dft_npj_2020}
Nagai,~R.; Akashi,~R.; Sugino,~O. {Completing Density Functional Theory by
  Machine Learning Hidden Messages from Molecules}. \emph{npj Comput. Mater.}
  \textbf{2020}, \emph{6}, 43\relax
\mciteBstWouldAddEndPuncttrue
\mciteSetBstMidEndSepPunct{\mcitedefaultmidpunct}
{\mcitedefaultendpunct}{\mcitedefaultseppunct}\relax
\EndOfBibitem
\bibitem[Chen \latin{et~al.}(2020)Chen, Zhang, Wang, and
  E]{weinan_e_deepks_arxiv_2020}
Chen,~Y.; Zhang,~L.; Wang,~H.; E,~W. {DeePKS: A Comprehensive Data-Driven
  Approach Towards Chemically Accurate Density Functional Theory}.
  \emph{arXiv:2008.00167} \textbf{2020}\relax
\mciteBstWouldAddEndPunctfalse
\mciteSetBstMidEndSepPunct{\mcitedefaultmidpunct}
{\mcitedefaultendpunct}{\mcitedefaultseppunct}\relax
\EndOfBibitem
\bibitem[Artrith \latin{et~al.}(2017)Artrith, Urban, and
  Ceder]{artrith_urban_ceder_ml_representation_prb_2017}
Artrith,~N.; Urban,~A.; Ceder,~G. {Efficient and Accurate Machine-Learning
  Interpolation of Atomic Energies in Compositions with Many Species}.
  \emph{Phys. Rev. B} \textbf{2017}, \emph{96}, 014112\relax
\mciteBstWouldAddEndPuncttrue
\mciteSetBstMidEndSepPunct{\mcitedefaultmidpunct}
{\mcitedefaultendpunct}{\mcitedefaultseppunct}\relax
\EndOfBibitem
\bibitem[Collins \latin{et~al.}(2018)Collins, Gordon, von Lilienfeld, and
  Yaron]{lilienfeld_yaron_ml_representations_jcp_2018}
Collins,~C.~R.; Gordon,~G.~J.; von Lilienfeld,~O.~A.; Yaron,~D.~J. {Constant
  Size Descriptors for Accurate Machine Learning Models of Molecular
  Properties}. \emph{J. Chem. Phys.} \textbf{2018}, \emph{148}, 241718\relax
\mciteBstWouldAddEndPuncttrue
\mciteSetBstMidEndSepPunct{\mcitedefaultmidpunct}
{\mcitedefaultendpunct}{\mcitedefaultseppunct}\relax
\EndOfBibitem
\bibitem[Schrier(2020)]{schrier_ml_representation_jcim_2020}
Schrier,~J. {Can One Hear the Shape of a Molecule (from its Coulomb Matrix
  Eigenvalues)?} \emph{J. Chem. Inf. Model.} \textbf{2020}, \emph{60},
  3804\relax
\mciteBstWouldAddEndPuncttrue
\mciteSetBstMidEndSepPunct{\mcitedefaultmidpunct}
{\mcitedefaultendpunct}{\mcitedefaultseppunct}\relax
\EndOfBibitem
\bibitem[Li \latin{et~al.}(2020)Li, Li, Seymour, Koziol, and
  Henkelman]{henkelman_ml_ann_jcp_2020}
Li,~L.; Li,~H.; Seymour,~I.~D.; Koziol,~L.; Henkelman,~G.
  {Pair-Distribution-Function Guided Optimization of Fingerprints for
  Atom-Centered Neural Network Potentials}. \emph{J. Chem. Phys.}
  \textbf{2020}, \emph{152}, 224102\relax
\mciteBstWouldAddEndPuncttrue
\mciteSetBstMidEndSepPunct{\mcitedefaultmidpunct}
{\mcitedefaultendpunct}{\mcitedefaultseppunct}\relax
\EndOfBibitem
\bibitem[Onat \latin{et~al.}(2020)Onat, Ortner, and
  Kermode]{onat_ortner_kermode_ml_representation_jcp_2020}
Onat,~B.; Ortner,~C.; Kermode,~J.~R. {Sensitivity and Dimensionality of Atomic
  Environment Representations Used for Machine Learning Interatomic
  Potentials}. \emph{J. Chem. Phys.} \textbf{2020}, \emph{153}, 144106\relax
\mciteBstWouldAddEndPuncttrue
\mciteSetBstMidEndSepPunct{\mcitedefaultmidpunct}
{\mcitedefaultendpunct}{\mcitedefaultseppunct}\relax
\EndOfBibitem
\bibitem[Nigam \latin{et~al.}(2020)Nigam, Pozdnyakov, and
  Ceriotti]{ceriotti_ml_representation_jcp_2020}
Nigam,~J.; Pozdnyakov,~S.; Ceriotti,~M. {Incompleteness of Atomic Structure
  Representations}. \emph{J. Chem. Phys.} \textbf{2020}, \emph{153},
  121101\relax
\mciteBstWouldAddEndPuncttrue
\mciteSetBstMidEndSepPunct{\mcitedefaultmidpunct}
{\mcitedefaultendpunct}{\mcitedefaultseppunct}\relax
\EndOfBibitem
\bibitem[Christiansen \latin{et~al.}(2020)Christiansen, Mortensen, Meldgaard,
  and Hammer]{hammer_ml_representation_jcp_2020}
Christiansen,~M.-P.~V.; Mortensen,~H.~L.; Meldgaard,~S.~A.; Hammer,~B.
  {Gaussian Representation for Image Recognition and Reinforcement Learning of
  Atomistic Structure}. \emph{J. Chem. Phys.} \textbf{2020}, \emph{153},
  044107\relax
\mciteBstWouldAddEndPuncttrue
\mciteSetBstMidEndSepPunct{\mcitedefaultmidpunct}
{\mcitedefaultendpunct}{\mcitedefaultseppunct}\relax
\EndOfBibitem
\bibitem[De \latin{et~al.}(2016)De, Bart{\'o}k, Cs{\'a}nyi, and
  Ceriotti]{ceriotti_chem_space_pccp_2016}
De,~S.; Bart{\'o}k,~A.~P.; Cs{\'a}nyi,~G.; Ceriotti,~M. {Comparing Molecules
  and Solids Across Structural and Alchemical Space}. \emph{Phys. Chem. Chem.
  Phys.} \textbf{2016}, \emph{18}, 13754\relax
\mciteBstWouldAddEndPuncttrue
\mciteSetBstMidEndSepPunct{\mcitedefaultmidpunct}
{\mcitedefaultendpunct}{\mcitedefaultseppunct}\relax
\EndOfBibitem
\bibitem[Smith \latin{et~al.}(2018)Smith, Nebgen, Lubbers, Isayev, and
  Roitberg]{isayev_roitberg_chem_space_jcp_2018}
Smith,~J.~S.; Nebgen,~B.; Lubbers,~N.; Isayev,~O.; Roitberg,~A.~E. {Less is
  More: Sampling Chemical Space with Active Learning}. \emph{J. Chem. Phys}
  \textbf{2018}, \emph{148}, 241733\relax
\mciteBstWouldAddEndPuncttrue
\mciteSetBstMidEndSepPunct{\mcitedefaultmidpunct}
{\mcitedefaultendpunct}{\mcitedefaultseppunct}\relax
\EndOfBibitem
\bibitem[von Lilienfeld(2018)]{lilienfeld_qml_angew_chem_2018}
von Lilienfeld,~O.~A. {Quantum Machine Learning in Chemical Compound Space}.
  \emph{Angew. Chem. Int. Ed.} \textbf{2018}, \emph{57}, 4164\relax
\mciteBstWouldAddEndPuncttrue
\mciteSetBstMidEndSepPunct{\mcitedefaultmidpunct}
{\mcitedefaultendpunct}{\mcitedefaultseppunct}\relax
\EndOfBibitem
\bibitem[von Lilienfeld \latin{et~al.}(2020)von Lilienfeld, M{\"u}ller, and
  Tkatchenko]{lilienfeld_mueller_tkatchenko_qml_nat_rev_chem_2020}
von Lilienfeld,~O.~A.; M{\"u}ller,~K.-R.; Tkatchenko,~A. {Exploring Chemical
  Compound Space with Quantum-Based Machine Learning}. \emph{Nat. Rev. Chem.}
  \textbf{2020}, \emph{4}, 347\relax
\mciteBstWouldAddEndPuncttrue
\mciteSetBstMidEndSepPunct{\mcitedefaultmidpunct}
{\mcitedefaultendpunct}{\mcitedefaultseppunct}\relax
\EndOfBibitem
\bibitem[Tkatchenko(2020)]{tkatchenko_ml_qc_nat_commun_2020}
Tkatchenko,~A. {Machine Learning for Chemical Discovery}. \emph{Nat. Commun.}
  \textbf{2020}, \emph{11}, 4125\relax
\mciteBstWouldAddEndPuncttrue
\mciteSetBstMidEndSepPunct{\mcitedefaultmidpunct}
{\mcitedefaultendpunct}{\mcitedefaultseppunct}\relax
\EndOfBibitem
\bibitem[Zhang \latin{et~al.}(2020)Zhang, Lei, Zhang, Chang, Li, Han, Yang,
  Yang, and Gao]{yang_gao_ml_qc_review_jpca_2020}
Zhang,~J.; Lei,~Y.-K.; Zhang,~Z.; Chang,~J.; Li,~M.; Han,~X.; Yang,~L.;
  Yang,~Y.~I.; Gao,~Y.~Q. {A Perspective on Deep Learning for Molecular
  Modeling and Simulations}. \emph{J. Phys. Chem. A} \textbf{2020}, \emph{124},
  6745\relax
\mciteBstWouldAddEndPuncttrue
\mciteSetBstMidEndSepPunct{\mcitedefaultmidpunct}
{\mcitedefaultendpunct}{\mcitedefaultseppunct}\relax
\EndOfBibitem
\bibitem[Sumpter and Noid(1992)Sumpter, and Noid]{sumpter_noid_ml_pes_cpl_1992}
Sumpter,~B.~G.; Noid,~D.~W. {Potential Energy Surfaces for Macromolecules. A
  Neural Network Technique}. \emph{Chem. Phys. Lett.} \textbf{1992},
  \emph{192}, 455\relax
\mciteBstWouldAddEndPuncttrue
\mciteSetBstMidEndSepPunct{\mcitedefaultmidpunct}
{\mcitedefaultendpunct}{\mcitedefaultseppunct}\relax
\EndOfBibitem
\bibitem[Ho and Rabitz(1996)Ho, and Rabitz]{ho_rabitz_ml_pes_jcp_1996}
Ho,~T.-S.; Rabitz,~H. {A General Method for Constructing Multidimensional
  Molecular Potential Energy Surfaces from {\textit{Ab Initio}} Calculations}.
  \emph{J. Chem. Phys.} \textbf{1996}, \emph{104}, 2584\relax
\mciteBstWouldAddEndPuncttrue
\mciteSetBstMidEndSepPunct{\mcitedefaultmidpunct}
{\mcitedefaultendpunct}{\mcitedefaultseppunct}\relax
\EndOfBibitem
\bibitem[Lorenz \latin{et~al.}(2004)Lorenz, Gro{\ss}, and
  Scheffler]{lorenz_gross_scheffler_ml_pes_cpl_2004}
Lorenz,~S.; Gro{\ss},~A.; Scheffler,~M. {Representing High-Dimensional
  Potential-Energy Surfaces for Reactions at Surfaces by Neural Networks}.
  \emph{Chem. Phys. Lett.} \textbf{2004}, \emph{395}, 210\relax
\mciteBstWouldAddEndPuncttrue
\mciteSetBstMidEndSepPunct{\mcitedefaultmidpunct}
{\mcitedefaultendpunct}{\mcitedefaultseppunct}\relax
\EndOfBibitem
\bibitem[Behler and Parrinello(2007)Behler, and
  Parrinello]{parrinello_ml_pes_prl_2007}
Behler,~J.; Parrinello,~M. {Generalized Neural-Network Representation of
  High-Dimensional Potential-Energy Surfaces}. \emph{Phys. Rev. Lett.}
  \textbf{2007}, \emph{98}, 146401\relax
\mciteBstWouldAddEndPuncttrue
\mciteSetBstMidEndSepPunct{\mcitedefaultmidpunct}
{\mcitedefaultendpunct}{\mcitedefaultseppunct}\relax
\EndOfBibitem
\bibitem[Braams and Bowman(2009)Braams, and
  Bowman]{braams_bowman_ml_pes_irpc_2009}
Braams,~B.~J.; Bowman,~J.~M. {Permutationally Invariant Potential Energy
  Surfaces in High Dimensionality}. \emph{Int. Rev. Phys. Chem.} \textbf{2009},
  \emph{28}, 577\relax
\mciteBstWouldAddEndPuncttrue
\mciteSetBstMidEndSepPunct{\mcitedefaultmidpunct}
{\mcitedefaultendpunct}{\mcitedefaultseppunct}\relax
\EndOfBibitem
\bibitem[Malshe \latin{et~al.}(2009)Malshe, Narulkar, Raff, Hagan, Bukkapatnam,
  Agrawal, and Komanduri]{agrawal_komanduri_ml_pes_jcp_2009}
Malshe,~M.; Narulkar,~R.; Raff,~L.~M.; Hagan,~M.; Bukkapatnam,~S.;
  Agrawal,~P.~M.; Komanduri,~R. {Development of Generalized Potential-Energy
  Surfaces Using Many-Body Expansions, Neural Networks, and Moiety Energy
  Approximations}. \emph{J. Chem. Phys.} \textbf{2009}, \emph{130},
  184102\relax
\mciteBstWouldAddEndPuncttrue
\mciteSetBstMidEndSepPunct{\mcitedefaultmidpunct}
{\mcitedefaultendpunct}{\mcitedefaultseppunct}\relax
\EndOfBibitem
\bibitem[Behler and Parrinello(2010)Behler, and
  Parrinello]{handley_popelier_ml_pes_jpca_2010}
Behler,~J.; Parrinello,~M. {Potential Energy Surfaces Fitted by Artificial
  Neural Networks}. \emph{J. Phys. Chem. A} \textbf{2010}, \emph{114},
  3371\relax
\mciteBstWouldAddEndPuncttrue
\mciteSetBstMidEndSepPunct{\mcitedefaultmidpunct}
{\mcitedefaultendpunct}{\mcitedefaultseppunct}\relax
\EndOfBibitem
\bibitem[Bart{\'o}k \latin{et~al.}(2010)Bart{\'o}k, Payne, Kondor, and
  Cs{\'a}nyi]{csanyi_ml_pes_prl_2010}
Bart{\'o}k,~A.~P.; Payne,~M.~C.; Kondor,~R.; Cs{\'a}nyi,~G. {Gaussian
  Approximation Potentials: The Accuracy of Quantum Mechanics, without the
  Electrons}. \emph{Phys. Rev. Lett.} \textbf{2010}, \emph{104}, 136403\relax
\mciteBstWouldAddEndPuncttrue
\mciteSetBstMidEndSepPunct{\mcitedefaultmidpunct}
{\mcitedefaultendpunct}{\mcitedefaultseppunct}\relax
\EndOfBibitem
\bibitem[Manzhos and {Carrington, Jr.}(2006)Manzhos, and {Carrington,
  Jr.}]{carrington_ml_pes_jcp_2006}
Manzhos,~S.; {Carrington, Jr.},~T. {A Random-Sampling High Dimensional Model
  Representation Neural Network for Building Potential Energy Surfaces}.
  \emph{J. Chem. Phys.} \textbf{2006}, \emph{125}, 084109\relax
\mciteBstWouldAddEndPuncttrue
\mciteSetBstMidEndSepPunct{\mcitedefaultmidpunct}
{\mcitedefaultendpunct}{\mcitedefaultseppunct}\relax
\EndOfBibitem
\bibitem[Manzhos and {Carrington, Jr.}(2007)Manzhos, and {Carrington,
  Jr.}]{carrington_ml_pes_jcp_2007}
Manzhos,~S.; {Carrington, Jr.},~T. {Using Redundant Coordinates to Represent
  Potential Energy Surfaces with Lower-Dimensional Functions}. \emph{J. Chem.
  Phys.} \textbf{2007}, \emph{127}, 014103\relax
\mciteBstWouldAddEndPuncttrue
\mciteSetBstMidEndSepPunct{\mcitedefaultmidpunct}
{\mcitedefaultendpunct}{\mcitedefaultseppunct}\relax
\EndOfBibitem
\bibitem[Manzhos \latin{et~al.}(2015)Manzhos, Dawes, and {Carrington,
  Jr.}]{carrington_ml_pes_ijqc_2015}
Manzhos,~S.; Dawes,~R.; {Carrington, Jr.},~T. {Neural Network-Based Approaches
  for Building High Dimensional and Quantum Dynamics-Friendly Potential Energy
  Surfaces}. \emph{Int. J. Quantum Chem.} \textbf{2015}, \emph{115}, 1012\relax
\mciteBstWouldAddEndPuncttrue
\mciteSetBstMidEndSepPunct{\mcitedefaultmidpunct}
{\mcitedefaultendpunct}{\mcitedefaultseppunct}\relax
\EndOfBibitem
\bibitem[Manzhos and {Carrington, Jr.}(2020)Manzhos, and {Carrington,
  Jr.}]{carrington_ml_pes_chem_rev_2020}
Manzhos,~S.; {Carrington, Jr.},~T. {Neural Network Potential Energy Surfaces
  for Small Molecules and Reactions}. \emph{Chem. Rev.} \textbf{2020}, DOI:
  10.1021/acs.chemrev.0c00665\relax
\mciteBstWouldAddEndPuncttrue
\mciteSetBstMidEndSepPunct{\mcitedefaultmidpunct}
{\mcitedefaultendpunct}{\mcitedefaultseppunct}\relax
\EndOfBibitem
\bibitem[Artrith \latin{et~al.}(2011)Artrith, Morawietz, and
  Behler]{behler_ml_pes_prb_2011}
Artrith,~N.; Morawietz,~T.; Behler,~J. {High-Dimensional Neural Network
  Potentials for Multicomponent Systems: Applications to Zinc Oxide}.
  \emph{Phys. Rev. B: Condens. Matter Mater. Phys.} \textbf{2011}, \emph{83},
  153101\relax
\mciteBstWouldAddEndPuncttrue
\mciteSetBstMidEndSepPunct{\mcitedefaultmidpunct}
{\mcitedefaultendpunct}{\mcitedefaultseppunct}\relax
\EndOfBibitem
\bibitem[Morawietz and Behler(2013)Morawietz, and
  Behler]{behler_ml_pes_jpca_2013}
Morawietz,~T.; Behler,~J. {A Density-Functional Theory-Based Neural Network
  Potential for Water Clusters Including van der Waals Corrections}. \emph{J.
  Phys. Chem. A} \textbf{2013}, \emph{117}, 7356\relax
\mciteBstWouldAddEndPuncttrue
\mciteSetBstMidEndSepPunct{\mcitedefaultmidpunct}
{\mcitedefaultendpunct}{\mcitedefaultseppunct}\relax
\EndOfBibitem
\bibitem[Behler(2016)]{behler_ml_qc_jcp_2016}
Behler,~J. {Perspective: Machine Learning Potentials for Atomistic
  Simulations}. \emph{J. Chem. Phys.} \textbf{2016}, \emph{145}, 170901\relax
\mciteBstWouldAddEndPuncttrue
\mciteSetBstMidEndSepPunct{\mcitedefaultmidpunct}
{\mcitedefaultendpunct}{\mcitedefaultseppunct}\relax
\EndOfBibitem
\bibitem[Schran \latin{et~al.}(2020)Schran, Behler, and
  Marx]{behler_marx_ml_qc_jctc_2020}
Schran,~C.; Behler,~J.; Marx,~D. {Automated Fitting of Neural Network
  Potentials at Coupled Cluster Accuracy: Protonated Water Clusters as Testing
  Ground}. \emph{J. Chem. Theory Comput.} \textbf{2020}, \emph{16}, 88\relax
\mciteBstWouldAddEndPuncttrue
\mciteSetBstMidEndSepPunct{\mcitedefaultmidpunct}
{\mcitedefaultendpunct}{\mcitedefaultseppunct}\relax
\EndOfBibitem
\bibitem[Ko \latin{et~al.}(2020)Ko, Finkler, Goedecker, and
  Behler]{behler_ml_qc_arxiv_2020}
Ko,~T.~W.; Finkler,~J.~A.; Goedecker,~S.; Behler,~J. {A Fourth-Generation
  High-Dimensional Neural Network Potential with Accurate Electrostatics
  Including Non-local Charge Transfer}. \emph{arXiv:2009.06484} \textbf{2020}\relax
\mciteBstWouldAddEndPunctfalse
\mciteSetBstMidEndSepPunct{\mcitedefaultmidpunct}
{\mcitedefaultendpunct}{\mcitedefaultseppunct}\relax
\EndOfBibitem
\bibitem[Gastegger \latin{et~al.}(2018)Gastegger, Schwiedrzik, Bittermann,
  Berzsenyi, and Marquetand]{marquetand_ml_wacsf_jcp_2018}
Gastegger,~M.; Schwiedrzik,~L.; Bittermann,~M.; Berzsenyi,~F.; Marquetand,~P.
  {wACSF---Weighted Atom-Centered Symmetry Functions as Descriptors in Machine
  Learning Potentials}. \emph{J. Chem. Phys.} \textbf{2018}, \emph{148},
  241709\relax
\mciteBstWouldAddEndPuncttrue
\mciteSetBstMidEndSepPunct{\mcitedefaultmidpunct}
{\mcitedefaultendpunct}{\mcitedefaultseppunct}\relax
\EndOfBibitem
\bibitem[Schmitz \latin{et~al.}(2019)Schmitz, Artiukhin, and
  Christiansen]{christiansen_ml_pes_jcp_2019_1}
Schmitz,~G.; Artiukhin,~D.~G.; Christiansen,~O. {Approximate High Mode Coupling
  Potentials Using Gaussian Process Regression and Adaptive Density Guided
  Sampling}. \emph{J. Chem. Phys.} \textbf{2019}, \emph{150}, 131102\relax
\mciteBstWouldAddEndPuncttrue
\mciteSetBstMidEndSepPunct{\mcitedefaultmidpunct}
{\mcitedefaultendpunct}{\mcitedefaultseppunct}\relax
\EndOfBibitem
\bibitem[Schmitz \latin{et~al.}(2019)Schmitz, Godtliebsen, and
  Christiansen]{christiansen_ml_pes_jcp_2019_2}
Schmitz,~G.; Godtliebsen,~I.~H.; Christiansen,~O. {Machine Learning for
  Potential Energy Surfaces: An Extensive Database and Assessment of Methods}.
  \emph{J. Chem. Phys.} \textbf{2019}, \emph{150}, 244113\relax
\mciteBstWouldAddEndPuncttrue
\mciteSetBstMidEndSepPunct{\mcitedefaultmidpunct}
{\mcitedefaultendpunct}{\mcitedefaultseppunct}\relax
\EndOfBibitem
\bibitem[Schmitz \latin{et~al.}(2020)Schmitz, Klinting, and
  Christiansen]{christiansen_ml_pes_jcp_2020}
Schmitz,~G.; Klinting,~E.~L.; Christiansen,~O. {A Gaussian Process Regression
  Adaptive Density Guided Approach for Potential Energy Surface Construction}.
  \emph{J. Chem. Phys.} \textbf{2020}, \emph{153}, 064105\relax
\mciteBstWouldAddEndPuncttrue
\mciteSetBstMidEndSepPunct{\mcitedefaultmidpunct}
{\mcitedefaultendpunct}{\mcitedefaultseppunct}\relax
\EndOfBibitem
\bibitem[Zhu \latin{et~al.}(2019)Zhu, Vuong, Sumpter, and
  Irle]{sumpter_irle_ml_pes_mrs_comm_2019}
Zhu,~J.; Vuong,~V.~Q.; Sumpter,~B.~G.; Irle,~S. {Artificial Neural Network
  Correction for Density-Functional Tight-Binding Molecular Dynamics
  Simulations}. \emph{MRS Comm.} \textbf{2019}, \emph{9}, 867\relax
\mciteBstWouldAddEndPuncttrue
\mciteSetBstMidEndSepPunct{\mcitedefaultmidpunct}
{\mcitedefaultendpunct}{\mcitedefaultseppunct}\relax
\EndOfBibitem
\bibitem[Amabilino \latin{et~al.}(2019)Amabilino, Bratholm, Bennie, Vaucher,
  Reiher, and Glowacki]{glowacki_nn_pes_jpca_2019}
Amabilino,~S.; Bratholm,~L.~A.; Bennie,~S.~J.; Vaucher,~A.~C.; Reiher,~M.;
  Glowacki,~D.~R. {Training Neural Nets To Learn Reactive Potential Energy
  Surfaces Using Interactive Quantum Chemistry in Virtual Reality}. \emph{J.
  Phys. Chem. A} \textbf{2019}, \emph{123}, 4486\relax
\mciteBstWouldAddEndPuncttrue
\mciteSetBstMidEndSepPunct{\mcitedefaultmidpunct}
{\mcitedefaultendpunct}{\mcitedefaultseppunct}\relax
\EndOfBibitem
\bibitem[Dral \latin{et~al.}(2017)Dral, Owens, Yurchenko, and
  Thiel]{dral_thiel_ml_pes_jcp_2017}
Dral,~P.~O.; Owens,~A.; Yurchenko,~S.~N.; Thiel,~W. {Structure-Based Sampling
  and Self-Correcting Machine Learning for Accurate Calculations of Potential
  Energy Surfaces and Vibrational Levels}. \emph{J. Chem. Phys.} \textbf{2017},
  \emph{146}, 244108\relax
\mciteBstWouldAddEndPuncttrue
\mciteSetBstMidEndSepPunct{\mcitedefaultmidpunct}
{\mcitedefaultendpunct}{\mcitedefaultseppunct}\relax
\EndOfBibitem
\bibitem[Dral \latin{et~al.}(2018)Dral, Barbatti, and
  Thiel]{dral_thiel_ml_pes_jpcl_2018}
Dral,~P.~O.; Barbatti,~M.; Thiel,~W. {Nonadiabatic Excited-State Dynamics with
  Machine Learning}. \emph{J. Phys. Chem. Lett.} \textbf{2018}, \emph{9},
  5660\relax
\mciteBstWouldAddEndPuncttrue
\mciteSetBstMidEndSepPunct{\mcitedefaultmidpunct}
{\mcitedefaultendpunct}{\mcitedefaultseppunct}\relax
\EndOfBibitem
\bibitem[Dral \latin{et~al.}(2020)Dral, Owens, Dral, and
  Cs{\'a}nyi]{dral_csanyi_ml_pes_jcp_2020}
Dral,~P.~O.; Owens,~A.; Dral,~A.; Cs{\'a}nyi,~G. {Hierarchical Machine Learning
  of Potential Energy Surfaces}. \emph{J. Chem. Phys.} \textbf{2020},
  \emph{152}, 204110\relax
\mciteBstWouldAddEndPuncttrue
\mciteSetBstMidEndSepPunct{\mcitedefaultmidpunct}
{\mcitedefaultendpunct}{\mcitedefaultseppunct}\relax
\EndOfBibitem
\bibitem[Bernstein \latin{et~al.}(2019)Bernstein, Cs{\'a}nyi, and
  Deringer]{bernstein_csanyi_deringer_npj_comp_mater_2019}
Bernstein,~N.; Cs{\'a}nyi,~G.; Deringer,~V.~L. {{\textit{De Novo}} Exploration
  and Self-Guided Learning of Potential-Energy Surfaces}. \emph{npj Comput.
  Mater.} \textbf{2019}, \emph{5}, 99\relax
\mciteBstWouldAddEndPuncttrue
\mciteSetBstMidEndSepPunct{\mcitedefaultmidpunct}
{\mcitedefaultendpunct}{\mcitedefaultseppunct}\relax
\EndOfBibitem
\bibitem[van~der Oord \latin{et~al.}(2020)van~der Oord, Dusson, Cs{\'a}nyi, and
  Ortner]{csanyi_ortner_ml_pes_mlst_2020}
van~der Oord,~C.; Dusson,~G.; Cs{\'a}nyi,~G.; Ortner,~C. {Regularised Atomic
  Body-Ordered Permutation-Invariant Polynomials for the Construction of
  Interatomic Potentials}. \emph{Mach. Learn.: Sci. Technol.} \textbf{2020},
  \emph{1}, 015004\relax
\mciteBstWouldAddEndPuncttrue
\mciteSetBstMidEndSepPunct{\mcitedefaultmidpunct}
{\mcitedefaultendpunct}{\mcitedefaultseppunct}\relax
\EndOfBibitem
\bibitem[Allen \latin{et~al.}(2020)Allen, Cs{\'a}nyi, Dusson, and
  Ortner]{csanyi_ortner_ml_pes_arxiv_2020}
Allen,~A.; Cs{\'a}nyi,~G.; Dusson,~G.; Ortner,~C. {Atomic Permutationally
  Invariant Polynomials for Fitting Molecular Force Fields}.
  \emph{arXiv:2010.12200} \textbf{2020}\relax
\mciteBstWouldAddEndPunctfalse
\mciteSetBstMidEndSepPunct{\mcitedefaultmidpunct}
{\mcitedefaultendpunct}{\mcitedefaultseppunct}\relax
\EndOfBibitem
\bibitem[Koner and Meuwly(2020)Koner, and
  Meuwly]{koner_meuwly_ml_pes_jctc_2020}
Koner,~D.; Meuwly,~M. {Permutationally Invariant, Reproducing Kernel-Based
  Potential Energy Surfaces for Polyatomic Molecules: From Formaldehyde to
  Acetone}. \emph{J. Chem. Theory Comput.} \textbf{2020}, \emph{16}, 5474\relax
\mciteBstWouldAddEndPuncttrue
\mciteSetBstMidEndSepPunct{\mcitedefaultmidpunct}
{\mcitedefaultendpunct}{\mcitedefaultseppunct}\relax
\EndOfBibitem
\bibitem[Evans and Coudert(2017)Evans, and
  Coudert]{coudert_ml_zeolites_chem_mater_2017}
Evans,~J.~D.; Coudert,~F.-X. {Predicting the Mechanical Properties of Zeolite
  Frameworks by Machine Learning}. \emph{Chem. Mater.} \textbf{2017},
  \emph{29}, 7833\relax
\mciteBstWouldAddEndPuncttrue
\mciteSetBstMidEndSepPunct{\mcitedefaultmidpunct}
{\mcitedefaultendpunct}{\mcitedefaultseppunct}\relax
\EndOfBibitem
\bibitem[Isayev \latin{et~al.}(2015)Isayev, Fourches, Muratov, Oses, Rasch,
  Tropsha, and Curtarolo]{isayev_curtarolo_crystals_chem_mater_2015}
Isayev,~O.; Fourches,~D.; Muratov,~E.~N.; Oses,~C.; Rasch,~K.; Tropsha,~A.;
  Curtarolo,~S. {Materials Cartography: Representing and Mining Materials Space
  Using Structural and Electronic Fingerprints}. \emph{Chem. Mater.}
  \textbf{2015}, \emph{27}, 735\relax
\mciteBstWouldAddEndPuncttrue
\mciteSetBstMidEndSepPunct{\mcitedefaultmidpunct}
{\mcitedefaultendpunct}{\mcitedefaultseppunct}\relax
\EndOfBibitem
\bibitem[Butler \latin{et~al.}(2018)Butler, Davies, Cartwright, Isayev, and
  Walsh]{walsh_ml_mat_nature_2018}
Butler,~K.~T.; Davies,~D.~W.; Cartwright,~H.; Isayev,~O.; Walsh,~A. {Machine
  Learning for Molecular and Materials Science}. \emph{Nature} \textbf{2018},
  \emph{559}, 547\relax
\mciteBstWouldAddEndPuncttrue
\mciteSetBstMidEndSepPunct{\mcitedefaultmidpunct}
{\mcitedefaultendpunct}{\mcitedefaultseppunct}\relax
\EndOfBibitem
\bibitem[Botu \latin{et~al.}(2017)Botu, Batra, Chapman, and
  Ramprasad]{botu_ml_ff_jpcc_2017}
Botu,~V.; Batra,~R.; Chapman,~J.; Ramprasad,~R. {Machine Learning Force Fields:
  Construction, Validation, and Outlook}. \emph{J Phys. Chem. C} \textbf{2017},
  \emph{121}, 511\relax
\mciteBstWouldAddEndPuncttrue
\mciteSetBstMidEndSepPunct{\mcitedefaultmidpunct}
{\mcitedefaultendpunct}{\mcitedefaultseppunct}\relax
\EndOfBibitem
\bibitem[Faber \latin{et~al.}(2016)Faber, Lindmaa, von Lilienfeld, and
  Armiento]{lilienfeld_armiento_ml_atom_energy_prl_2016}
Faber,~F.; Lindmaa,~A.; von Lilienfeld,~O.~A.; Armiento,~R. {Machine Learning
  Energies of 2 Million Elpasolite ($ABC_2D_6$) Crystals}. \emph{Phys. Rev.
  Lett.} \textbf{2016}, \emph{117}, 135502\relax
\mciteBstWouldAddEndPuncttrue
\mciteSetBstMidEndSepPunct{\mcitedefaultmidpunct}
{\mcitedefaultendpunct}{\mcitedefaultseppunct}\relax
\EndOfBibitem
\bibitem[Chmiela \latin{et~al.}(2017)Chmiela, Tkatchenko, Sauceda, Poltavsky,
  Sch{\"u}tt, and M{\"u}ller]{tkatchenko_mueller_ml_ff_sci_adv_2017}
Chmiela,~S.; Tkatchenko,~A.; Sauceda,~H.~E.; Poltavsky,~I.; Sch{\"u}tt,~K.~T.;
  M{\"u}ller,~K.-R. {Machine Learning of Accurate Energy-Conserving Molecular
  Force Fields}. \emph{Sci. Adv.} \textbf{2017}, \emph{3}, e1603015\relax
\mciteBstWouldAddEndPuncttrue
\mciteSetBstMidEndSepPunct{\mcitedefaultmidpunct}
{\mcitedefaultendpunct}{\mcitedefaultseppunct}\relax
\EndOfBibitem
\bibitem[Chmiela \latin{et~al.}(2018)Chmiela, Sauceda, M{\"u}ller, and
  Tkatchenko]{mueller_tkatchenko_ml_ff_nature_comm_2018}
Chmiela,~S.; Sauceda,~H.~E.; M{\"u}ller,~K.-R.; Tkatchenko,~A. {Towards Exact
  Molecular Dynamics Simulations with Machine-Learned Force Fields}. \emph{Nat.
  Commun.} \textbf{2018}, \emph{9}, 3887\relax
\mciteBstWouldAddEndPuncttrue
\mciteSetBstMidEndSepPunct{\mcitedefaultmidpunct}
{\mcitedefaultendpunct}{\mcitedefaultseppunct}\relax
\EndOfBibitem
\bibitem[Unke \latin{et~al.}(2020)Unke, Chmiela, Sauceda, Gastegger, Poltavsky,
  Sch{\"u}tt, Tkatchenko, and M{\"u}ller]{tkatchenko_mueller_ml_ff_arxiv_2020}
Unke,~O.~T.; Chmiela,~S.; Sauceda,~H.~E.; Gastegger,~M.; Poltavsky,~I.;
  Sch{\"u}tt,~K.~T.; Tkatchenko,~A.; M{\"u}ller,~K.-R. {Machine Learning Force
  Fields}. \emph{arXiv:2010.07067} \textbf{2020}\relax
\mciteBstWouldAddEndPunctfalse
\mciteSetBstMidEndSepPunct{\mcitedefaultmidpunct}
{\mcitedefaultendpunct}{\mcitedefaultseppunct}\relax
\EndOfBibitem
\bibitem[Li \latin{et~al.}(2015)Li, Kermode, and
  De~Vita]{li_kermode_de_vita_ml_ff_prl_2015}
Li,~Z.; Kermode,~J.~R.; De~Vita,~A. {Molecular Dynamics with On-the-Fly Machine
  Learning of Quantum-Mechanical Forces}. \emph{Phys. Rev. Lett.}
  \textbf{2015}, \emph{114}, 096405\relax
\mciteBstWouldAddEndPuncttrue
\mciteSetBstMidEndSepPunct{\mcitedefaultmidpunct}
{\mcitedefaultendpunct}{\mcitedefaultseppunct}\relax
\EndOfBibitem
\bibitem[Musil \latin{et~al.}(2018)Musil, De, Yang, Campbell, Day, and
  Ceriotti]{day_ceriotti_ml_crystals_chem_sci_2018}
Musil,~F.; De,~S.; Yang,~J.; Campbell,~J.~E.; Day,~G.~M.; Ceriotti,~M. {Machine
  Learning for the Structure-Energy-Property Landscapes of Molecular Crystals}.
  \emph{Chem. Sci.} \textbf{2018}, \emph{9}, 1289\relax
\mciteBstWouldAddEndPuncttrue
\mciteSetBstMidEndSepPunct{\mcitedefaultmidpunct}
{\mcitedefaultendpunct}{\mcitedefaultseppunct}\relax
\EndOfBibitem
\bibitem[Bart{\'o}k \latin{et~al.}(2018)Bart{\'o}k, Kermode, Bernstein, and
  Cs{\'a}nyi]{csanyi_ml_ff_prx_2018}
Bart{\'o}k,~A.; Kermode,~J.; Bernstein,~N.; Cs{\'a}nyi,~G. {Machine Learning a
  General-Purpose Interatomic Potential for Silicon}. \emph{Phys. Rev. X}
  \textbf{2018}, \emph{8}, 041048\relax
\mciteBstWouldAddEndPuncttrue
\mciteSetBstMidEndSepPunct{\mcitedefaultmidpunct}
{\mcitedefaultendpunct}{\mcitedefaultseppunct}\relax
\EndOfBibitem
\bibitem[Deringer \latin{et~al.}(2018)Deringer, Bernstein, Bart{\'o}k, Cliffe,
  Kerber, Marbella, Grey, Elliot, and
  Cs{\'a}nyi]{deringer_csanyi_silicon_jpcl_2018}
Deringer,~V.~L.; Bernstein,~N.; Bart{\'o}k,~A.~P.; Cliffe,~M.~J.;
  Kerber,~R.~N.; Marbella,~L.~E.; Grey,~C.~P.; Elliot,~S.~R.; Cs{\'a}nyi,~G.
  {Realistic Atomistic Structure of Amorphous Silicon from
  Machine-Learning-Driven Molecular Dynamics}. \emph{J. Phys. Chem. Lett.}
  \textbf{2018}, \emph{9}, 2879\relax
\mciteBstWouldAddEndPuncttrue
\mciteSetBstMidEndSepPunct{\mcitedefaultmidpunct}
{\mcitedefaultendpunct}{\mcitedefaultseppunct}\relax
\EndOfBibitem
\bibitem[Bernstein \latin{et~al.}(2019)Bernstein, Bhattarai, Cs{\'a}nyi,
  Drabold, Elliott, and Deringer]{csanyi_deringer_silicon_angew_chem_2019}
Bernstein,~N.; Bhattarai,~B.; Cs{\'a}nyi,~G.; Drabold,~D.~A.; Elliott,~S.~R.;
  Deringer,~V.~L. {Quantifying Chemical Structure and Machine-Learned Atomic
  Energies in Amorphous and Liquid Silicon}. \emph{Angew. Chem, Int. Ed.}
  \textbf{2019}, \emph{58}, 7057\relax
\mciteBstWouldAddEndPuncttrue
\mciteSetBstMidEndSepPunct{\mcitedefaultmidpunct}
{\mcitedefaultendpunct}{\mcitedefaultseppunct}\relax
\EndOfBibitem
\bibitem[Veit \latin{et~al.}(2019)Veit, Jain, Bonakala, Rudra, Hohl, and
  Cs{\'a}nyi]{csanyi_ml_pes_jctc_2019}
Veit,~M.; Jain,~S.~K.; Bonakala,~S.; Rudra,~I.; Hohl,~D.; Cs{\'a}nyi,~G.
  {Equation of State of Fluid Methane from First Principles with Machine
  Learning Potentials}. \emph{J. Chem. Theory Comput.} \textbf{2019},
  \emph{15}, 2574\relax
\mciteBstWouldAddEndPuncttrue
\mciteSetBstMidEndSepPunct{\mcitedefaultmidpunct}
{\mcitedefaultendpunct}{\mcitedefaultseppunct}\relax
\EndOfBibitem
\bibitem[Podryabinkin \latin{et~al.}(2019)Podryabinkin, Tikhonov, Shapeev, and
  Oganov]{podryabinkin_ml_crystals_prb_2019}
Podryabinkin,~E.~V.; Tikhonov,~E.~V.; Shapeev,~A.~V.; Oganov,~A.~R.
  {Accelerating Crystal Structure Prediction by Machine-Learning Interatomic
  Potentials with Active Learning}. \emph{Phys. Rev. B} \textbf{2019},
  \emph{99}, 064114\relax
\mciteBstWouldAddEndPuncttrue
\mciteSetBstMidEndSepPunct{\mcitedefaultmidpunct}
{\mcitedefaultendpunct}{\mcitedefaultseppunct}\relax
\EndOfBibitem
\bibitem[Cole \latin{et~al.}(2020)Cole, Mones, and
  Cs{\'a}nyi]{csanyi_ml_ff_faraday_2020}
Cole,~D.~J.; Mones,~L.; Cs{\'a}nyi,~G. {A Machine Learning Based Intramolecular
  Potential for a Flexible Organic Molecule}. \emph{Faraday Discuss.}
  \textbf{2020}, DOI: 10.1039/D0FD00028K\relax
\mciteBstWouldAddEndPuncttrue
\mciteSetBstMidEndSepPunct{\mcitedefaultmidpunct}
{\mcitedefaultendpunct}{\mcitedefaultseppunct}\relax
\EndOfBibitem
\bibitem[Jinnouchi \latin{et~al.}(2019)Jinnouchi, Karsai, and
  Kresse]{kresse_ml_ff_prb_2019}
Jinnouchi,~R.; Karsai,~F.; Kresse,~G. {On-the-Fly Machine Learning Force Field
  Generation: Application to Melting Points}. \emph{Phys. Rev. B}
  \textbf{2019}, \emph{100}, 014105\relax
\mciteBstWouldAddEndPuncttrue
\mciteSetBstMidEndSepPunct{\mcitedefaultmidpunct}
{\mcitedefaultendpunct}{\mcitedefaultseppunct}\relax
\EndOfBibitem
\bibitem[Coupry \latin{et~al.}(2016)Coupry, Addicoat, and
  Heine]{heine_uff4mof_jctc_2016}
Coupry,~D.~E.; Addicoat,~M.~A.; Heine,~T. {Extension of the Universal Force
  Field for Metal-Organic Frameworks}. \emph{J. Chem. Theory Comput.}
  \textbf{2016}, \emph{12}, 5215\relax
\mciteBstWouldAddEndPuncttrue
\mciteSetBstMidEndSepPunct{\mcitedefaultmidpunct}
{\mcitedefaultendpunct}{\mcitedefaultseppunct}\relax
\EndOfBibitem
\bibitem[Haldar \latin{et~al.}(2019)Haldar, Batra, Marschner, Kuc, Zahn,
  Fischer, Br{\"a}se, Heine, and W{\"o}ll]{heine_woell_mof_chem_eur_journ_2019}
Haldar,~R.; Batra,~K.; Marschner,~S.~M.; Kuc,~A.~B.; Zahn,~S.; Fischer,~R.~A.;
  Br{\"a}se,~S.; Heine,~T.; W{\"o}ll,~C. {Bridging the Green Gap: Metal-Organic
  Framework Heteromultilayers Assembled from Porphyrinic Linkers Identified by
  Using Computational Screening}. \emph{Chem.: Eur. J.} \textbf{2019},
  \emph{25}, 7847\relax
\mciteBstWouldAddEndPuncttrue
\mciteSetBstMidEndSepPunct{\mcitedefaultmidpunct}
{\mcitedefaultendpunct}{\mcitedefaultseppunct}\relax
\EndOfBibitem
\bibitem[Jinnouchi \latin{et~al.}(2020)Jinnouchi, Miwa, Karsai, Kresse, and
  Asahi]{kresse_asahi_ml_qc_review_jpcl_2020}
Jinnouchi,~R.; Miwa,~K.; Karsai,~F.; Kresse,~G.; Asahi,~R. {On-the-Fly Active
  Learning of Interatomic Potentials for Large-Scale Atomistic Simulations}.
  \emph{J. Phys. Chem. Lett.} \textbf{2020}, \emph{11}, 6946\relax
\mciteBstWouldAddEndPuncttrue
\mciteSetBstMidEndSepPunct{\mcitedefaultmidpunct}
{\mcitedefaultendpunct}{\mcitedefaultseppunct}\relax
\EndOfBibitem
\bibitem[Langer \latin{et~al.}(2020)Langer, Goe{\ss}mann, and
  Rupp]{rupp_ml_qc_review_arxiv_2020}
Langer,~M.~F.; Goe{\ss}mann,~A.; Rupp,~M. {Representations of Molecules and
  Materials for Interpolation of Quantum-Mechanical Simulations via Machine
  Learning}. \emph{arXiv:2003.12081} \textbf{2020}\relax
\mciteBstWouldAddEndPunctfalse
\mciteSetBstMidEndSepPunct{\mcitedefaultmidpunct}
{\mcitedefaultendpunct}{\mcitedefaultseppunct}\relax
\EndOfBibitem
\bibitem[Behler(2011)]{behler_ml_acsf_jcp_2011}
Behler,~J. {Atom-Centered Symmetry Functions for Constructing High-Dimensional
  Neural Networks Potentials}. \emph{J. Chem. Phys.} \textbf{2011}, \emph{134},
  074106\relax
\mciteBstWouldAddEndPuncttrue
\mciteSetBstMidEndSepPunct{\mcitedefaultmidpunct}
{\mcitedefaultendpunct}{\mcitedefaultseppunct}\relax
\EndOfBibitem
\bibitem[Rupp \latin{et~al.}(2012)Rupp, Tkatchenko, M{\"u}ller, and von
  Lilienfeld]{lilienfeld_ml_atom_energy_prl_2012}
Rupp,~M.; Tkatchenko,~A.; M{\"u}ller,~K.-R.; von Lilienfeld,~O.~A. {Fast and
  Accurate Modeling of Molecular Atomization Energies with Machine Learning}.
  \emph{Phys. Rev. Lett.} \textbf{2012}, \emph{108}, 058301\relax
\mciteBstWouldAddEndPuncttrue
\mciteSetBstMidEndSepPunct{\mcitedefaultmidpunct}
{\mcitedefaultendpunct}{\mcitedefaultseppunct}\relax
\EndOfBibitem
\bibitem[Montavon \latin{et~al.}(2013)Montavon, Rupp, Gobre,
  Vazquez-Mayagoitia, Hansen, Tkatchenko, M{\"u}ller, and von
  Lilienfeld]{tkatchenko_mueller_lilienfeld_ml_atom_energy_njp_2013}
Montavon,~G.; Rupp,~M.; Gobre,~V.; Vazquez-Mayagoitia,~A.; Hansen,~K.;
  Tkatchenko,~A.; M{\"u}ller,~K.-R.; von Lilienfeld,~O.~A. {Machine Learning of
  Molecular Electronic Properties in Chemical Compound Space}. \emph{New J.
  Phys.} \textbf{2013}, \emph{15}, 095003\relax
\mciteBstWouldAddEndPuncttrue
\mciteSetBstMidEndSepPunct{\mcitedefaultmidpunct}
{\mcitedefaultendpunct}{\mcitedefaultseppunct}\relax
\EndOfBibitem
\bibitem[Hansen \latin{et~al.}(2015)Hansen, Biegler, Ramakrishnan, Pronobis,
  von Lilienfeld, M{\"u}ller, and
  Tkatchenko]{lilienfeld_mueller_tkatchenko_ml_atom_energy_jpcl_2015}
Hansen,~K.; Biegler,~F.; Ramakrishnan,~R.; Pronobis,~W.; von Lilienfeld,~O.~A.;
  M{\"u}ller,~K.-R.; Tkatchenko,~A. {Machine Learning Predictions of Molecular
  Properties: Accurate Many-Body Potentials and Nonlocality in Chemical Space}.
  \emph{J. Phys. Chem. Lett.} \textbf{2015}, \emph{6}, 2326\relax
\mciteBstWouldAddEndPuncttrue
\mciteSetBstMidEndSepPunct{\mcitedefaultmidpunct}
{\mcitedefaultendpunct}{\mcitedefaultseppunct}\relax
\EndOfBibitem
\bibitem[Faber \latin{et~al.}(2018)Faber, Christensen, Huang, and von
  Lilienfeld]{lilienfeld_fchl_jcp_2018}
Faber,~F.~A.; Christensen,~A.~S.; Huang,~B.; von Lilienfeld,~O.~A. {Alchemical
  and Structural Distribution Based Representation for Universal Quantum
  Machine Learning}. \emph{J. Chem. Phys.} \textbf{2018}, \emph{148},
  241717\relax
\mciteBstWouldAddEndPuncttrue
\mciteSetBstMidEndSepPunct{\mcitedefaultmidpunct}
{\mcitedefaultendpunct}{\mcitedefaultseppunct}\relax
\EndOfBibitem
\bibitem[Christensen \latin{et~al.}(2020)Christensen, Bratholm, Faber,
  Glowacki, and von Lilienfeld]{lilienfeld_fchl_jcp_2020}
Christensen,~A.~S.; Bratholm,~L.~A.; Faber,~F.~A.; Glowacki,~D.~R.; von
  Lilienfeld,~O.~A. {FCHL Revisited: Faster and More Accurate Quantum Machine
  Learning}. \emph{J. Chem. Phys.} \textbf{2020}, \emph{152}, 044107\relax
\mciteBstWouldAddEndPuncttrue
\mciteSetBstMidEndSepPunct{\mcitedefaultmidpunct}
{\mcitedefaultendpunct}{\mcitedefaultseppunct}\relax
\EndOfBibitem
\bibitem[Christensen and von Lilienfeld(2020)Christensen, and von
  Lilienfeld]{lilienfeld_ml_qc_mlst_2020}
Christensen,~A.~S.; von Lilienfeld,~O.~A. {On the Role of Gradients for Machine
  Learning of Molecular Energies and Forces}. \emph{Mach. Learn.: Sci.
  Technol.} \textbf{2020}, \emph{1}, 045018\relax
\mciteBstWouldAddEndPuncttrue
\mciteSetBstMidEndSepPunct{\mcitedefaultmidpunct}
{\mcitedefaultendpunct}{\mcitedefaultseppunct}\relax
\EndOfBibitem
\bibitem[Bart{\'o}k \latin{et~al.}(2013)Bart{\'o}k, Kondor, and
  Cs{\'a}nyi]{csanyi_ml_soap_prb_2013}
Bart{\'o}k,~A.~P.; Kondor,~R.; Cs{\'a}nyi,~G. {On Representing Chemical
  Environments}. \emph{Phys. Rev. B} \textbf{2013}, \emph{87}, 184115\relax
\mciteBstWouldAddEndPuncttrue
\mciteSetBstMidEndSepPunct{\mcitedefaultmidpunct}
{\mcitedefaultendpunct}{\mcitedefaultseppunct}\relax
\EndOfBibitem
\bibitem[Bart{\'o}k \latin{et~al.}(2017)Bart{\'o}k, De, Poelking, Bernstein,
  Kermode, Cs{\'a}nyi, and Ceriotti]{ceriotti_ml_materials_sci_adv_2017}
Bart{\'o}k,~A.~P.; De,~S.; Poelking,~C.; Bernstein,~N.; Kermode,~J.~R.;
  Cs{\'a}nyi,~G.; Ceriotti,~M. {Machine Learning Unifies the Modeling of
  Materials and Molecules}. \emph{Sci. Adv.} \textbf{2017}, \emph{3},
  e1701816\relax
\mciteBstWouldAddEndPuncttrue
\mciteSetBstMidEndSepPunct{\mcitedefaultmidpunct}
{\mcitedefaultendpunct}{\mcitedefaultseppunct}\relax
\EndOfBibitem
\bibitem[Grisafi and Ceriotti(2019)Grisafi, and
  Ceriotti]{ceriotti_ml_long_range_jcp_2019}
Grisafi,~A.; Ceriotti,~M. {Incorporating Long-Range Physics in Atomic-Scale
  Machine Learning}. \emph{J. Chem. Phys.} \textbf{2019}, \emph{151},
  204105\relax
\mciteBstWouldAddEndPuncttrue
\mciteSetBstMidEndSepPunct{\mcitedefaultmidpunct}
{\mcitedefaultendpunct}{\mcitedefaultseppunct}\relax
\EndOfBibitem
\bibitem[Willatt \latin{et~al.}(2019)Willatt, Musli, and
  Ceriotti]{ceriotti_ml_qc_jcp_2019}
Willatt,~M.~J.; Musli,~F.; Ceriotti,~M. {Atom-Density Representations for
  Machine Learning}. \emph{J. Chem. Phys.} \textbf{2019}, \emph{150},
  154110\relax
\mciteBstWouldAddEndPuncttrue
\mciteSetBstMidEndSepPunct{\mcitedefaultmidpunct}
{\mcitedefaultendpunct}{\mcitedefaultseppunct}\relax
\EndOfBibitem
\bibitem[Zhu \latin{et~al.}(2016)Zhu, Amsler, Fuhrer, Schaefer, Faraji,
  Rostami, Ghasemi, Sadeghi, Grauzinyte, Wolverton, and
  Goedecker]{goedecker_ml_crystal_jcp_2016}
Zhu,~L.; Amsler,~M.; Fuhrer,~T.; Schaefer,~B.; Faraji,~S.; Rostami,~S.;
  Ghasemi,~S.~A.; Sadeghi,~A.; Grauzinyte,~M.; Wolverton,~C.; Goedecker,~S. {A
  Fingerprint Based Metric for Measuring Similarities of Crystalline
  Structures}. \emph{J. Chem. Phys.} \textbf{2016}, \emph{144}, 034203\relax
\mciteBstWouldAddEndPuncttrue
\mciteSetBstMidEndSepPunct{\mcitedefaultmidpunct}
{\mcitedefaultendpunct}{\mcitedefaultseppunct}\relax
\EndOfBibitem
\bibitem[Townsend \latin{et~al.}(2020)Townsend, Micucci, Hymel, Maroulas, and
  Vogiatzis]{vogiatzis_ml_qc_nat_commun_2020}
Townsend,~J.; Micucci,~C.~P.; Hymel,~J.~H.; Maroulas,~V.; Vogiatzis,~K.~D.
  {Representation of Molecular Structures with Persistent Homology for Machine
  Learning Applications in Chemistry}. \emph{Nat. Commun.} \textbf{2020},
  \emph{11}, 3230\relax
\mciteBstWouldAddEndPuncttrue
\mciteSetBstMidEndSepPunct{\mcitedefaultmidpunct}
{\mcitedefaultendpunct}{\mcitedefaultseppunct}\relax
\EndOfBibitem
\bibitem[Huang and von Lilienfeld(2016)Huang, and von
  Lilienfeld]{lilienfeld_qml_jcp_2016}
Huang,~B.; von Lilienfeld,~O.~A. {Communication: Understanding Molecular
  Representations in Machine Learning: The Role of Uniqueness and Target
  Similarity}. \emph{J. Chem. Phys.} \textbf{2016}, \emph{145}, 161102\relax
\mciteBstWouldAddEndPuncttrue
\mciteSetBstMidEndSepPunct{\mcitedefaultmidpunct}
{\mcitedefaultendpunct}{\mcitedefaultseppunct}\relax
\EndOfBibitem
\bibitem[Pronobis \latin{et~al.}(2018)Pronobis, Tkatchenko, and
  M{\"u}ller]{tkatchenko_mueller_ml_many_body_jctc_2018}
Pronobis,~W.; Tkatchenko,~A.; M{\"u}ller,~K.-R. {Many-Body Descriptors for
  Predicting Molecular Properties with Machine Learning: Analysis of Pairwise
  and Three-Body Interactions in Molecules}. \emph{J. Chem. Theory Comput.}
  \textbf{2018}, \emph{14}, 2991\relax
\mciteBstWouldAddEndPuncttrue
\mciteSetBstMidEndSepPunct{\mcitedefaultmidpunct}
{\mcitedefaultendpunct}{\mcitedefaultseppunct}\relax
\EndOfBibitem
\bibitem[Huang and von Lilienfeld(2020)Huang, and von
  Lilienfeld]{huang_lilienfeld_ml_amons_nat_chem_2020}
Huang,~B.; von Lilienfeld,~O.~A. {Quantum Machine Learning Using
  Atom-in-Molecule-Based Fragments Selected On the Fly}. \emph{Nat. Chem.}
  \textbf{2020}, \emph{12}, 945\relax
\mciteBstWouldAddEndPuncttrue
\mciteSetBstMidEndSepPunct{\mcitedefaultmidpunct}
{\mcitedefaultendpunct}{\mcitedefaultseppunct}\relax
\EndOfBibitem
\bibitem[Huang and von Lilienfeld(2020)Huang, and von
  Lilienfeld]{huang_lilienfeld_ml_amons_arxiv_2020}
Huang,~B.; von Lilienfeld,~O.~A. {Dictionary of 140k GDB and ZINC derived
  AMONs}. \emph{arXiv:2008.05260} \textbf{2020}\relax
\mciteBstWouldAddEndPunctfalse
\mciteSetBstMidEndSepPunct{\mcitedefaultmidpunct}
{\mcitedefaultendpunct}{\mcitedefaultseppunct}\relax
\EndOfBibitem
\bibitem[Gordon \latin{et~al.}(2012)Gordon, Fedorov, Pruitt, and
  Slipchenko]{gordon_slipchenko_chem_rev_2012}
Gordon,~M.~S.; Fedorov,~D.~G.; Pruitt,~S.~R.; Slipchenko,~L.~V. {Fragmentation
  Methods: A Route to Accurate Calculations on Large Systems.} \emph{{C}hem.
  {R}ev.} \textbf{2012}, \emph{112}, 632\relax
\mciteBstWouldAddEndPuncttrue
\mciteSetBstMidEndSepPunct{\mcitedefaultmidpunct}
{\mcitedefaultendpunct}{\mcitedefaultseppunct}\relax
\EndOfBibitem
\bibitem[Isayev \latin{et~al.}(2017)Isayev, Oses, Toher, Gossett, Curtarolo,
  and Tropsha]{isayev_tropsha_crystals_nat_commun_2017}
Isayev,~O.; Oses,~C.; Toher,~C.; Gossett,~E.; Curtarolo,~S.; Tropsha,~A.
  {Universal Fragment Descriptors for Predicting Properties of Inorganic
  Crystals}. \emph{Nat. Commun.} \textbf{2017}, \emph{8}, 15679\relax
\mciteBstWouldAddEndPuncttrue
\mciteSetBstMidEndSepPunct{\mcitedefaultmidpunct}
{\mcitedefaultendpunct}{\mcitedefaultseppunct}\relax
\EndOfBibitem
\bibitem[Eckhoff and Behler(2019)Eckhoff, and
  Behler]{eckhoff_behler_mof_jctc_2019}
Eckhoff,~M.; Behler,~J. {From Molecular Fragments to the Bulk: Development of a
  Neural Network Potential for MOF-5}. \emph{J. Chem. Theory Comput.}
  \textbf{2019}, \emph{15}, 3793\relax
\mciteBstWouldAddEndPuncttrue
\mciteSetBstMidEndSepPunct{\mcitedefaultmidpunct}
{\mcitedefaultendpunct}{\mcitedefaultseppunct}\relax
\EndOfBibitem
\bibitem[McDonagh \latin{et~al.}(2019)McDonagh, Skylaris, and
  Day]{skylaris_day_ml_crystals_jctc_2019}
McDonagh,~D.; Skylaris,~C.-K.; Day,~G.~M. {Machine-Learned Fragment-Based
  Energies for Crystal Structure Prediction}. \emph{J. Chem. Theory Comput.}
  \textbf{2019}, \emph{15}, 2743\relax
\mciteBstWouldAddEndPuncttrue
\mciteSetBstMidEndSepPunct{\mcitedefaultmidpunct}
{\mcitedefaultendpunct}{\mcitedefaultseppunct}\relax
\EndOfBibitem
\bibitem[Smith \latin{et~al.}(2017)Smith, Isayev, and
  Roitberg]{isayev_roitberg_ml_qc_chem_sci_2017}
Smith,~J.~S.; Isayev,~O.; Roitberg,~A.~E. {ANI-1: An Extensible Neural Network
  Potential with DFT Accuracy at Force Field Computational Cost}. \emph{Chem.
  Sci.} \textbf{2017}, \emph{8}, 3192\relax
\mciteBstWouldAddEndPuncttrue
\mciteSetBstMidEndSepPunct{\mcitedefaultmidpunct}
{\mcitedefaultendpunct}{\mcitedefaultseppunct}\relax
\EndOfBibitem
\bibitem[Smith \latin{et~al.}(2019)Smith, Isayev, and
  Roitberg]{isayev_roitberg_ml_qc_nat_commun_2019}
Smith,~J.~S.; Isayev,~O.; Roitberg,~A.~E. {Approaching Coupled Cluster Accuracy
  with a General-Purpose Neural Network Potential Through Transfer Learning}.
  \emph{Nat. Commun.} \textbf{2019}, \emph{10}, 2903\relax
\mciteBstWouldAddEndPuncttrue
\mciteSetBstMidEndSepPunct{\mcitedefaultmidpunct}
{\mcitedefaultendpunct}{\mcitedefaultseppunct}\relax
\EndOfBibitem
\bibitem[Chen \latin{et~al.}(2018)Chen, J{\o}rgensen, Li, and
  Hammer]{hammer_ml_qc_jctc_2018}
Chen,~X.; J{\o}rgensen,~M.~S.; Li,~J.; Hammer,~B. {Atomic Energies from a
  Convolutional Neural Network}. \emph{J. Chem. Theory Comput.} \textbf{2018},
  \emph{14}, 3933\relax
\mciteBstWouldAddEndPuncttrue
\mciteSetBstMidEndSepPunct{\mcitedefaultmidpunct}
{\mcitedefaultendpunct}{\mcitedefaultseppunct}\relax
\EndOfBibitem
\bibitem[Meldgaard \latin{et~al.}(2018)Meldgaard, Kolsbjerg, and
  Hammer]{hammer_ml_qc_jcp_2018}
Meldgaard,~S.~A.; Kolsbjerg,~E.~L.; Hammer,~B. {Machine Learning Enhanced
  Global Optimization by Clustering Local Environments to Enable Bundled Atomic
  Energies}. \emph{J. Chem. Phys.} \textbf{2018}, \emph{149}, 134104\relax
\mciteBstWouldAddEndPuncttrue
\mciteSetBstMidEndSepPunct{\mcitedefaultmidpunct}
{\mcitedefaultendpunct}{\mcitedefaultseppunct}\relax
\EndOfBibitem
\bibitem[J{\o}rgensen \latin{et~al.}(2019)J{\o}rgensen, Mortensen, Meldgaard,
  Kolsbjerg, Jacobsen, S{\o}rensen, and Hammer]{hammer_ml_qc_jcp_2019}
J{\o}rgensen,~M.~S.; Mortensen,~H.~L.; Meldgaard,~S.~A.; Kolsbjerg,~E.~L.;
  Jacobsen,~T.~L.; S{\o}rensen,~K.~H.; Hammer,~B. {Atomistic Structure
  Learning}. \emph{J. Chem. Phys.} \textbf{2019}, \emph{151}, 054111\relax
\mciteBstWouldAddEndPuncttrue
\mciteSetBstMidEndSepPunct{\mcitedefaultmidpunct}
{\mcitedefaultendpunct}{\mcitedefaultseppunct}\relax
\EndOfBibitem
\bibitem[Unke and Meuwly(2018)Unke, and Meuwly]{unke_meuwly_ml_nn_jcp_2018}
Unke,~O.~T.; Meuwly,~M. {A Reactive, Scalable, and Transferable Model for
  Molecular Energies from a Neural Network Approach Based on Local
  Information}. \emph{J. Chem. Phys.} \textbf{2018}, \emph{148}, 241708\relax
\mciteBstWouldAddEndPuncttrue
\mciteSetBstMidEndSepPunct{\mcitedefaultmidpunct}
{\mcitedefaultendpunct}{\mcitedefaultseppunct}\relax
\EndOfBibitem
\bibitem[Kang and Wang(2017)Kang, and Wang]{kang_wang_dft_decomp_prb_2017}
Kang,~J.; Wang,~L.-W. {First-Principles Green-Kubo Method for Thermal
  Conductivity Calculations}. \emph{Phys. Rev. B} \textbf{2017}, \emph{96},
  020302(R)\relax
\mciteBstWouldAddEndPuncttrue
\mciteSetBstMidEndSepPunct{\mcitedefaultmidpunct}
{\mcitedefaultendpunct}{\mcitedefaultseppunct}\relax
\EndOfBibitem
\bibitem[Huang \latin{et~al.}(2019)Huang, Kang, Goddard, and
  Wang]{huang_wang_nnff_prb_2019}
Huang,~Y.; Kang,~J.; Goddard,~W.~A.,~III; Wang,~L.-W. {Density Functional
  Theory Based Neural Network Force Fields from Energy Decompositions}.
  \emph{Phys. Rev. B} \textbf{2019}, \emph{99}, 064103\relax
\mciteBstWouldAddEndPuncttrue
\mciteSetBstMidEndSepPunct{\mcitedefaultmidpunct}
{\mcitedefaultendpunct}{\mcitedefaultseppunct}\relax
\EndOfBibitem
\bibitem[Perdew and Schmidt(2001)Perdew, and
  Schmidt]{perdew_jacobs_ladder_aip_conf_proc_2001}
Perdew,~J.~P.; Schmidt,~K. {Jacob's Ladder of Density Functional Approximations
  for the Exchange-Correlation Energy}. \emph{AIP Conf. Proc.} \textbf{2001},
  \emph{577}, 1\relax
\mciteBstWouldAddEndPuncttrue
\mciteSetBstMidEndSepPunct{\mcitedefaultmidpunct}
{\mcitedefaultendpunct}{\mcitedefaultseppunct}\relax
\EndOfBibitem
\bibitem[Mayer \latin{et~al.}(2017)Mayer, P{\'a}pai, Bak{\'o}, and
  Nagy]{mayer_nagy_fci_dft_rdm_jctc_2017}
Mayer,~I.; P{\'a}pai,~I.; Bak{\'o},~I.; Nagy,~A. {Conceptual Problem with
  Calculating Electron Densities in Finite Basis Density Functional Theory}.
  \emph{J. Chem. Theory Comput.} \textbf{2017}, \emph{13}, 3961\relax
\mciteBstWouldAddEndPuncttrue
\mciteSetBstMidEndSepPunct{\mcitedefaultmidpunct}
{\mcitedefaultendpunct}{\mcitedefaultseppunct}\relax
\EndOfBibitem
\bibitem[Medvedev \latin{et~al.}(2017)Medvedev, Bushmarinov, Sun, Perdew, and
  Lyssenko]{medvedev_perdew_lyssenko_density_errors_science_2017}
Medvedev,~M.~G.; Bushmarinov,~I.~S.; Sun,~J.; Perdew,~J.~P.; Lyssenko,~K.~A.
  {Density Functional Theory is Straying From the Path Toward the Exact
  Functional}. \emph{Science} \textbf{2017}, \emph{355}, 49\relax
\mciteBstWouldAddEndPuncttrue
\mciteSetBstMidEndSepPunct{\mcitedefaultmidpunct}
{\mcitedefaultendpunct}{\mcitedefaultseppunct}\relax
\EndOfBibitem
\bibitem[Sim \latin{et~al.}(2018)Sim, Song, and
  Burke]{sim_song_burke_density_errors_jpcl_2018}
Sim,~E.; Song,~S.; Burke,~K. {Quantifying Density Errors in DFT}. \emph{J.
  Phys. Chem. Lett.} \textbf{2018}, \emph{9}, 6385\relax
\mciteBstWouldAddEndPuncttrue
\mciteSetBstMidEndSepPunct{\mcitedefaultmidpunct}
{\mcitedefaultendpunct}{\mcitedefaultseppunct}\relax
\EndOfBibitem
\bibitem[Edmiston and Ruedenberg(1963)Edmiston, and
  Ruedenberg]{edmiston_ruedenberg_rev_mod_phys_1963}
Edmiston,~C.; Ruedenberg,~K. {Localized Atomic and Molecular Orbitals}.
  \emph{{R}ev. {M}od. {P}hys.} \textbf{1963}, \emph{35}, 457\relax
\mciteBstWouldAddEndPuncttrue
\mciteSetBstMidEndSepPunct{\mcitedefaultmidpunct}
{\mcitedefaultendpunct}{\mcitedefaultseppunct}\relax
\EndOfBibitem
\bibitem[Lehtola and J{\'o}nsson(2013)Lehtola, and
  J{\'o}nsson]{lehtola_jonsson_loc_orbs_jctc_2013}
Lehtola,~S.; J{\'o}nsson,~H. {Unitary Optimization of Localized Molecular
  Orbitals}. \emph{{J}. {C}hem. {T}heory {C}omput.} \textbf{2013}, \emph{9},
  5365\relax
\mciteBstWouldAddEndPuncttrue
\mciteSetBstMidEndSepPunct{\mcitedefaultmidpunct}
{\mcitedefaultendpunct}{\mcitedefaultseppunct}\relax
\EndOfBibitem
\bibitem[Yao \latin{et~al.}(2017)Yao, Herr, Brown, and
  Parkhill]{parkhill_intrinsic_bond_energies_jpcl_2017}
Yao,~K.; Herr,~J.~E.; Brown,~S.~N.; Parkhill,~J. {Intrinsic Bond Energies from
  a Bonds-in-Molecules Neural Network}. \emph{J. Phys. Chem. Lett.}
  \textbf{2017}, \emph{8}, 2689\relax
\mciteBstWouldAddEndPuncttrue
\mciteSetBstMidEndSepPunct{\mcitedefaultmidpunct}
{\mcitedefaultendpunct}{\mcitedefaultseppunct}\relax
\EndOfBibitem
\bibitem[Tao \latin{et~al.}(2003)Tao, Perdew, Staroverov, and
  Scuseria]{perdew_scuseria_tpssh_functional_prl_2003}
Tao,~J.; Perdew,~J.~P.; Staroverov,~V.~N.; Scuseria,~G.~E. {Climbing the
  Density Functional Ladder: Nonempirical Meta–Generalized Gradient
  Approximation Designed for Molecules and Solids}. \emph{Phys. Rev. Lett.}
  \textbf{2003}, \emph{91}, 146401\relax
\mciteBstWouldAddEndPuncttrue
\mciteSetBstMidEndSepPunct{\mcitedefaultmidpunct}
{\mcitedefaultendpunct}{\mcitedefaultseppunct}\relax
\EndOfBibitem
\bibitem[Staroverov \latin{et~al.}(2003)Staroverov, Scuseria, Tao, and
  Perdew]{scuseria_perdew_tpssh_functional_jcp_2003}
Staroverov,~V.~N.; Scuseria,~G.~E.; Tao,~J.; Perdew,~J.~P. {Comparative
  Assessment of a New Nonempirical Density Functional: Molecules and
  Hydrogen-Bonded Complexes}. \emph{J. Chem. Phys.} \textbf{2003}, \emph{119},
  12129\relax
\mciteBstWouldAddEndPuncttrue
\mciteSetBstMidEndSepPunct{\mcitedefaultmidpunct}
{\mcitedefaultendpunct}{\mcitedefaultseppunct}\relax
\EndOfBibitem
\bibitem[Staroverov \latin{et~al.}(2004)Staroverov, Scuseria, Tao, and
  Perdew]{scuseria_perdew_tpssh_functional_jcp_2004_erratum}
Staroverov,~V.~N.; Scuseria,~G.~E.; Tao,~J.; Perdew,~J.~P. {Erratum:
  ``Comparative Assessment of a New Nonempirical Density Functional: Molecules
  and Hydrogen-Bonded Complexes" [J. Chem. Phys. 119, 12129 (2003)]}. \emph{J.
  Chem. Phys.} \textbf{2004}, \emph{121}, 11507\relax
\mciteBstWouldAddEndPuncttrue
\mciteSetBstMidEndSepPunct{\mcitedefaultmidpunct}
{\mcitedefaultendpunct}{\mcitedefaultseppunct}\relax
\EndOfBibitem
\bibitem[Li and Parr(1986)Li, and Parr]{li_parr_partitioning_jcp_1986}
Li,~L.; Parr,~R.~G. {The Atom in a Molecule: A Density Matrix Approach}.
  \emph{J. Chem. Phys.} \textbf{1986}, \emph{84}, 1704\relax
\mciteBstWouldAddEndPuncttrue
\mciteSetBstMidEndSepPunct{\mcitedefaultmidpunct}
{\mcitedefaultendpunct}{\mcitedefaultseppunct}\relax
\EndOfBibitem
\bibitem[Ichikawa and Yoshida(1999)Ichikawa, and
  Yoshida]{ichikawa_yoshida_partitioning_ijqc_1999}
Ichikawa,~H.; Yoshida,~A. {Complete One‐ and Two‐Center Partitioning Scheme
  for the Total Energy in the Hartree-Fock Theory}. \emph{Int. J. Quant. Chem.}
  \textbf{1999}, \emph{71}, 35\relax
\mciteBstWouldAddEndPuncttrue
\mciteSetBstMidEndSepPunct{\mcitedefaultmidpunct}
{\mcitedefaultendpunct}{\mcitedefaultseppunct}\relax
\EndOfBibitem
\bibitem[Mayer(2000)]{mayer_partitioning_cpl_2000}
Mayer,~I. {A Chemical Energy Component Analysis}. \emph{Chem. Phys. Lett.}
  \textbf{2000}, \emph{332}, 381\relax
\mciteBstWouldAddEndPuncttrue
\mciteSetBstMidEndSepPunct{\mcitedefaultmidpunct}
{\mcitedefaultendpunct}{\mcitedefaultseppunct}\relax
\EndOfBibitem
\bibitem[Mayer(2003)]{mayer_partitioning_cpl_2003}
Mayer,~I. {An Exact Chemical Decomposition Scheme for the Molecular Energy}.
  \emph{Chem. Phys. Lett.} \textbf{2003}, \emph{382}, 265\relax
\mciteBstWouldAddEndPuncttrue
\mciteSetBstMidEndSepPunct{\mcitedefaultmidpunct}
{\mcitedefaultendpunct}{\mcitedefaultseppunct}\relax
\EndOfBibitem
\bibitem[Mandado \latin{et~al.}(2006)Mandado, Van~Alsenoy, Geerlings, De~Proft,
  and Mosquera]{mandado_van_alsenoy_partitioning_cpc_2006}
Mandado,~M.; Van~Alsenoy,~C.; Geerlings,~P.; De~Proft,~F.; Mosquera,~R.~A.
  {Hartree-Fock Energy Partitioning in Terms of Hirshfeld Atoms}. \emph{Chem.
  Phys. Chem.} \textbf{2006}, \emph{7}, 1294\relax
\mciteBstWouldAddEndPuncttrue
\mciteSetBstMidEndSepPunct{\mcitedefaultmidpunct}
{\mcitedefaultendpunct}{\mcitedefaultseppunct}\relax
\EndOfBibitem
\bibitem[Blanco \latin{et~al.}(2005)Blanco, Pend{\'a}s, and
  Francisco]{blanco_pendas_francisco_partitioning_jctc_2005}
Blanco,~M.~A.; Pend{\'a}s,~A.~M.; Francisco,~E. {Interacting Quantum Atoms: A
  Correlated Energy Decomposition Scheme Based on the Quantum Theory of Atoms
  in Molecules}. \emph{J. Chem. Theory Comput.} \textbf{2005}, \emph{1},
  1096\relax
\mciteBstWouldAddEndPuncttrue
\mciteSetBstMidEndSepPunct{\mcitedefaultmidpunct}
{\mcitedefaultendpunct}{\mcitedefaultseppunct}\relax
\EndOfBibitem
\bibitem[Francisco \latin{et~al.}(2006)Francisco, Pend{\'a}s, and
  Blanco]{francisco_pendas_blanco_partitioning_jctc_2006}
Francisco,~E.; Pend{\'a}s,~A.~M.; Blanco,~M.~A. {A Molecular Energy
  Decomposition Scheme for Atoms in Molecules}. \emph{J. Chem. Theory Comput.}
  \textbf{2006}, \emph{2}, 90\relax
\mciteBstWouldAddEndPuncttrue
\mciteSetBstMidEndSepPunct{\mcitedefaultmidpunct}
{\mcitedefaultendpunct}{\mcitedefaultseppunct}\relax
\EndOfBibitem
\bibitem[Nalewajski and Parr(2000)Nalewajski, and
  Parr]{nalewajski_parr_partitioning_pnas_2000}
Nalewajski,~R.~F.; Parr,~R.~G. {Information Theory, Atoms in Molecules, and
  Molecular Similarity}. \emph{Proc. Nat. Acad. Sci.} \textbf{2000}, \emph{97},
  8879\relax
\mciteBstWouldAddEndPuncttrue
\mciteSetBstMidEndSepPunct{\mcitedefaultmidpunct}
{\mcitedefaultendpunct}{\mcitedefaultseppunct}\relax
\EndOfBibitem
\bibitem[Parr \latin{et~al.}(2005)Parr, Ayers, and
  Nalewajski]{parr_ayers_nalewajski_partitioning_jpca_2005}
Parr,~R.~G.; Ayers,~P.~W.; Nalewajski,~R.~F. {What Is an Atom in a Molecule?}
  \emph{J. Phys. Chem. A} \textbf{2005}, \emph{109}, 3957\relax
\mciteBstWouldAddEndPuncttrue
\mciteSetBstMidEndSepPunct{\mcitedefaultmidpunct}
{\mcitedefaultendpunct}{\mcitedefaultseppunct}\relax
\EndOfBibitem
\bibitem[von Rudorff and von Lilienfeld(2019)von Rudorff, and von
  Lilienfeld]{lilienfeld_alchemical_pt_jpcb_2019}
von Rudorff,~G.~F.; von Lilienfeld,~O.~A. {Atoms in Molecules from Alchemical
  Perturbation Density Functional Theory}. \emph{J. Phys. Chem. B}
  \textbf{2019}, \emph{123}, 10073\relax
\mciteBstWouldAddEndPuncttrue
\mciteSetBstMidEndSepPunct{\mcitedefaultmidpunct}
{\mcitedefaultendpunct}{\mcitedefaultseppunct}\relax
\EndOfBibitem
\bibitem[Nakai(2002)]{nakai_eda_partitioning_cpl_2002}
Nakai,~H. {Energy Density Analysis with Kohn-Sham Orbitals}. \emph{Chem. Phys.
  Lett.} \textbf{2002}, \emph{363}, 73\relax
\mciteBstWouldAddEndPuncttrue
\mciteSetBstMidEndSepPunct{\mcitedefaultmidpunct}
{\mcitedefaultendpunct}{\mcitedefaultseppunct}\relax
\EndOfBibitem
\bibitem[Kikuchi \latin{et~al.}(2009)Kikuchi, Imamura, and
  Nakai]{nakai_eda_partitioning_ijqc_2009}
Kikuchi,~Y.; Imamura,~Y.; Nakai,~H. {One-Body Energy Decomposition Schemes
  Revisited: Assessment of Mulliken-, Grid-, and Conventional Energy Density
  Analyses}. \emph{Int. J. Quant. Chem.} \textbf{2009}, \emph{109}, 2464\relax
\mciteBstWouldAddEndPuncttrue
\mciteSetBstMidEndSepPunct{\mcitedefaultmidpunct}
{\mcitedefaultendpunct}{\mcitedefaultseppunct}\relax
\EndOfBibitem
\bibitem[Mulliken(1955)]{mulliken_population_jcp_1955}
Mulliken,~R.~S. {Electronic Population Analysis on LCAO-MO Molecular Wave
  Functions. I}. \emph{J. Chem. Phys} \textbf{1955}, \emph{23}, 1833\relax
\mciteBstWouldAddEndPuncttrue
\mciteSetBstMidEndSepPunct{\mcitedefaultmidpunct}
{\mcitedefaultendpunct}{\mcitedefaultseppunct}\relax
\EndOfBibitem
\bibitem[Knizia(2013)]{knizia_iao_ibo_jctc_2013}
Knizia,~G. {Intrinsic Atomic Orbitals: An Unbiased Bridge Between Quantum
  Theory and Chemical Concepts}. \emph{{J}. {C}hem. {T}heory {C}omput.}
  \textbf{2013}, \emph{9}, 4834\relax
\mciteBstWouldAddEndPuncttrue
\mciteSetBstMidEndSepPunct{\mcitedefaultmidpunct}
{\mcitedefaultendpunct}{\mcitedefaultseppunct}\relax
\EndOfBibitem
\bibitem[Senjean \latin{et~al.}(2020)Senjean, Sen, Repisky, Knizia, and
  Visscher]{knizia_visscher_iao_arxiv_2020}
Senjean,~B.; Sen,~S.; Repisky,~M.; Knizia,~G.; Visscher,~L. {Generalization of
  Intrinsic Orbitals to Kramers-Paired Quaternion Spinors, Molecular Fragments
  and Valence Virtual Spinors}. \emph{arXiv:2009.08671} \textbf{2020}\relax
\mciteBstWouldAddEndPunctfalse
\mciteSetBstMidEndSepPunct{\mcitedefaultmidpunct}
{\mcitedefaultendpunct}{\mcitedefaultseppunct}\relax
\EndOfBibitem
\bibitem[Christensen \latin{et~al.}(2016)Christensen, Kuba{\v{r}}, Cui, and
  Elstner]{christensen_elstner_semiempirical_qc_chem_rev_2016}
Christensen,~A.~S.; Kuba{\v{r}},~T.; Cui,~Q.; Elstner,~M. {Semiempirical
  Quantum Mechanical Methods for Noncovalent Interactions for Chemical and
  Biochemical Applications}. \emph{Chem. Rev.} \textbf{2016}, \emph{116},
  5301\relax
\mciteBstWouldAddEndPuncttrue
\mciteSetBstMidEndSepPunct{\mcitedefaultmidpunct}
{\mcitedefaultendpunct}{\mcitedefaultseppunct}\relax
\EndOfBibitem
\bibitem[Lehtola and J{\'o}nsson(2014)Lehtola, and
  J{\'o}nsson]{lehtola_jonsson_pm_jctc_2014}
Lehtola,~S.; J{\'o}nsson,~H. {Pipek-Mezey Orbital Localization Using Various
  Partial Charge Estimates}. \emph{{J}. {C}hem. {T}heory {C}omput.}
  \textbf{2014}, \emph{10}, 642\relax
\mciteBstWouldAddEndPuncttrue
\mciteSetBstMidEndSepPunct{\mcitedefaultmidpunct}
{\mcitedefaultendpunct}{\mcitedefaultseppunct}\relax
\EndOfBibitem
\bibitem[Hirshfeld(1977)]{hirshfeld_partitioning_tca_1977}
Hirshfeld,~F.~L. {Bonded-Atom Fragments for Describing Molecular Charge
  Densities}. \emph{Theor. Chim. Acta} \textbf{1977}, \emph{44}, 129\relax
\mciteBstWouldAddEndPuncttrue
\mciteSetBstMidEndSepPunct{\mcitedefaultmidpunct}
{\mcitedefaultendpunct}{\mcitedefaultseppunct}\relax
\EndOfBibitem
\bibitem[Becke(1988)]{becke_partitioning_jcp_1988}
Becke,~A.~D. {A Multicenter Numerical Integration Scheme for Polyatomic
  Molecules}. \emph{J. Chem. Phys.} \textbf{1988}, \emph{88}, 2547\relax
\mciteBstWouldAddEndPuncttrue
\mciteSetBstMidEndSepPunct{\mcitedefaultmidpunct}
{\mcitedefaultendpunct}{\mcitedefaultseppunct}\relax
\EndOfBibitem
\bibitem[Bader(1990)]{bader_partitioning_book_1990}
Bader,~R. F.~W. \emph{Atoms in Molecules: A Quantum Theory}, 1st ed.; Oxford
  University Press: New York, USA, 1990\relax
\mciteBstWouldAddEndPuncttrue
\mciteSetBstMidEndSepPunct{\mcitedefaultmidpunct}
{\mcitedefaultendpunct}{\mcitedefaultseppunct}\relax
\EndOfBibitem
\bibitem[M{\"u}ller \latin{et~al.}(2001)M{\"u}ller, Mika, R{\"a}tsch, Tsuda,
  and Sch{\"o}lkopf]{mueller_schoelkopf_ieee_trans_nn_2001}
M{\"u}ller,~K.-R.; Mika,~S.; R{\"a}tsch,~G.; Tsuda,~K.; Sch{\"o}lkopf,~B. {An
  Introduction to Kernel-Based Learning Algorithms}. \emph{IEEE Trans. Neural
  Networks} \textbf{2001}, \emph{12}, 181\relax
\mciteBstWouldAddEndPuncttrue
\mciteSetBstMidEndSepPunct{\mcitedefaultmidpunct}
{\mcitedefaultendpunct}{\mcitedefaultseppunct}\relax
\EndOfBibitem
\bibitem[Tikhonov and Arsenin(1977)Tikhonov, and
  Arsenin]{tikhonov_ill_posed_prob_1977}
Tikhonov,~A.~N.; Arsenin,~V.~Y. \emph{{Solutions of Ill-Posed Problems (Scripta
  Series in Mathematics)}}, 1st ed.; John Wiley \& Sons, Inc., 1977\relax
\mciteBstWouldAddEndPuncttrue
\mciteSetBstMidEndSepPunct{\mcitedefaultmidpunct}
{\mcitedefaultendpunct}{\mcitedefaultseppunct}\relax
\EndOfBibitem
\bibitem[dec()]{decodense}
{\texttt{DECODENSE}}: A Decomposed Mean-Field Theory Code,
  See~{\url{https://github.com/januseriksen/decodense}}\relax
\mciteBstWouldAddEndPuncttrue
\mciteSetBstMidEndSepPunct{\mcitedefaultmidpunct}
{\mcitedefaultendpunct}{\mcitedefaultseppunct}\relax
\EndOfBibitem
\bibitem[Sun \latin{et~al.}(2018)Sun, Berkelbach, Blunt, Booth, Guo, Li, Liu,
  McClain, Sayfutyarova, Sharma, Wouters, and Chan]{pyscf_wires_2018}
Sun,~Q.; Berkelbach,~T.~C.; Blunt,~N.~S.; Booth,~G.~H.; Guo,~S.; Li,~Z.;
  Liu,~J.; McClain,~J.~D.; Sayfutyarova,~E.~R.; Sharma,~S.; Wouters,~S.;
  Chan,~G. K.-L. {PySCF: The Python-Based Simulations of Chemistry Framework}.
  \emph{{W}IREs {C}omput. {M}ol. {S}ci.} \textbf{2018}, \emph{8}, e1340\relax
\mciteBstWouldAddEndPuncttrue
\mciteSetBstMidEndSepPunct{\mcitedefaultmidpunct}
{\mcitedefaultendpunct}{\mcitedefaultseppunct}\relax
\EndOfBibitem
\bibitem[Sun \latin{et~al.}(2020)Sun, Zhang, Banerjee, Bao, Barbry, Blunt,
  Bogdanov, Booth, Chen, Cui, Eriksen, Gao, Guo, Hermann, Hermes, Koh, Koval,
  Lehtola, Li, Liu, Mardirossian, McClain, Motta, Mussard, Pham, Pulkin,
  Purwanto, Robinson, Ronca, Sayfutyarova, Scheurer, Schurkus, Smith, Sun, Sun,
  Upadhyay, Wagner, Wang, White, Whitfield, Williamson, Wouters, Yang, Yu, Zhu,
  Berkelbach, Sharma, Sokolov, and Chan]{pyscf_jcp_2020}
Sun,~Q.; Zhang,~X.; Banerjee,~S.; Bao,~P.; Barbry,~M.; Blunt,~N.~S.;
  Bogdanov,~N.~A.; Booth,~G.~H.; Chen,~J.; Cui,~Z.-H.; Eriksen,~J.~J.; Gao,~Y.;
  Guo,~S.; Hermann,~J.; Hermes,~M.~R.; Koh,~K.; Koval,~P.; Lehtola,~S.; Li,~Z.;
  Liu,~J.; Mardirossian,~N.; McClain,~J.~D.; Motta,~M.; Mussard,~B.;
  Pham,~H.~Q.; Pulkin,~A.; Purwanto,~W.; Robinson,~P.~J.; Ronca,~E.;
  Sayfutyarova,~E.~R.; Scheurer,~M.; Schurkus,~H.~F.; Smith,~J. E.~T.; Sun,~C.;
  Sun,~S.-N.; Upadhyay,~S.; Wagner,~L.~K.; Wang,~X.; White,~A.;
  Whitfield,~J.~D.; Williamson,~M.~J.; Wouters,~S.; Yang,~J.; Yu,~J.~M.;
  Zhu,~T.; Berkelbach,~T.~C.; Sharma,~S.; Sokolov,~A.~Y.; Chan,~G. K.-L.
  {Recent Developments in the PySCF Program Package}. \emph{J. Chem. Phys.}
  \textbf{2020}, \emph{153}, 024109\relax
\mciteBstWouldAddEndPuncttrue
\mciteSetBstMidEndSepPunct{\mcitedefaultmidpunct}
{\mcitedefaultendpunct}{\mcitedefaultseppunct}\relax
\EndOfBibitem
\bibitem[Chai and Head-Gordon(2008)Chai, and
  Head-Gordon]{chai_head_gordon_wb97x_d_functional_pccp_2008}
Chai,~J.-D.; Head-Gordon,~M. {Long-Range Corrected Hybrid Density Functionals
  with Damped Atom-Atom Dispersion Corrections}. \emph{Phys. Chem. Chem. Phys.}
  \textbf{2008}, \emph{10}, 6615\relax
\mciteBstWouldAddEndPuncttrue
\mciteSetBstMidEndSepPunct{\mcitedefaultmidpunct}
{\mcitedefaultendpunct}{\mcitedefaultseppunct}\relax
\EndOfBibitem
\bibitem[Zhao and Truhlar(2008)Zhao, and
  Truhlar]{zhao_truhlar_m06_functional_tca_2008}
Zhao,~Y.; Truhlar,~D.~G. {The M06 Suite of Density Functionals for Main Group
  Thermochemistry, Thermochemical Kinetics, Noncovalent Interactions, Excited
  States, and Transition Elements: Two New Functionals and Systematic Testing
  of Four M06-Class Functionals and 12 Other Functionals}. \emph{Theor. Chem.
  Account} \textbf{2008}, \emph{120}, 215\relax
\mciteBstWouldAddEndPuncttrue
\mciteSetBstMidEndSepPunct{\mcitedefaultmidpunct}
{\mcitedefaultendpunct}{\mcitedefaultseppunct}\relax
\EndOfBibitem
\bibitem[Jensen(2001)]{jensen_pc_basis_sets_jcp_2001}
Jensen,~F. {Polarization Consistent Basis Sets: Principles}. \emph{{J}. {C}hem.
  {P}hys.} \textbf{2001}, \emph{115}, 9113\relax
\mciteBstWouldAddEndPuncttrue
\mciteSetBstMidEndSepPunct{\mcitedefaultmidpunct}
{\mcitedefaultendpunct}{\mcitedefaultseppunct}\relax
\EndOfBibitem
\bibitem[Not()]{Note-1}
The reason for this is that the centre of the electronic density
  associated with a given atom (of which the partial charge is the essential
  measured quantity) will not necessarily coincide with the corresponding
  nuclear coordinates, e.g., when constituent atoms carry lone pairs.\relax
\mciteBstWouldAddEndPunctfalse
\mciteSetBstMidEndSepPunct{\mcitedefaultmidpunct}
{}{\mcitedefaultseppunct}\relax
\EndOfBibitem
\bibitem[Gilbert \latin{et~al.}(2008)Gilbert, Besley, and
  Gill]{delta_scf_mom_gill_jpca_2008}
Gilbert,~A. T.~B.; Besley,~N.~A.; Gill,~P. M.~W. {Self-Consistent Field
  Calculations of Excited States Using the Maximum Overlap Method (MOM)}.
  \emph{{J}. {P}hys. {C}hem. A} \textbf{2008}, \emph{112}, 13164\relax
\mciteBstWouldAddEndPuncttrue
\mciteSetBstMidEndSepPunct{\mcitedefaultmidpunct}
{\mcitedefaultendpunct}{\mcitedefaultseppunct}\relax
\EndOfBibitem
\bibitem[Urban and Sadlej(1990)Urban, and
  Sadlej]{urban_sadlej_h2o_dipole_tca_1990}
Urban,~M.; Sadlej,~A.~J. {Molecular Electric Properties in Electronic Excited
  States: Multipole Moments and Polarizabilities of H$_2$O in the Lowest
  ${^{1}}B_1$ and ${^{3}}B_1$ Excited States}. \emph{Theor. Chim. Acta.}
  \textbf{1990}, \emph{78}, 189\relax
\mciteBstWouldAddEndPuncttrue
\mciteSetBstMidEndSepPunct{\mcitedefaultmidpunct}
{\mcitedefaultendpunct}{\mcitedefaultseppunct}\relax
\EndOfBibitem
\bibitem[P{\'a}len{\i}kov{\'a} \latin{et~al.}(2008)P{\'a}len{\i}kov{\'a},
  Kraus, Neogr{\'a}dy, Kell{\"o}, and Urban]{urban_h2o_dipole_mol_phys_2008}
P{\'a}len{\i}kov{\'a},~J.; Kraus,~M.; Neogr{\'a}dy,~P.; Kell{\"o},~V.;
  Urban,~M. {Theoretical Study of Molecular Properties of Low-Lying Electronic
  Excited States of H$_2$O and H$_2$S}. \emph{Mol. Phys.} \textbf{2008},
  \emph{106}, 2333\relax
\mciteBstWouldAddEndPuncttrue
\mciteSetBstMidEndSepPunct{\mcitedefaultmidpunct}
{\mcitedefaultendpunct}{\mcitedefaultseppunct}\relax
\EndOfBibitem
\bibitem[Sharma \latin{et~al.}(2016)Sharma, Bernales, Knecht, Truhlar, and
  Gagliardi]{gagliardi_polyacetylenes_chem_sci_2019}
Sharma,~P.; Bernales,~V.; Knecht,~S.; Truhlar,~D.~G.; Gagliardi,~L. {Density
  Matrix Renormalization Group Pair-Density Functional Theory (DMRG-PDFT):
  Singlet-Triplet Gaps in Polyacenes and Polyacetylenes}. \emph{{C}hem. {S}ci.}
  \textbf{2016}, \emph{10}, 1716\relax
\mciteBstWouldAddEndPuncttrue
\mciteSetBstMidEndSepPunct{\mcitedefaultmidpunct}
{\mcitedefaultendpunct}{\mcitedefaultseppunct}\relax
\EndOfBibitem
\bibitem[Perdew \latin{et~al.}(1996)Perdew, Burke, and
  Ernzerhof]{perdew_burke_ernzerhof_pbe_functional_prl_1996}
Perdew,~J.~P.; Burke,~K.; Ernzerhof,~M. {Generalized Gradient Approximation
  Made Simple}. \emph{Phys. Rev. Lett.} \textbf{1996}, \emph{77}, 3865\relax
\mciteBstWouldAddEndPuncttrue
\mciteSetBstMidEndSepPunct{\mcitedefaultmidpunct}
{\mcitedefaultendpunct}{\mcitedefaultseppunct}\relax
\EndOfBibitem
\bibitem[Hehre \latin{et~al.}(1972)Hehre, Ditchfield, and Pople]{pople_1}
Hehre,~W.~J.; Ditchfield,~R.; Pople,~J.~A. {Self-Consistent Molecular Orbital
  Methods. XII. Further Extensions of Gaussian-Type Basis Sets for Use in
  Molecular Orbital Studies of Organic Molecules}. \emph{{J}. {C}hem. {P}hys.}
  \textbf{1972}, \emph{56}, 2257\relax
\mciteBstWouldAddEndPuncttrue
\mciteSetBstMidEndSepPunct{\mcitedefaultmidpunct}
{\mcitedefaultendpunct}{\mcitedefaultseppunct}\relax
\EndOfBibitem
\bibitem[Woerly \latin{et~al.}(2014)Woerly, Roy, and
  Burke]{burke_polyacetylenes_nat_chem_2014}
Woerly,~E.~M.; Roy,~J.; Burke,~M.~D. {Synthesis of Most Polyene Natural Product
  Motifs Using Just 12 Building Blocks and One Coupling Reaction}. \emph{Nat.
  Chem.} \textbf{2014}, \emph{6}, 484\relax
\mciteBstWouldAddEndPuncttrue
\mciteSetBstMidEndSepPunct{\mcitedefaultmidpunct}
{\mcitedefaultendpunct}{\mcitedefaultseppunct}\relax
\EndOfBibitem
\bibitem[Ghosh \latin{et~al.}(2008)Ghosh, Hachmann, Yanai, and
  Chan]{chan_polyacetylenes_jcp_2008}
Ghosh,~D.; Hachmann,~J.; Yanai,~T.; Chan,~G. K.-L. {Orbital Optimization in the
  Density Matrix Renormalization Group, with Applications to Polyenes and
  $\beta$-Carotene}. \emph{J. Chem. Phys.} \textbf{2008}, \emph{128},
  144117\relax
\mciteBstWouldAddEndPuncttrue
\mciteSetBstMidEndSepPunct{\mcitedefaultmidpunct}
{\mcitedefaultendpunct}{\mcitedefaultseppunct}\relax
\EndOfBibitem
\bibitem[Foster and Boys(1960)Foster, and Boys]{foster_boys_rev_mod_phys_1960}
Foster,~J.~M.; Boys,~S.~F. {Canonical Configurational Interaction Procedure}.
  \emph{{R}ev. {M}od. {P}hys.} \textbf{1960}, \emph{32}, 300\relax
\mciteBstWouldAddEndPuncttrue
\mciteSetBstMidEndSepPunct{\mcitedefaultmidpunct}
{\mcitedefaultendpunct}{\mcitedefaultseppunct}\relax
\EndOfBibitem
\bibitem[Pipek and Mezey(1989)Pipek, and Mezey]{pipek_mezey_jcp_1989}
Pipek,~J.; Mezey,~P.~G. {A Fast Intrinsic Localization Procedure Applicable for
  {\it{Ab Initio}} and Semiempirical Linear Combination of Atomic Orbital Wave
  Functions}. \emph{{J}. {C}hem. {P}hys.} \textbf{1989}, \emph{90}, 4916\relax
\mciteBstWouldAddEndPuncttrue
\mciteSetBstMidEndSepPunct{\mcitedefaultmidpunct}
{\mcitedefaultendpunct}{\mcitedefaultseppunct}\relax
\EndOfBibitem
\bibitem[Not()]{Note-2}
In the iterative optimization of IBOs used in the present work, a PM
  localization power ($p=2$) has been used throughout, cf. Appendix D of Ref.
  \citenum{knizia_iao_ibo_jctc_2013}\relax
\mciteBstWouldAddEndPuncttrue
\mciteSetBstMidEndSepPunct{\mcitedefaultmidpunct}
{\mcitedefaultendpunct}{\mcitedefaultseppunct}\relax
\EndOfBibitem
\bibitem[Garner \latin{et~al.}(2018)Garner, Hoffmann, Rettrup, and
  Solomon]{hoffmann_solomon_helical_orbs_acs_cent_sci_2018}
Garner,~M.~H.; Hoffmann,~R.; Rettrup,~S.; Solomon,~G.~C. {Coarctate and
  M{\"o}bius: The Helical Orbitals of Allene and Other Cumulenes}. \emph{ACS
  Cent. Sci.} \textbf{2018}, \emph{4}, 688\relax
\mciteBstWouldAddEndPuncttrue
\mciteSetBstMidEndSepPunct{\mcitedefaultmidpunct}
{\mcitedefaultendpunct}{\mcitedefaultseppunct}\relax
\EndOfBibitem
\bibitem[Garner \latin{et~al.}(2019)Garner, Jensen, Hyllested, and
  Solomon]{solomon_helical_orbs_chem_sci_2019}
Garner,~M.~H.; Jensen,~A.; Hyllested,~L. O.~H.; Solomon,~G.~C. {Helical
  Orbitals and Circular Currents in Linear Carbon Wires}. \emph{Chem. Sci.}
  \textbf{2019}, \emph{10}, 4598\relax
\mciteBstWouldAddEndPuncttrue
\mciteSetBstMidEndSepPunct{\mcitedefaultmidpunct}
{\mcitedefaultendpunct}{\mcitedefaultseppunct}\relax
\EndOfBibitem
\bibitem[Garner \latin{et~al.}(2020)Garner, Bro-J{\o}rgensen, and
  Solomon]{solomon_elec_trans_jpcc_2020}
Garner,~M.~H.; Bro-J{\o}rgensen,~W.; Solomon,~G.~C. {Three Distinct Torsion
  Profiles of Electronic Transmission through Linear Carbon Wires}. \emph{J.
  Phys. Chem. C} \textbf{2020}, \emph{124}, 18968\relax
\mciteBstWouldAddEndPuncttrue
\mciteSetBstMidEndSepPunct{\mcitedefaultmidpunct}
{\mcitedefaultendpunct}{\mcitedefaultseppunct}\relax
\EndOfBibitem
\bibitem[Becke(1993)]{becke_b3lyp_functional_jcp_1993}
Becke,~A.~D. {Density-Functional Thermochemistry. III. The Role of Exact
  Exchange}. \emph{J. Chem. Phys.} \textbf{1993}, \emph{98}, 5648\relax
\mciteBstWouldAddEndPuncttrue
\mciteSetBstMidEndSepPunct{\mcitedefaultmidpunct}
{\mcitedefaultendpunct}{\mcitedefaultseppunct}\relax
\EndOfBibitem
\bibitem[Stephens \latin{et~al.}(1994)Stephens, Devlin, Chabalowski, and
  Frisch]{frisch_b3lyp_functional_jpc_1994}
Stephens,~P.~J.; Devlin,~F.~J.; Chabalowski,~C.~F.; Frisch,~M.~J. {{\it{Ab
  Initio}} Calculation of Vibrational Absorption and Circular Dichroism Spectra
  Using Density Functional Force Fields}. \emph{J. Phys. Chem.} \textbf{1994},
  \emph{98}, 11623\relax
\mciteBstWouldAddEndPuncttrue
\mciteSetBstMidEndSepPunct{\mcitedefaultmidpunct}
{\mcitedefaultendpunct}{\mcitedefaultseppunct}\relax
\EndOfBibitem
\bibitem[Adamo and Barone(1999)Adamo, and
  Barone]{adamo_barone_pbe0_functional_jcp_1999}
Adamo,~C.; Barone,~V. {Toward Reliable Density Functional Methods without
  Adjustable Parameters: The PBE0 Model}. \emph{J. Chem. Phys.} \textbf{1999},
  \emph{110}, 6158\relax
\mciteBstWouldAddEndPuncttrue
\mciteSetBstMidEndSepPunct{\mcitedefaultmidpunct}
{\mcitedefaultendpunct}{\mcitedefaultseppunct}\relax
\EndOfBibitem
\bibitem[sea()]{seaborn}
{{\texttt{seaborn}}, Statistical Data Visualization,
  See~{\url{https://seaborn.pydata.org/index.html}}}\relax
\mciteBstWouldAddEndPuncttrue
\mciteSetBstMidEndSepPunct{\mcitedefaultmidpunct}
{\mcitedefaultendpunct}{\mcitedefaultseppunct}\relax
\EndOfBibitem
\bibitem[Cheng \latin{et~al.}(2019)Cheng, Welborn, Christensen, and
  Miller~III]{miller_ml_qc_data_2019}
Cheng,~L.; Welborn,~M.; Christensen,~A.~S.; Miller~III,~T.~F. {Thermalized
  (350K) QM7b, GDB-13, Water, and Short Alkane Quantum Chemistry Dataset
  Including MOB-ML Features}. 2019;
  \url{https://data.caltech.edu/records/1177}, DOI: 10.22002/D1.1177\relax
\mciteBstWouldAddEndPuncttrue
\mciteSetBstMidEndSepPunct{\mcitedefaultmidpunct}
{\mcitedefaultendpunct}{\mcitedefaultseppunct}\relax
\EndOfBibitem
\bibitem[Christensen and von Lilienfeld(2019)Christensen, and von
  Lilienfeld]{lilienfeld_ml_qc_data_2019}
Christensen,~A.~S.; von Lilienfeld,~O.~A. {The Water40 10K Dataset}. 2019;
  \url{https://figshare.com/articles/dataset/Water40_10K/8058497}, DOI:
  10.6084/m9.figshare.8058497\relax
\mciteBstWouldAddEndPuncttrue
\mciteSetBstMidEndSepPunct{\mcitedefaultmidpunct}
{\mcitedefaultendpunct}{\mcitedefaultseppunct}\relax
\EndOfBibitem
\bibitem[Gerrard \latin{et~al.}(2020)Gerrard, Bratholm, Packer, Mulholland,
  Glowacki, and Butts]{glowacki_butts_impression_chem_sci_2020}
Gerrard,~W.; Bratholm,~L.~A.; Packer,~M.~J.; Mulholland,~A.~J.; Glowacki,~D.~R.;
  Butts,~C.~P. {IMPRESSION -- Prediction of NMR Parameters for 3-Dimensional
  Chemical Structures Using Machine Learning with Near Quantum Chemical
  Accuracy}. \emph{Chem. Sci.} \textbf{2020}, \emph{11}, 508\relax
\mciteBstWouldAddEndPuncttrue
\mciteSetBstMidEndSepPunct{\mcitedefaultmidpunct}
{\mcitedefaultendpunct}{\mcitedefaultseppunct}\relax
\EndOfBibitem
\bibitem[Gupta \latin{et~al.}(2020)Gupta, Chakraborty, and
  Ramakrishnan]{ramakrishnan_ml_nmr_arxiv_2020}
Gupta,~A.; Chakraborty,~S.; Ramakrishnan,~R. {Revving Up 13C NMR Shielding
  Predictions Across Chemical Space: Benchmarks for Atoms-in-Molecules Kernel
  Machine Learning with New Data for 134 Kilo Molecules}.
  \emph{arXiv:2009.06814} \textbf{2020}\relax
\mciteBstWouldAddEndPunctfalse
\mciteSetBstMidEndSepPunct{\mcitedefaultmidpunct}
{\mcitedefaultendpunct}{\mcitedefaultseppunct}\relax
\EndOfBibitem
\bibitem[qml()]{qml_prog}
Christensen, A. S.; Faber, F. A.; Huang, B.; Bratholm, L. A.; Tkatchenko, A.;
  M{\"u}ller, K.-R.; von Lilienfeld, O. A. {\texttt{QML}}: A Python Toolkit for
  Quantum Machine Learning, See~{\url{https://github.com/qmlcode/qml}}\relax
\mciteBstWouldAddEndPuncttrue
\mciteSetBstMidEndSepPunct{\mcitedefaultmidpunct}
{\mcitedefaultendpunct}{\mcitedefaultseppunct}\relax
\EndOfBibitem
\bibitem[Not()]{Note-3}
The choice of IAO over Mulliken partial charges are justified in Figure S6 of
  the SI, in which the former are shown to vary continuously along the
  symmetric stretch in H$_2$O while also giving rise to smoothly varying
  atom-RDM1s\relax
\mciteBstWouldAddEndPuncttrue
\mciteSetBstMidEndSepPunct{\mcitedefaultmidpunct}
{\mcitedefaultendpunct}{\mcitedefaultseppunct}\relax
\EndOfBibitem
\bibitem[Olsen \latin{et~al.}(1996)Olsen, J{\o}rgensen, Koch, Balkova, and
  Bartlett]{olsen_bond_break_h2o_jcp_1996}
Olsen,~J.; J{\o}rgensen,~P.; Koch,~H.; Balkova,~A.; Bartlett,~R.~J. {Full
  Configuration-Interaction and State of the Art Correlation Calculations on
  Water in a Valence Double-Zeta Basis with Polarization Functions}. \emph{{J}.
  {C}hem. {P}hys.} \textbf{1996}, \emph{104}, 8007\relax
\mciteBstWouldAddEndPuncttrue
\mciteSetBstMidEndSepPunct{\mcitedefaultmidpunct}
{\mcitedefaultendpunct}{\mcitedefaultseppunct}\relax
\EndOfBibitem
\bibitem[Chan and Head-Gordon(2003)Chan, and
  Head-Gordon]{chan_bond_break_h2o_jcp_2003}
Chan,~G. K.-L.; Head-Gordon,~M. {Exact Solution (Within a Triple-Zeta, Double
  Polarization Basis Set) of the Electronic Schr{\"o}dinger Equation for
  Water}. \emph{{J}. {C}hem. {P}hys.} \textbf{2003}, \emph{118}, 8551\relax
\mciteBstWouldAddEndPuncttrue
\mciteSetBstMidEndSepPunct{\mcitedefaultmidpunct}
{\mcitedefaultendpunct}{\mcitedefaultseppunct}\relax
\EndOfBibitem
\bibitem[Yao \latin{et~al.}(2018)Yao, Herr, Toth, Mckintyre, and
  Parkhill]{parkhill_tensormol_chem_sci_2018}
Yao,~K.; Herr,~J.~E.; Toth,~D.~W.; Mckintyre,~R.; Parkhill,~J. {The
  {\texttt{TensorMol-0.1}} Model Chemistry: A Neural Network Augmented with
  Long-Range Physics}. \emph{Chem. Sci.} \textbf{2018}, \emph{9}, 2261\relax
\mciteBstWouldAddEndPuncttrue
\mciteSetBstMidEndSepPunct{\mcitedefaultmidpunct}
{\mcitedefaultendpunct}{\mcitedefaultseppunct}\relax
\EndOfBibitem
\bibitem[Gao \latin{et~al.}(2020)Gao, Ramezanghorbani, Isayev, Smith, and
  Roitberg]{isayev_roitberg_torchani_jcim_2020}
Gao,~X.; Ramezanghorbani,~F.; Isayev,~O.; Smith,~J.~S.; Roitberg,~A.~E.
  {{\texttt{TorchANI}}: A Free and Open Source PyTorch-Based Deep Learning
  Implementation of the ANI Neural Network Potentials}. \emph{J. Chem. Inf.
  Model.} \textbf{2020}, \emph{60}, 3408\relax
\mciteBstWouldAddEndPuncttrue
\mciteSetBstMidEndSepPunct{\mcitedefaultmidpunct}
{\mcitedefaultendpunct}{\mcitedefaultseppunct}\relax
\EndOfBibitem
\bibitem[Gastegger \latin{et~al.}(2017)Gastegger, Behler, and
  Marquetand]{marquetand_ml_infrared_chem_sci_2017}
Gastegger,~M.; Behler,~J.; Marquetand,~P. {Machine Learning Molecular Dynamics
  for the Simulation of Infrared Spectra}. \emph{Chem. Sci.} \textbf{2017},
  \emph{8}, 6924\relax
\mciteBstWouldAddEndPuncttrue
\mciteSetBstMidEndSepPunct{\mcitedefaultmidpunct}
{\mcitedefaultendpunct}{\mcitedefaultseppunct}\relax
\EndOfBibitem
\bibitem[Pereira and Aires-de Sousa(2018)Pereira, and Aires-de
  Sousa]{pereira_aires_de_sousa_ml_dipole_jc_2018}
Pereira,~F.; Aires-de Sousa,~J. {Machine Learning for the Prediction of
  Molecular Dipole Moments Obtained by Density Functional Theory}. \emph{J.
  Cheminformatics} \textbf{2018}, \emph{10}, 43\relax
\mciteBstWouldAddEndPuncttrue
\mciteSetBstMidEndSepPunct{\mcitedefaultmidpunct}
{\mcitedefaultendpunct}{\mcitedefaultseppunct}\relax
\EndOfBibitem
\bibitem[Christensen \latin{et~al.}(2019)Christensen, Faber, and von
  Lilienfeld]{lilienfeld_ml_qc_jcp_2019}
Christensen,~A.~S.; Faber,~F.~A.; von Lilienfeld,~O.~A. {Operators in Quantum
  Machine Learning: Response Properties in Chemical Space}. \emph{J. Chem.
  Phys.} \textbf{2019}, \emph{150}, 064105\relax
\mciteBstWouldAddEndPuncttrue
\mciteSetBstMidEndSepPunct{\mcitedefaultmidpunct}
{\mcitedefaultendpunct}{\mcitedefaultseppunct}\relax
\EndOfBibitem
\bibitem[Grisafi \latin{et~al.}(2018)Grisafi, Wilkins, Cs{\'a}nyi, and
  Ceriotti]{ceriotti_ml_resp_prop_prl_2018}
Grisafi,~A.; Wilkins,~D.~M.; Cs{\'a}nyi,~G.; Ceriotti,~M. {Symmetry-Adapted
  Machine Learning for Tensorial Properties of Atomistic Systems}. \emph{Phys.
  Rev. Lett.} \textbf{2018}, \emph{120}, 036002\relax
\mciteBstWouldAddEndPuncttrue
\mciteSetBstMidEndSepPunct{\mcitedefaultmidpunct}
{\mcitedefaultendpunct}{\mcitedefaultseppunct}\relax
\EndOfBibitem
\bibitem[Wilkins \latin{et~al.}(2019)Wilkins, Grisafi, Yang, Lao, DiStasio~Jr.,
  and Ceriotti]{ceriotti_ml_polarizability_pnas_2019}
Wilkins,~D.~M.; Grisafi,~A.; Yang,~Y.; Lao,~K.~U.; DiStasio~Jr.,~R.~A.;
  Ceriotti,~M. {Accurate Molecular Polarizabilities with Coupled Cluster Theory
  and Machine Learning}. \emph{Proc. Nat. Acad. Sci.} \textbf{2019},
  \emph{116}, 3401\relax
\mciteBstWouldAddEndPuncttrue
\mciteSetBstMidEndSepPunct{\mcitedefaultmidpunct}
{\mcitedefaultendpunct}{\mcitedefaultseppunct}\relax
\EndOfBibitem
\bibitem[Veit \latin{et~al.}(2020)Veit, Wilkins, Yang, DiStasio~Jr., and
  Ceriotti]{ceriotti_ml_dipole_jcp_2020}
Veit,~M.; Wilkins,~D.~M.; Yang,~Y.; DiStasio~Jr.,~R.~A.; Ceriotti,~M.
  {Predicting Molecular Dipole Moments by Combining Atomic Partial Charges and
  Atomic Dipoles}. \emph{J. Chem. Phys.} \textbf{2020}, \emph{153},
  024113\relax
\mciteBstWouldAddEndPuncttrue
\mciteSetBstMidEndSepPunct{\mcitedefaultmidpunct}
{\mcitedefaultendpunct}{\mcitedefaultseppunct}\relax
\EndOfBibitem
\bibitem[Parsaeifard \latin{et~al.}(2020)Parsaeifard, De, Christensen, Faber,
  Kocer, De, Behler, von Lilienfeld, and
  Goedecker]{behler_lilienfeld_goedecker_ml_representation_mlst_2020}
Parsaeifard,~B.; De,~D.~S.; Christensen,~A.~S.; Faber,~F.~A.; Kocer,~E.;
  De,~S.; Behler,~J.; von Lilienfeld,~A.; Goedecker,~S. {An Assessment of
  the Structural Resolution of Various Fingerprints Commonly Used in Machine
  Learning}. \emph{Mach. Learn.: Sci. Technol.} \textbf{2020}, DOI:
  10.1088/2632-2153/abb212\relax
\mciteBstWouldAddEndPuncttrue
\mciteSetBstMidEndSepPunct{\mcitedefaultmidpunct}
{\mcitedefaultendpunct}{\mcitedefaultseppunct}\relax
\EndOfBibitem
\bibitem[Glielmo \latin{et~al.}(2018)Glielmo, Zeni, and
  De~Vita]{de_vita_ml_represent_prb_2018}
Glielmo,~A.; Zeni,~C.; De~Vita,~A. {Efficient Nonparametric $n$-Body Force
  Fields from Machine Learning}. \emph{Phys. Rev. B} \textbf{2018}, \emph{97},
  184307\relax
\mciteBstWouldAddEndPuncttrue
\mciteSetBstMidEndSepPunct{\mcitedefaultmidpunct}
{\mcitedefaultendpunct}{\mcitedefaultseppunct}\relax
\EndOfBibitem
\bibitem[Jinnouchi \latin{et~al.}(2020)Jinnouchi, Karsai, Verdi, Asahi, and
  Kresse]{kresse_ml_represent_jcp_2020}
Jinnouchi,~R.; Karsai,~F.; Verdi,~C.; Asahi,~R.; Kresse,~G. {Descriptors
  Representing Two- and Three-Body Atomic Distributions and Their Effects on
  the Accuracy of Machine-Learned Inter-Atomic Potentials}. \emph{J. Chem.
  Phys.} \textbf{2020}, \emph{152}, 234102\relax
\mciteBstWouldAddEndPuncttrue
\mciteSetBstMidEndSepPunct{\mcitedefaultmidpunct}
{\mcitedefaultendpunct}{\mcitedefaultseppunct}\relax
\EndOfBibitem
\bibitem[Pozdnyakov \latin{et~al.}(2020)Pozdnyakov, Willatt, Bart{\'o}k,
  Ortner, Cs{\'a}nyi, and Ceriotti]{ceriotti_ml_representation_prl_2020}
Pozdnyakov,~S.~N.; Willatt,~M.~J.; Bart{\'o}k,~A.~P.; Ortner,~C.;
  Cs{\'a}nyi,~G.; Ceriotti,~M. {Incompleteness of Atomic Structure
  Representations}. \emph{Phys. Rev. Lett.} \textbf{2020}, \emph{125},
  166001\relax
\mciteBstWouldAddEndPuncttrue
\mciteSetBstMidEndSepPunct{\mcitedefaultmidpunct}
{\mcitedefaultendpunct}{\mcitedefaultseppunct}\relax
\EndOfBibitem
\bibitem[von Lilienfeld \latin{et~al.}(2015)von Lilienfeld, Ramakrishnan, Rupp,
  and Knoll]{lilienfeld_ml_representation_ijqc_2015}
von Lilienfeld,~O.~A.; Ramakrishnan,~R.; Rupp,~M.; Knoll,~A. {Fourier Series of
  Atomic Radial Distribution Functions: A Molecular Fingerprint for Machine
  Learning Models of Quantum Chemical Properties}. \emph{Int. J. Quantum Chem.}
  \textbf{2015}, \emph{115}, 1084\relax
\mciteBstWouldAddEndPuncttrue
\mciteSetBstMidEndSepPunct{\mcitedefaultmidpunct}
{\mcitedefaultendpunct}{\mcitedefaultseppunct}\relax
\EndOfBibitem
\bibitem[Goscinski \latin{et~al.}(2020)Goscinski, Fraux, and
  Ceriotti]{ceriotti_ml_representation_arxiv_2020}
Goscinski,~A.; Fraux,~G.; Ceriotti,~M. {The Role of Feature Space in Atomistic
  Learning}. \emph{arXiv:2009.02741} \textbf{2020}\relax
\mciteBstWouldAddEndPunctfalse
\mciteSetBstMidEndSepPunct{\mcitedefaultmidpunct}
{\mcitedefaultendpunct}{\mcitedefaultseppunct}\relax
\EndOfBibitem
\bibitem[Carleo and Troyer(2017)Carleo, and
  Troyer]{carleo_troyer_ml_qc_science_2017}
Carleo,~G.; Troyer,~M. {Solving the Quantum Many-Body Problem with Artificial
  Neural Networks}. \emph{Science} \textbf{2017}, \emph{355}, 602\relax
\mciteBstWouldAddEndPuncttrue
\mciteSetBstMidEndSepPunct{\mcitedefaultmidpunct}
{\mcitedefaultendpunct}{\mcitedefaultseppunct}\relax
\EndOfBibitem
\bibitem[Carleo \latin{et~al.}(2018)Carleo, Nomura, and
  Imada]{carleo_nomura_imada_ml_qc_nat_commun_2018}
Carleo,~G.; Nomura,~Y.; Imada,~M. {Constructing Exact Representations of
  Quantum Many-Body Systems with Deep Neural Networks}. \emph{Nat. Commun.}
  \textbf{2018}, \emph{9}, 5322\relax
\mciteBstWouldAddEndPuncttrue
\mciteSetBstMidEndSepPunct{\mcitedefaultmidpunct}
{\mcitedefaultendpunct}{\mcitedefaultseppunct}\relax
\EndOfBibitem
\bibitem[Sch{\"u}tt \latin{et~al.}(2017)Sch{\"u}tt, Arbabzadah, Chmiela,
  M{\"u}ller, and Tkatchenko]{mueller_tkatchenko_ml_qc_nature_comm_2017}
Sch{\"u}tt,~K.~T.; Arbabzadah,~F.; Chmiela,~S.; M{\"u}ller,~K.-R.;
  Tkatchenko,~A. {Quantum-Chemical Insights from Deep Tensor Neural Networks}.
  \emph{Nat. Commun.} \textbf{2017}, \emph{8}, 13890\relax
\mciteBstWouldAddEndPuncttrue
\mciteSetBstMidEndSepPunct{\mcitedefaultmidpunct}
{\mcitedefaultendpunct}{\mcitedefaultseppunct}\relax
\EndOfBibitem
\bibitem[Sch{\"u}tt \latin{et~al.}(2019)Sch{\"u}tt, Gastegger, Tkatchenko,
  M{\"u}ller, and Maurer]{tkatchenko_mueller_maurer_ml_qc_nat_commun_2019}
Sch{\"u}tt,~K.~T.; Gastegger,~M.; Tkatchenko,~A.; M{\"u}ller,~K.-R.;
  Maurer,~R.~J. {Unifying Machine Learning and Quantum Chemistry With a Deep
  Neural Network for Molecular Wavefunctions}. \emph{Nat. Commun.}
  \textbf{2019}, \emph{10}, 5024\relax
\mciteBstWouldAddEndPuncttrue
\mciteSetBstMidEndSepPunct{\mcitedefaultmidpunct}
{\mcitedefaultendpunct}{\mcitedefaultseppunct}\relax
\EndOfBibitem
\bibitem[Gastegger \latin{et~al.}(2020)Gastegger, McSloy, Luya, Sch{\"u}tt, and
  Maurer]{gastegger_maurer_ml_qc_jcp_2020}
Gastegger,~M.; McSloy,~A.; Luya,~M.; Sch{\"u}tt,~K.~T.; Maurer,~R.~J. {A Deep
  Neural Network for Molecular Wave Functions in Quasi-Atomic Minimal Basis
  Representation}. \emph{J. Chem. Phys.} \textbf{2020}, \emph{153},
  044123\relax
\mciteBstWouldAddEndPuncttrue
\mciteSetBstMidEndSepPunct{\mcitedefaultmidpunct}
{\mcitedefaultendpunct}{\mcitedefaultseppunct}\relax
\EndOfBibitem
\bibitem[Lubbers \latin{et~al.}(2018)Lubbers, Smith, and
  Barros]{lubbers_smith_barros_ml_hip_nn_jcp_2018}
Lubbers,~N.; Smith,~J.~S.; Barros,~K. {Hierarchical Modeling of Molecular
  Energies Using a Deep Neural Network}. \emph{J. Chem. Phys.} \textbf{2018},
  \emph{148}, 241715\relax
\mciteBstWouldAddEndPuncttrue
\mciteSetBstMidEndSepPunct{\mcitedefaultmidpunct}
{\mcitedefaultendpunct}{\mcitedefaultseppunct}\relax
\EndOfBibitem
\bibitem[Sch{\"u}tt \latin{et~al.}(2017)Sch{\"u}tt, Kindermans, Sauceda,
  Chmiela, Tkatchenko, and
  M{\"u}ller]{tkatchenko_mueller_ml_qc_adv_neural_2017}
Sch{\"u}tt,~K.~T.; Kindermans,~P.-J.; Sauceda,~H.~E.; Chmiela,~S.;
  Tkatchenko,~A.; M{\"u}ller,~K.-R. {{\texttt{SchNet}}: A Continuous-Filter
  Convolutional Neural Network for Modeling Quantum Interactions}. \emph{Adv.
  Neural Inf. Process. Syst.} \textbf{2017}, \emph{30}, 1\relax
\mciteBstWouldAddEndPuncttrue
\mciteSetBstMidEndSepPunct{\mcitedefaultmidpunct}
{\mcitedefaultendpunct}{\mcitedefaultseppunct}\relax
\EndOfBibitem
\bibitem[Sch{\"u}tt \latin{et~al.}(2018)Sch{\"u}tt, Sauceda, Kindermans,
  Tkatchenko, and M{\"u}ller]{tkatchenko_mueller_ml_qc_jcp_2018}
Sch{\"u}tt,~K.~T.; Sauceda,~H.~E.; Kindermans,~P.-J.; Tkatchenko,~A.;
  M{\"u}ller,~K.-R. {{\texttt{SchNet}} -- A Deep Learning Architecture for
  Molecules and Materials}. \emph{J. Chem. Phys.} \textbf{2018}, \emph{148},
  241722\relax
\mciteBstWouldAddEndPuncttrue
\mciteSetBstMidEndSepPunct{\mcitedefaultmidpunct}
{\mcitedefaultendpunct}{\mcitedefaultseppunct}\relax
\EndOfBibitem
\bibitem[Unke and Meuwly(2019)Unke, and
  Meuwly]{unke_meuwly_ml_physnet_jctc_2019}
Unke,~O.~T.; Meuwly,~M. {{\texttt{PhysNet}}: A Neural Network for Predicting
  Energies, Forces, Dipole Moments, and Partial Charges}. \emph{J. Chem. Theory
  Comput.} \textbf{2019}, \emph{15}, 3678\relax
\mciteBstWouldAddEndPuncttrue
\mciteSetBstMidEndSepPunct{\mcitedefaultmidpunct}
{\mcitedefaultendpunct}{\mcitedefaultseppunct}\relax
\EndOfBibitem
\bibitem[Zubatyuk \latin{et~al.}(2019)Zubatyuk, Smith, Leszczynski, and
  Isayev]{isayev_ml_qc_sci_adv_2019}
Zubatyuk,~R.; Smith,~J.~S.; Leszczynski,~J.; Isayev,~O. {Accurate and
  Transferable Multitask Prediction of Chemical Properties with an
  Atoms-in-Molecules Neural Network}. \emph{Sci. Adv.} \textbf{2019}, \emph{5},
  eaav6490\relax
\mciteBstWouldAddEndPuncttrue
\mciteSetBstMidEndSepPunct{\mcitedefaultmidpunct}
{\mcitedefaultendpunct}{\mcitedefaultseppunct}\relax
\EndOfBibitem
\bibitem[Sch{\"u}tt \latin{et~al.}(2019)Sch{\"u}tt, Kessel, Gastegger, Nicoli,
  Tkatchenko, and M{\"u}ller]{tkatchenko_mueller_ml_qc_jctc_2019}
Sch{\"u}tt,~K.~T.; Kessel,~P.; Gastegger,~M.; Nicoli,~K.~A.; Tkatchenko,~A.;
  M{\"u}ller,~K.-R. {{\texttt{SchNetPack}}: A Deep Learning Toolbox For
  Atomistic Systems}. \emph{J. Chem. Theory Comput.} \textbf{2019}, \emph{15},
  448\relax
\mciteBstWouldAddEndPuncttrue
\mciteSetBstMidEndSepPunct{\mcitedefaultmidpunct}
{\mcitedefaultendpunct}{\mcitedefaultseppunct}\relax
\EndOfBibitem
\bibitem[Anderson \latin{et~al.}(2019)Anderson, Hy, and
  Kondor]{anderson_hy_kondor_cormorant_nips_2019}
Anderson,~B.; Hy,~T.~S.; Kondor,~R. In \emph{Adv. Neural Inf. Process. Syst.};
  Wallach,~H., Larochelle,~H., Beygelzimer,~A., {d'Alch\'{e}-Buc},~F., Fox,~E.,
  Garnett,~R., Eds.; 2019; Vol.~32; p 14537\relax
\mciteBstWouldAddEndPuncttrue
\mciteSetBstMidEndSepPunct{\mcitedefaultmidpunct}
{\mcitedefaultendpunct}{\mcitedefaultseppunct}\relax
\EndOfBibitem
\bibitem[Klicpera \latin{et~al.}(2020)Klicpera, Gro{\ss}, and
  G{\"u}nnemann]{klicpera_gross_gruennemann_dimenet_2020}
Klicpera,~J.; Gro{\ss},~J.; G{\"u}nnemann,~S. {Directional Message Passing for
  Molecular Graphs}. \emph{arXiv:2003.03123} \textbf{2020}\relax
\mciteBstWouldAddEndPunctfalse
\mciteSetBstMidEndSepPunct{\mcitedefaultmidpunct}
{\mcitedefaultendpunct}{\mcitedefaultseppunct}\relax
\EndOfBibitem
\bibitem[Miller \latin{et~al.}(2020)Miller, Geiger, Smidt, and
  No{\'e}]{noe_e3nn_arxiv_2020}
Miller,~B.~K.; Geiger,~M.; Smidt,~T.~E.; No{\'e},~F. {Relevance of Rotationally
  Equivariant Convolutions for Predicting Molecular Properties}.
  \emph{arXiv:2008.08461} \textbf{2020}\relax
\mciteBstWouldAddEndPunctfalse
\mciteSetBstMidEndSepPunct{\mcitedefaultmidpunct}
{\mcitedefaultendpunct}{\mcitedefaultseppunct}\relax
\EndOfBibitem
\bibitem[Smidt(2020)]{smidt_equivariant_chemrxiv_2020}
Smidt,~T.~E. {Euclidean Symmetry and Equivariance in Machine Learning}.
  \textbf{2020}, DOI: 10.26434/chemrxiv.12935198.v1\relax
\mciteBstWouldAddEndPuncttrue
\mciteSetBstMidEndSepPunct{\mcitedefaultmidpunct}
{\mcitedefaultendpunct}{\mcitedefaultseppunct}\relax
\EndOfBibitem
\bibitem[Welborn \latin{et~al.}(2018)Welborn, Cheng, and
  Miller~III]{miller_ml_qc_jctc_2018}
Welborn,~M.; Cheng,~L.; Miller~III,~T.~F. {Transferability in Machine Learning
  for Electronic Structure via the Molecular Orbital Basis}. \emph{J. Chem.
  Theory Comput.} \textbf{2018}, \emph{14}, 4772\relax
\mciteBstWouldAddEndPuncttrue
\mciteSetBstMidEndSepPunct{\mcitedefaultmidpunct}
{\mcitedefaultendpunct}{\mcitedefaultseppunct}\relax
\EndOfBibitem
\bibitem[Cheng \latin{et~al.}(2019)Cheng, Welborn, Christensen, and
  Miller~III]{miller_ml_qc_jcp_2019}
Cheng,~L.; Welborn,~M.; Christensen,~A.~S.; Miller~III,~T.~F. {A Universal
  Density Matrix Functional From Molecular Orbital-Based Machine Learning:
  Transferability Across Organic Molecules}. \emph{J. Chem. Phys.}
  \textbf{2019}, \emph{150}, 131103\relax
\mciteBstWouldAddEndPuncttrue
\mciteSetBstMidEndSepPunct{\mcitedefaultmidpunct}
{\mcitedefaultendpunct}{\mcitedefaultseppunct}\relax
\EndOfBibitem
\bibitem[Cheng \latin{et~al.}(2019)Cheng, Kovachki, Welborn, and
  Miller~III]{miller_ml_qc_jctc_2019}
Cheng,~L.; Kovachki,~N.~B.; Welborn,~M.; Miller~III,~T.~F.
  {Regression-Clustering for Improved Accuracy and Training Cost with
  Molecular-Orbital-Based Machine Learning}. \emph{J. Chem. Theory Comput.}
  \textbf{2019}, \emph{15}, 6668\relax
\mciteBstWouldAddEndPuncttrue
\mciteSetBstMidEndSepPunct{\mcitedefaultmidpunct}
{\mcitedefaultendpunct}{\mcitedefaultseppunct}\relax
\EndOfBibitem
\bibitem[Husch \latin{et~al.}(2020)Husch, Sun, Cheng, Lee, and
  Miller~III]{miller_ml_qc_arxiv_2020}
Husch,~T.; Sun,~J.; Cheng,~L.; Lee,~S. J.~R.; Miller~III,~T.~F. {Improved
  Accuracy and Transferability of Molecular-Orbital-Based Machine Learning:
  Organics, Transition-Metal Complexes, Non-Covalent Interactions, and
  Transition States}. \emph{arXiv:2010.03626} \textbf{2020}\relax
\mciteBstWouldAddEndPunctfalse
\mciteSetBstMidEndSepPunct{\mcitedefaultmidpunct}
{\mcitedefaultendpunct}{\mcitedefaultseppunct}\relax
\EndOfBibitem
\bibitem[Chen \latin{et~al.}(2020)Chen, Zhang, Wang, and
  E]{weinan_e_deephf_jpca_2020}
Chen,~Y.; Zhang,~L.; Wang,~H.; E,~W. {Ground State Energy Functional with
  Hartree-Fock Efficiency and Chemical Accuracy}. \emph{J. Phys. Chem. A}
  \textbf{2020}, \emph{124}, 7155\relax
\mciteBstWouldAddEndPuncttrue
\mciteSetBstMidEndSepPunct{\mcitedefaultmidpunct}
{\mcitedefaultendpunct}{\mcitedefaultseppunct}\relax
\EndOfBibitem
\bibitem[Grimme \latin{et~al.}(2017)Grimme, Bannwarth, and
  Shushkov]{grimme_gfn1_xtb_jctc_2017}
Grimme,~S.; Bannwarth,~C.; Shushkov,~P. {A Robust and Accurate Tight-Binding
  Quantum Chemical Method for Structures, Vibrational Frequencies, and
  Noncovalent Interactions of Large Molecular Systems Parameterized for All
  spd-Block Elements ($Z = 1-86$)}. \emph{J. Chem. Theory Comput.}
  \textbf{2017}, \emph{13}, 1989\relax
\mciteBstWouldAddEndPuncttrue
\mciteSetBstMidEndSepPunct{\mcitedefaultmidpunct}
{\mcitedefaultendpunct}{\mcitedefaultseppunct}\relax
\EndOfBibitem
\bibitem[Bannwarth \latin{et~al.}(2019)Bannwarth, Ehlert, and
  Grimme]{grimme_gfn2_xtb_jctc_2019}
Bannwarth,~C.; Ehlert,~S.; Grimme,~S. {GFN2-xTB---An Accurate and Broadly
  Parametrized Self-Consistent Tight-Binding Quantum Chemical Method with
  Multipole Electrostatics and Density-Dependent Dispersion Contributions}.
  \emph{J. Chem. Theory Comput.} \textbf{2019}, \emph{15}, 1652\relax
\mciteBstWouldAddEndPuncttrue
\mciteSetBstMidEndSepPunct{\mcitedefaultmidpunct}
{\mcitedefaultendpunct}{\mcitedefaultseppunct}\relax
\EndOfBibitem
\bibitem[Bannwarth \latin{et~al.}(2020)Bannwarth, Caldeweyher, Ehlert, Pracht,
  Seibert, Spicher, and Grimme]{grimme_gfn_xtb_review_wires_2020}
Bannwarth,~C.; Caldeweyher,~E.; Ehlert,~S.; Hansen,~A.; Pracht,~P.; Seibert,~J.;
  Spicher,~S.; Grimme,~S. {Extended Tight‐Binding Quantum Chemistry Methods}.
  \emph{{W}IREs {C}omput. {M}ol. {S}ci.} \textbf{2020}, e01493, DOI:
  10.1002/wcms.1493\relax
\mciteBstWouldAddEndPuncttrue
\mciteSetBstMidEndSepPunct{\mcitedefaultmidpunct}
{\mcitedefaultendpunct}{\mcitedefaultseppunct}\relax
\EndOfBibitem
\bibitem[Qiao \latin{et~al.}(2020)Qiao, Welborn, Anandkumar, Manby, and
  Miller~III]{miller_orbnet_jcp_2020}
Qiao,~Z.; Welborn,~M.; Anandkumar,~A.; Manby,~F.~R.; Miller~III,~T.~F.
  {{\texttt{OrbNet}}: Deep Learning for Quantum Chemistry Using
  Symmetry-Adapted Atomic-Orbital Features}. \emph{J. Chem. Phys.}
  \textbf{2020}, \emph{153}, 124111\relax
\mciteBstWouldAddEndPuncttrue
\mciteSetBstMidEndSepPunct{\mcitedefaultmidpunct}
{\mcitedefaultendpunct}{\mcitedefaultseppunct}\relax
\EndOfBibitem
\bibitem[Parasuk \latin{et~al.}(1991)Parasuk, Alml{\"o}f, and
  Feyereisen]{almlof_feyereisen_c18_jacs_1991}
Parasuk,~V.; Alml{\"o}f,~J.; Feyereisen,~M.~W. {The [18] All-Carbon Molecule:
  Cumulene or Polyacetylene?} \emph{J. Am. Chem. Soc.} \textbf{1991},
  \emph{113}, 1049\relax
\mciteBstWouldAddEndPuncttrue
\mciteSetBstMidEndSepPunct{\mcitedefaultmidpunct}
{\mcitedefaultendpunct}{\mcitedefaultseppunct}\relax
\EndOfBibitem
\bibitem[Plattner and Houk(1995)Plattner, and
  Houk]{plattner_houk_c18_jacs_1995}
Plattner,~D.~A.; Houk,~K.~N. {C$_{18}$ is a Polyyne}. \emph{J. Am. Chem. Soc.}
  \textbf{1995}, \emph{117}, 4405\relax
\mciteBstWouldAddEndPuncttrue
\mciteSetBstMidEndSepPunct{\mcitedefaultmidpunct}
{\mcitedefaultendpunct}{\mcitedefaultseppunct}\relax
\EndOfBibitem
\bibitem[Neiss \latin{et~al.}(2014)Neiss, Trushin, and
  G{\"o}rling]{neiss_gorling_c18_chem_phys_chem_2014}
Neiss,~C.; Trushin,~E.; G{\"o}rling,~A. {The Nature of One‐Dimensional
  Carbon: Polyynic versus Cumulenic}. \emph{ChemPhysChem.} \textbf{2014},
  \emph{15}, 2497\relax
\mciteBstWouldAddEndPuncttrue
\mciteSetBstMidEndSepPunct{\mcitedefaultmidpunct}
{\mcitedefaultendpunct}{\mcitedefaultseppunct}\relax
\EndOfBibitem
\bibitem[Kaiser \latin{et~al.}(2019)Kaiser, Scriven, Schulz, Gawel, Gross, and
  Anderson]{kaiser_gross_anderson_c18_science_2019}
Kaiser,~K.; Scriven,~L.~M.; Schulz,~F.; Gawel,~P.; Gross,~L.; Anderson,~H.~L.
  {An sp-Hybridized Molecular Carbon Allotrope, Cyclo[18]carbon}.
  \emph{Science} \textbf{2019}, \emph{365}, 1299\relax
\mciteBstWouldAddEndPuncttrue
\mciteSetBstMidEndSepPunct{\mcitedefaultmidpunct}
{\mcitedefaultendpunct}{\mcitedefaultseppunct}\relax
\EndOfBibitem
\bibitem[Br{\'e}mond \latin{et~al.}(2019)Br{\'e}mond, P{\'e}rez-Jim{\'e}nez,
  Adamo, and Sancho-Garc{\'i}a]{bremond_sancho_garcia_c18_jcp_2019}
Br{\'e}mond,~{\'E}.; P{\'e}rez-Jim{\'e}nez,~{\'A}.~J.; Adamo,~C.;
  Sancho-Garc{\'i}a,~J.~C. {sp-Hybridized Carbon Allotrope Molecular
  Structures: An Ongoing Challenge for Density-Functional Approximations}.
  \emph{J. Chem. Phys.} \textbf{2019}, \emph{151}, 211104\relax
\mciteBstWouldAddEndPuncttrue
\mciteSetBstMidEndSepPunct{\mcitedefaultmidpunct}
{\mcitedefaultendpunct}{\mcitedefaultseppunct}\relax
\EndOfBibitem
\bibitem[Baryshnikov \latin{et~al.}(2019)Baryshnikov, Valiev, Kuklin, Sundholm,
  and {\AA}gren]{baryshnikov_aagren_c18_jpcl_2019}
Baryshnikov,~G.~V.; Valiev,~R.~R.; Kuklin,~A.~V.; Sundholm,~D.; {\AA}gren,~H.
  {Cyclo[18]carbon: Insight into Electronic Structure, Aromaticity, and Surface
  Coupling}. \emph{J. Phys. Chem. Lett.} \textbf{2019}, \emph{10}, 6701\relax
\mciteBstWouldAddEndPuncttrue
\mciteSetBstMidEndSepPunct{\mcitedefaultmidpunct}
{\mcitedefaultendpunct}{\mcitedefaultseppunct}\relax
\EndOfBibitem
\bibitem[Stasyuk \latin{et~al.}(2020)Stasyuk, Stasyuk, Sol{\`a}, and
  Voityuk]{stasyuk_voityuk_c18_chem_commun_2020}
Stasyuk,~A.~J.; Stasyuk,~O.~A.; Sol{\`a},~M.; Voityuk,~A.~A. {Cyclo[18]carbon:
  The Smallest All-Carbon Electron Acceptor}. \emph{Chem. Commun.}
  \textbf{2020}, \emph{56}, 352\relax
\mciteBstWouldAddEndPuncttrue
\mciteSetBstMidEndSepPunct{\mcitedefaultmidpunct}
{\mcitedefaultendpunct}{\mcitedefaultseppunct}\relax
\EndOfBibitem
\bibitem[Raeber and Mazziotti(2020)Raeber, and
  Mazziotti]{mazziotti_c18_pccp_2020}
Raeber,~A.~E.; Mazziotti,~D.~A. {Non-Equilibrium Steady State Conductivity in
  Cyclo[18]carbon and Its Boron Nitride Analogue}. \emph{Phys. Chem. Chem. Phys.}
  \textbf{2020}, \emph{22}, 23998\relax
\mciteBstWouldAddEndPunctfalse
\mciteSetBstMidEndSepPunct{\mcitedefaultmidpunct}
{\mcitedefaultendpunct}{\mcitedefaultseppunct}\relax
\EndOfBibitem
\bibitem[Solomon \latin{et~al.}(2008)Solomon, Andrews, Hansen, Goldsmith,
  Wasielewski, Van~Duyne, and Ratner]{solomon_ratner_interference_jcp_2008}
Solomon,~G.~C.; Andrews,~D.~Q.; Hansen,~T.; Goldsmith,~R.~H.;
  Wasielewski,~M.~R.; Van~Duyne,~R.~P.; Ratner,~M.~A. {Understanding Quantum
  Interference in Coherent Molecular Conduction}. \emph{J. Chem. Phys.}
  \textbf{2008}, \emph{129}, 054701\relax
\mciteBstWouldAddEndPuncttrue
\mciteSetBstMidEndSepPunct{\mcitedefaultmidpunct}
{\mcitedefaultendpunct}{\mcitedefaultseppunct}\relax
\EndOfBibitem
\bibitem[Solomon \latin{et~al.}(2010)Solomon, Herrmann, Hansen, Mujica, and
  Ratner]{solomon_ratner_current_nat_chem_2010}
Solomon,~G.~C.; Herrmann,~C.; Hansen,~T.; Mujica,~V.; Ratner,~M.~A. {Exploring
  Local Currents in Molecular Junctions}. \emph{Nat. Chem.} \textbf{2010},
  \emph{2}, 223\relax
\mciteBstWouldAddEndPuncttrue
\mciteSetBstMidEndSepPunct{\mcitedefaultmidpunct}
{\mcitedefaultendpunct}{\mcitedefaultseppunct}\relax
\EndOfBibitem
\bibitem[Solomon(2015)]{solomon_interference_nat_chem_2015}
Solomon,~G.~C. {Interfering with Interference}. \emph{Nat. Chem.}
  \textbf{2015}, \emph{7}, 621\relax
\mciteBstWouldAddEndPuncttrue
\mciteSetBstMidEndSepPunct{\mcitedefaultmidpunct}
{\mcitedefaultendpunct}{\mcitedefaultseppunct}\relax
\EndOfBibitem
\bibitem[Gorczak \latin{et~al.}(2015)Gorczak, Renaud, Tarku{\c{c}}, Houtepen,
  Eelkema, Siebbeles, and Grozema]{grozema_interference_chem_sci_2015}
Gorczak,~N.; Renaud,~N.; Tarku{\c{c}},~S.; Houtepen,~A.~J.; Eelkema,~R.;
  Siebbeles,~L. D.~A.; Grozema,~F.~C. {Charge Transfer versus Molecular
  Conductance: Molecular Orbital Symmetry Turns Quantum Interference Rules
  Upside Down}. \emph{Chem. Sci.} \textbf{2015}, \emph{6}, 4196\relax
\mciteBstWouldAddEndPuncttrue
\mciteSetBstMidEndSepPunct{\mcitedefaultmidpunct}
{\mcitedefaultendpunct}{\mcitedefaultseppunct}\relax
\EndOfBibitem
\bibitem[Maggio \latin{et~al.}(2013)Maggio, Martsinovich, and
  Troisi]{troisi_orb_sym_acie_2013}
Maggio,~E.; Martsinovich,~N.; Troisi,~A. {Using Orbital Symmetry to Minimize
  Charge Recombination in Dye‐Sensitized Solar Cells}. \emph{Angew. Chem.
  Int. Ed.} \textbf{2013}, \emph{52}, 973\relax
\mciteBstWouldAddEndPuncttrue
\mciteSetBstMidEndSepPunct{\mcitedefaultmidpunct}
{\mcitedefaultendpunct}{\mcitedefaultseppunct}\relax
\EndOfBibitem
\bibitem[Hendon \latin{et~al.}(2013)Hendon, Tiana, Murray, Carbery, and
  Walsh]{walsh_helical_orbs_chem_sci_2013}
Hendon,~C.~H.; Tiana,~D.; Murray,~A.~T.; Carbery,~D.~R.; Walsh,~A. {Helical
  Frontier Orbitals of Conjugated Linear Molecules}. \emph{Chem. Sci.}
  \textbf{2013}, \emph{4}, 4278\relax
\mciteBstWouldAddEndPuncttrue
\mciteSetBstMidEndSepPunct{\mcitedefaultmidpunct}
{\mcitedefaultendpunct}{\mcitedefaultseppunct}\relax
\EndOfBibitem
\bibitem[Januszewskia and Tykwinski(2014)Januszewskia, and
  Tykwinski]{januszewskia_tykwinski_cumulenes_chem_soc_rev_2014}
Januszewskia,~J.~A.; Tykwinski,~R.~R. {Synthesis and Properties of Long
  $[n]$Cumulenes ($n \geq 5$)}. \emph{Chem. Soc. Rev.} \textbf{2014},
  \emph{43}, 3184\relax
\mciteBstWouldAddEndPuncttrue
\mciteSetBstMidEndSepPunct{\mcitedefaultmidpunct}
{\mcitedefaultendpunct}{\mcitedefaultseppunct}\relax
\EndOfBibitem
\bibitem[Tsuji \latin{et~al.}(2015)Tsuji, Movassagh, Datta, and
  Hoffmann]{hoffmann_bond_trans_acs_nano_2015}
Tsuji,~Y.; Movassagh,~R.; Datta,~S.; Hoffmann,~R. {Exponential Attenuation of
  Through-Bond Transmission in a Polyene: Theory and Potential Realizations}.
  \emph{ACS Nano} \textbf{2015}, \emph{9}, 11109\relax
\mciteBstWouldAddEndPuncttrue
\mciteSetBstMidEndSepPunct{\mcitedefaultmidpunct}
{\mcitedefaultendpunct}{\mcitedefaultseppunct}\relax
\EndOfBibitem
\bibitem[Tsuji \latin{et~al.}(2016)Tsuji, Hoffmann, Strange, and
  Solomon]{hoffmann_solomon_interference_pnas_2016}
Tsuji,~Y.; Hoffmann,~R.; Strange,~M.; Solomon,~G.~C. {Close Relation Between
  Quantum Interference in Molecular Conductance and Diradical Existence}.
  \emph{Proc. Nat. Acad. Sci.} \textbf{2016}, \emph{113}, E413\relax
\mciteBstWouldAddEndPuncttrue
\mciteSetBstMidEndSepPunct{\mcitedefaultmidpunct}
{\mcitedefaultendpunct}{\mcitedefaultseppunct}\relax
\EndOfBibitem
\bibitem[Garner \latin{et~al.}(2018)Garner, Bro-J{\o}rgensen, Pedersen, and
  Solomon]{solomon_elec_trans_jpcc_2018}
Garner,~M.~H.; Bro-J{\o}rgensen,~W.; Pedersen,~P.~D.; Solomon,~G.~C. {Reverse
  Bond-Length Alternation in Cumulenes: Candidates for Increasing Electronic
  Transmission with Length}. \emph{J. Phys. Chem. C} \textbf{2018}, \emph{122},
  26777\relax
\mciteBstWouldAddEndPuncttrue
\mciteSetBstMidEndSepPunct{\mcitedefaultmidpunct}
{\mcitedefaultendpunct}{\mcitedefaultseppunct}\relax
\EndOfBibitem
\bibitem[Jensen \latin{et~al.}(2019)Jensen, Garner, and
  Solomon]{solomon_current_dens_jpcc_2019}
Jensen,~A.; Garner,~M.~H.; Solomon,~G.~C. {When Current Does Not Follow Bonds:
  Current Density in Saturated Molecules}. \emph{J. Phys. Chem. C}
  \textbf{2019}, \emph{123}, 12042\relax
\mciteBstWouldAddEndPuncttrue
\mciteSetBstMidEndSepPunct{\mcitedefaultmidpunct}
{\mcitedefaultendpunct}{\mcitedefaultseppunct}\relax
\EndOfBibitem
\bibitem[Garner and Solomon(2020)Garner, and
  Solomon]{solomon_elec_trans_jpcl_2020}
Garner,~M.~H.; Solomon,~G.~C. {Simultaneous Suppression of $\pi$- and
  $\sigma$-Transmission in $\pi$-Conjugated Molecules}. \emph{J. Phys. Chem.
  Lett.} \textbf{2020}, \emph{11}, 7400\relax
\mciteBstWouldAddEndPuncttrue
\mciteSetBstMidEndSepPunct{\mcitedefaultmidpunct}
{\mcitedefaultendpunct}{\mcitedefaultseppunct}\relax
\EndOfBibitem
\end{mcitethebibliography}
\end{document}